\documentclass[12pt]{iopart}

\usepackage{iopams}
\usepackage{amssymb}
\usepackage{graphicx}
\usepackage{fancybox,color}

\usepackage[latin1]{inputenc} % Umalaute

\usepackage{amssymb}
\usepackage{latexsym}
\usepackage{amsfonts}
\usepackage{epsfig}
\usepackage{psfrag}
\usepackage{graphicx}

\newcommand{\ket}[1]{\left|#1\right>}
\newcommand{\bra}[1]{\left<#1\right|}   

\newcommand{\nn}{\nonumber\\}

\newcommand{\fk}[1]{\mbox{\boldmath$\scriptstyle#1$}}

\newcommand{\bea}{\begin{eqnarray}}
\newcommand{\ea}{\end{eqnarray}}
\newcommand{\eea}{\end{eqnarray}}
\newcommand{\ord}{\,{\cal O}}
\newcommand{\li}{\,{\cal L}}

\begin{document}

\title{Correlations in the Bose \& Fermi Hubbard Model}

\author{F Queisser$^1$, K V Krutitsky$^1$, P Navez$^{2,1}$ 
and R Sch{\"u}tzhold$^{1,*}$}

\address{$^1$Fakult{\"a}t f{\"u}r Physik, Universit{\"a}t Duisburg-Essen, 
Duisburg, Germany}

\address{$^2$Institut f\"ur Theoretische Physik, TU Dresden, 01062 Dresden, 
Germany}

\eads{$^*$ralf.schuetzhold@uni-due.de}

\date{\today}

\begin{abstract}
We study the Bose-Hubbard and Fermi-Hubbard model in the limit of large 
coordination numbers $Z$ (i.e., many tunnelling partners). 
Via a controlled expansion into powers of $1/Z$, we establish a 
hierarchy of correlations, which facilitates an approximate analytic 
solution of the quantum evolution.
For the Bose-Hubbard model, we derive the growth of phase coherence after 
a quench from the Mott to the superfluid phase.
For a quench within the Mott phase, we find that various local observables 
approach a quasi-equilibrium state after a finite period of time. 
However, this state is not thermal, i.e., real thermalisation -- 
if it occurs -- requires much longer time scales. 
For a tilted lattice in the Mott state, we calculate the tunnelling 
probability and find a remarkable analogy to the Sauter-Schwinger effect
(i.e., electron-positron pair creation out of the vacuum due to a strong 
electric field). 
These analytical results are compared to numerical simulations for finite 
lattices in one and two dimensions and we find qualitative agreement. 
Finally, we generalise these studies to the more involved case of the 
Fermi-Hubbard model. 
\end{abstract}


\maketitle

%%%%%%%%%%%%%%%%%%%%%%%%%%%%%%%%%%%%%%%%%%%%%%%%%%%%%%%%%%%%%%%%%%%%%%%%%%%%%%%
\section{Introduction}\label{section-Introduction}
%%%%%%%%%%%%%%%%%%%%%%%%%%%%%%%%%%%%%%%%%%%%%%%%%%%%%%%%%%%%%%%%%%%%%%%%%%%%%%%

The theoretical description of strongly interacting many-particle quantum 
systems in condensed matter physics is a difficult undertaking in general. 
Even in those cases where it is possible to derive a Hamiltonian which 
contains all relevant parts for a complete description of the many-particle 
system under consideration, a complete solution of the dynamics or an exact 
evaluation of the ground state is often out of reach. 

However, essential features such as phase transitions or quasi-particle 
excitations are already contained in drastically simplified models.
One of the most studied models which describes strongly interacting 
electrons in a solid is the Fermi-Hubbard model \cite{H63,H64a,H64b}.
Although it describes only the physics of a single energy band without 
long-ranged Coulomb-interactions, it is believed to exhibit various 
interesting phenomena such as the metal-insulator transition, 
anti-ferromagnetism, ferrimagnetism, ferromagnetism, Tomonaga-Luttinger 
liquid, and superconductivity \cite{T98}.
The Fermi-Hubbard model was first established by J.~Hubbard in order to 
describe correlations of electrons in narrow energy bands \cite{H63,H64a,H64b}.
It connects the Heitler-London theory of strongly interacting electrons on 
one side and the band theory, valid for weak interactions, on the other side.

Apart from electrons in solids, there are also other physical systems which 
can be described (within suitable approximations) by the Fermi-Hubbard model,
for example ultra-cold atoms in optical lattices. 
Since optical lattice systems allow for a tuning of the model parameters 
over a wide range (in contrast to most condensed matter systems), they are 
predestinated for the study of the fermionic Hubbard model.     
This has been experimentally realised by trapping fermionic Potassium atoms 
in an optical lattice \cite{KMS05,JSG08,SHW08}.
By increasing the on-site interaction among the atoms, the transition from 
a metallic phase to a Mott insulating phase was deduced from compressibility 
measurements and in situ imaging \cite{SHW08}.

After replacing the fermionic by bosonic atoms, the optical lattice system 
corresponds to the Bose-Hubbard model, which describes interacting bosons 
in periodic potentials.
The Bose-Hubbard model was first motivated by experimental realisations
such as Helium-4 absorbed in porous media or Cooper pairs in granular 
media \cite{FWGF89}.
Within the last decade, the Bose-Hubbard model has been attracting 
increasing attention due to its experimental realisation with interacting 
bosons, for example Rubidium atoms, in optical lattices \cite{G02,B10}.
The phase diagram, which is much better understood than in the fermionic 
case, contains (for vanishing disorder) a superfluid regime and Mott 
insulating phases.
The transition between the two phases is characterised by a natural order 
parameter, such as the superfluid density.
This second-order quantum phase transition, which results from the 
competition between kinetic energy and on-site interaction, has been 
observed for bosons in optical lattices \cite{G02}.

Though at first glance seemingly very simple, the fermionic as well as the 
bosonic lattice models are not exactly solvable in general, the only 
exceptions being the Fermi-Hubbard model in one dimension \cite{LW68,EFGKK05}
and trivial situations, such as the case of vanishing or infinitely strong 
interaction, or an empty lattice, for example.  
Thus, various analytical and numerical techniques have been developed 
to study their physical properties.
Widely used numerical methods, ignoring distance-dependent quantum 
correlations between different lattice sites, are mean-field approaches 
such as the dynamical mean-field theory (DMFT) \cite{AP98,GK96}. 
Exact diagonalisation (analytical or numerical) is only possible for small 
system sizes \cite{LHS88,KLA07}. 
Monte Carlo methods can improve the situation \cite{SBG91}, but they can 
sample only a small part of the whole Hilbert space and have problems with 
general time-dependent situations. 

An important and widely used analytical approach is the Gutzwiller
ansatz \cite{G63} -- for the bosonic and the fermionic case. 
However, a serious drawback of this Gutzwiller mean-field approach is again 
the neglect of distance-dependent quantum correlations between different 
lattice sites. 
This simplification leads sometimes to unphysical behaviour -- 
for example, the 
Mott-insulator state would not react to an external force within the 
Gutzwiller mean-field approach. 

The main goal of the present work is to study these non-local quantum 
correlations in the Bose and Fermi-Hubbard models.
To this end, we develop an analytical expansion in powers of the inverse
coordination number $Z$ (i.e., the number of tunnelling partners of a 
given lattice site). 
For comparison, we also calculated these non-local quantum correlations
by means of exact (numerical) diagonalisation for one and two-dimensional 
lattices.
Although exact diagonalisation is possible only for small systems, the 
results are in good agreement with those obtained by the analytical 
expansion and the Density Matrix Renormalisation Group technique (DMRG) 
\cite{CBPE12,E11}.

The paper is organised as follows.
After a brief introduction into the Bose-Hubbard model in 
Sec.~\ref{section-Bose-Hubbard-Model}, we present our analytical method 
for general Hubbard type Hamiltonians in Section~\ref{hierarchyofcorr}.
With this method, the ground state properties of the bosonic Mott 
insulator phase are derived in Sec.~\ref{mottsection}. 
Taking this Mott insulator as the initial state, the temporal evolution of 
the correlations after a quantum quench to the superfluid regime as well 
as within the Mott phase are studied in Sections~\ref{superquench} and 
\ref{equilibration}.
In Section~\ref{tiltlattice}, we investigate particle-hole pair creation 
out of the (bosonic) Mott state induced by a weak tilt of the lattice and 
establish a quantitative analogy to electron-positron pair creation due to 
a strong electric field (Sauter-Schwinger effect) in 
Sec.~\ref{section-Analogue}. 
The results of Secs.~\ref{mottsection}--\ref{section-Analogue} are derived 
in first order $1/Z$ (where $Z$ is the coordination number).
In Section~\ref{Z2}, we show how to extend our analytical method to higher 
orders in $1/Z$. 
Subsequently, we compare our analytical results with exact numerical studies 
of finite bosonic lattices in Section~\ref{numerical} and find qualitative 
agreement. 

In the second part of the paper, we consider similar problems for fermions.
After a brief introduction into the Fermi-Hubbard model in 
Section~\ref{Fermi-Hubbard Model}, we discuss its ground state correlations
(in the fermionic Mott state) and quench dynamics Secs.~\ref{chargemodes}
and \ref{Fermi-Mott}.
Section~\ref{fermitilt} is then devoted to particle-hole pair creation 
induced by a weak lattice tilt. 
Finally, we address resonant tunnelling in the Bose and Fermi-Hubbard 
model due to a large lattice tilt in Section \ref{restun}. 

%%%%%%%%%%%%%%%%%%%%%%%%%%%%%%%%%%%%%%%%%%%%%%%%%%%%%%%%%%%%%%%%%%%%%%%%%%%%%%%
\section{Bose-Hubbard Model}\label{section-Bose-Hubbard-Model}
%%%%%%%%%%%%%%%%%%%%%%%%%%%%%%%%%%%%%%%%%%%%%%%%%%%%%%%%%%%%%%%%%%%%%%%%%%%%%%%

The Bose-Hubbard model is one of the most simple and yet non-trivial models 
in condensed matter theory {\cite{FWGF89,J98,Z03}.
It describes identical bosons hopping on a lattice with the tunnelling 
rate $J$.
In addition, two (or more) bosons at the same lattice site repel each other 
with the interaction energy $U$.
The Hamiltonian reads 
\bea
\label{Bose-Hubbard-Hamiltonian}
\hat H
=
-\frac{J}{Z}\sum_{\mu\nu} T_{\mu\nu} \hat b^\dagger_\mu \hat b_\nu
+\frac{U}{2}\sum_{\mu} 
\hat n_{\mu}(\hat n_{\mu}-1)
\,.
\ea
Here $\hat b^\dagger_\mu$ and $\hat b_\nu$ are the creation and annihilation 
operators at the lattice sites $\mu$ and $\nu$, respectively, which obey 
the usual commutation relations 
\bea
\left[\hat b_\nu,\hat b_\mu^\dagger\right]=\delta_{\mu\nu}
\;,\,
\left[\hat b_\nu^\dagger,\hat b_\mu^\dagger\right]=
\left[\hat b_\nu,\hat b_\mu\right]=0
\,.
\ea
The lattice structure is encoded in the adjacency matrix $T_{\mu\nu}$ 
which equals unity if $\mu$ and $\nu$ are tunnelling neighbours 
(i.e., if a particle can hop from $\mu$ to $\nu$) and zero otherwise. 
The number of tunnelling neighbours at a given site $\mu$ yields the 
coordination number $Z=\sum_\nu T_{\mu\nu}$
(we assume a translationally invariant lattice). 
Finally, $\hat n_{\mu}=\hat b^\dagger_\mu \hat b_\mu$ is the number 
operator and we assume unit filling $\langle\hat n_{\mu}\rangle=1$ 
in the following. 
Note that the total particle number 
$\hat N=\sum_\mu \hat n_\mu$ is conserved $[\hat H,\hat N]=0$. 

The Bose-Hubbard model is considered {\cite{sachdev}}
one of the prototypical examples for a quantum phase transition. 
If the interaction term dominates $U\gg J$, the bosons are pinned to 
their lattice sites and we have the Mott insulator state 
\bea
\label{Mott-state}
\ket{\Psi_{\rm Mott}}_{J=0}
=
\bigotimes\limits_\mu\ket{1}_\mu
=
\prod\limits_\mu\hat b_\mu^\dagger\ket{0}
\;\leadsto\;
\hat H\ket{\Psi_{\rm Mott}}_{J=0}=0
\,,
\ea
which is fully localised. 
If the hopping rate dominates $U\ll J$, on the other hand, the particles 
can propagate freely across the lattice and become completely delocalised 
\bea
\label{superfluid-state}
\ket{\Psi_{\rm superfluid}}_{U=0}
=
\frac{1}{\sqrt{N!}}
\left(\hat b_{\fk{k}=0}^\dagger\right)^N\ket{0}
=
\frac{1}{\sqrt{N! N^N}}
\left(\sum_{\mu}\hat b_\mu^\dagger\right)^N\ket{0}
\,,
\ea
which is the superfluid phase. 
Obviously, the Mott state (\ref{Mott-state}) does not have any correlations, 
for example 
$\langle\hat b^\dagger_\mu \hat b_\nu\rangle_{\rm Mott}=\delta_{\mu\nu}$, 
whereas the superfluid state in (\ref{superfluid-state}) shows correlations 
across the whole lattice 
$\langle\hat b^\dagger_\mu \hat b_\nu\rangle_{\rm superfluid}=1$. 
Furthermore, the Mott insulator state is separated by a finite energy gap 
from the lowest excited state, while the superfluid state possesses 
sound-like modes with arbitrarily low energies 
(for an infinitely large lattice $N\uparrow\infty$). 
Finally, the Bose-Hubbard model can be realised experimentally 
(to a very good approximation) with ultra-cold atoms in optical lattices 
\cite{B05,S07,RSN97} and it was even possible to observe the aforementioned 
phase transition in these systems \cite{G02}.  

In spite of its simplicity, the Bose-Hubbard model 
(\ref{Bose-Hubbard-Hamiltonian}) cannot be solved analytically. 
Numerical simulations are limited to reduced sub-spaces or 
small systems sizes, see Section \ref{numerical} below.
Analytical approaches are based on suitable approximations.
In order to control the error of these approximations, they should be 
based on an expansion in term of some large or small control parameter. 
For the Bose-Hubbard model (\ref{Bose-Hubbard-Hamiltonian}), one could 
consider the limit of large $\langle\hat n_{\mu}\rangle\gg1$  
or small $\langle\hat n_{\mu}\rangle\ll1$ filling \cite{SUXF06,FSU08},
for example, or the limit of weak coupling $U\ll J$ or strong coupling 
$U\gg J$ \cite{FM94,FM96,DZ06}. 
However, none of these limits is particularly well suited for studying 
the Mott--superfluid phase transition. 
To this end, we consider the limit $Z\gg1$ in the following and employ 
an expansion into powers of $1/Z$ as small control parameter. 
Note that an expansion in powers of $1/Z$ was also used to derive bosonic 
dynamical mean-field equations (which were then solved numerically) in  
\cite{HSH09,LBHH11,LBHH12}.

%%%%%%%%%%%%%%%%%%%%%%%%%%%%%%%%%%%%%%%%%%%%%%%%%%%%%%%%%%%%%%%%%%%%%%%%%%%%%%%
\section{Hierarchy of Correlations}\label{hierarchyofcorr}
%%%%%%%%%%%%%%%%%%%%%%%%%%%%%%%%%%%%%%%%%%%%%%%%%%%%%%%%%%%%%%%%%%%%%%%%%%%%%%%

Let us consider general Hamiltonians of the form 
\bea
\hat H=\frac1Z\sum_{\mu\nu}\hat H_{\mu \nu}+\sum_\mu\hat H_\mu
\,,
\ea
which includes the Bose-Hubbard model (\ref{Bose-Hubbard-Hamiltonian})
as a special case. 
The quantum evolution of the density operator $\hat\rho$ describing the 
state of the full lattice can be written as 
\bea
\label{Liouville}
i\partial_t\hat\rho 
= \left[\hat H,\hat\rho\right] 
&=&
\frac1Z\sum_{\mu\nu}\left[\hat H_{\mu \nu},\hat\rho\right] 
+\sum_\mu\left[\hat H_\mu,\hat\rho\right] 
\nn
&=& 
\frac1Z\sum_{\mu\nu}\li_{\mu \nu}\hat\rho 
+
\sum_\mu\li_\mu\hat\rho
\,,
\ea
where we have introduced the Liouville super-operators $\li_{\mu \nu}$
and $\li_\mu$ as short-hand notation.
As the next step, we introduce the reduced density matrices for one 
or more lattice sites via averaging (tracing) over all other sites
\bea
\label{reduced-density-matrices}
\hat\rho_\mu 
&=&
\tr_{\not\mu}\{\hat\rho\}
\nn
\hat\rho_{\mu\nu}
&=&
\tr_{\not\mu\not\nu}\{\hat\rho\}
\,,
\ea
and so on. 
Note that $\tr\{\hat\rho\}=1$ implies $\tr_\mu\{\hat\rho_\mu\}=1$
and $\tr_{\mu\nu}\{\hat\rho_{\mu\nu}\}=1$ etc. 
Since we are interested in the (quantum) correlations, we separate the 
correlated and uncorrelated parts of the reduced density matrices via  
\bea
\label{correlated-parts}
\hat\rho_{\mu\nu}
&=&
\hat\rho_{\mu\nu}^{\rm corr}+\hat\rho_{\mu}\hat\rho_{\nu}
\nn
\hat\rho_{\mu\nu\lambda}
&=&
\hat\rho_{\mu\nu\lambda}^{\rm corr}+
\hat\rho_{\mu\nu}^{\rm corr}\hat\rho_{\lambda}+
\hat\rho_{\mu\lambda}^{\rm corr}\hat\rho_{\nu}+
\hat\rho_{\nu\lambda}^{\rm corr}\hat\rho_{\mu}+
\hat\rho_{\mu}\hat\rho_{\nu}\hat\rho_{\lambda}
\,,
\ea
and analogously for more lattice sites. 
As a consequence, we obtain from (\ref{Liouville}) the evolution 
equation for the on-site density matrix 
\bea
\label{one-site}
i\partial_t\hat\rho_{\mu}
=
\frac{1}{Z}
\sum_{\kappa\neq\mu}\tr_{\kappa}\left\{
\li^S_{\mu \kappa}
(\hat\rho^{\rm corr}_{\mu \kappa}+\hat\rho_\mu \hat\rho_\kappa)\right\}
+
\li_\mu\hat\rho_{\mu}
\,,
\eea
where $\li_{\mu \nu}^S=\li_{\mu \nu}+\li_{\nu \mu}$ denotes the 
symmetrised form. 
Obviously, solving this equation exactly requires knowledge of the 
two-point correlation $\hat\rho^{\rm corr}_{\mu \kappa}$.
The time-evolution of this quantity can also obtained from 
(\ref{Liouville}) and reads 
\bea
\label{two-sites}
i \partial_t \hat\rho^{\rm corr}_{\mu \nu}
&=&
\li_\mu\hat\rho^{\rm corr}_{\mu\nu}
+
\frac1Z\li_{\mu\nu}
(\hat\rho^{\rm corr}_{\mu\nu}+\hat\rho_\mu\hat\rho_\nu)
-
\frac{\hat\rho_{\mu}}{Z}
\tr_{\mu}
\left\{\li^S_{\mu\nu}
(\hat\rho^{\rm corr}_{\mu\nu}+\hat\rho_\mu\hat\rho_\nu)
\right\}
\nn
&&+
\frac1Z
\sum_{\kappa\not=\mu,\nu} 
\tr_{\kappa}
\left\{
\li^S_{\mu \kappa}
(\hat\rho^{\rm corr}_{\mu\nu\kappa}+
\hat\rho^{\rm corr}_{\mu\nu}\hat\rho_{\kappa}+
\hat\rho^{\rm corr}_{\nu\kappa}\hat\rho_{\mu})
\right\}
+(\mu\leftrightarrow\nu)
\,.
\eea
As one would expect, this equation contains the three-point 
correlator $\hat\rho^{\rm corr}_{\mu\nu\kappa}$, and similarly 
the evolution equation for $\hat\rho^{\rm corr}_{\mu\nu\kappa\lambda}$
contains the four-point correlator etc. 
Consequently, one can never exactly solve this set of equations, 
truncated at any finite order. 

However, the limit $Z\gg1$ facilitates an approximate solution:
Let us imagine that we start from an initial state 
$\hat\rho^{\rm in}=\bigotimes_\mu\hat\rho^{\rm in}_\mu$
without any correlations 
(i.e., $\hat\rho^{\rm corr}_{\mu \nu}=0$ and 
$\hat\rho^{\rm corr}_{\mu\nu\kappa}=0$, etc.)
such as the Mott state (\ref{Mott-state}).
In this case, the right-hand side of (\ref{two-sites}) 
is suppressed by $\ord(1/Z)$ and thus the time evolution 
creates only small correlations 
$i\partial_t\hat\rho^{\rm corr}_{\mu \nu}$. 
Moreover, if these correlations are small initially 
$\hat\rho^{\rm corr}_{\mu \nu}=\ord(1/Z)$, they remain small 
-- at least for a finite amount of time -- because there is no 
term in (\ref{two-sites}) to compensate the $\ord(1/Z)$ suppression.
Note that the sum $\sum_\kappa$ in (\ref{two-sites}) might scale with $Z$, 
but this is compensated by the $1/Z$ factor in front of it. 
On the other hand, if we insert $\hat\rho^{\rm corr}_{\mu \nu}=\ord(1/Z)$ 
into (\ref{one-site}), we find that we can neglect this term and arrive at 
an approximate equation containing on-site density matrices only 
\bea
\label{one-site-approx}
i\partial_t\hat\rho_{\mu}
&=&
\frac{1}{Z}
\sum_{\kappa\neq\mu}\tr_{\kappa}\left\{
\li^S_{\mu \kappa}
\hat\rho_\mu \hat\rho_\kappa\right\}
+
\li_\mu\hat\rho_{\mu}
+\ord(1/Z)
\nn
&\approx&
\frac{1}{Z}
\sum_{\kappa\neq\mu}\tr_{\kappa}\left\{
\li^S_{\mu \kappa}
\hat\rho_\mu \hat\rho_\kappa\right\}
+
\li_\mu\hat\rho_{\mu}
\,,
\eea
The approximate solution $\hat\rho_\mu^0$ of this self-consistent equation 
is valid to lowest order in $1/Z$, i.e., 
$\hat\rho_\mu=\hat\rho_\mu^0+\ord(1/Z)$ 
and reproduces the well-known Gutzwiller ansatz \cite{G63,J98,RK91}.
If we now insert this approximate solution $\hat\rho_\mu^0$ into 
(\ref{two-sites}), we get an approximate evolution equation for 
the two-point correlator 
\bea
\label{two-sites-approx}
i \partial_t \hat\rho^{\rm corr}_{\mu \nu}
&=&
\li_\mu\hat\rho^{\rm corr}_{\mu\nu}
+
\frac1Z\li_{\mu\nu}\hat\rho^0_\mu\hat\rho^0_\nu
+
\frac1Z
\sum_{\kappa\not=\mu,\nu} 
\tr_{\kappa}
\left\{
\li^S_{\mu \kappa}
(\hat\rho^{\rm corr}_{\mu\nu}\hat\rho^0_{\kappa}+
\hat\rho^{\rm corr}_{\nu\kappa}\hat\rho^0_{\mu})
\right\}
\nn
&&-
\frac{\hat\rho^0_{\mu}}{Z}
\tr_{\mu}
\left\{\li^S_{\mu\nu}\hat\rho^0_\mu\hat\rho^0_\nu\right\}
+(\mu\leftrightarrow\nu)
+\ord(1/Z^2)
\,.
\eea
Note that we have assumed that the three-point correlations 
$\hat\rho^{\rm corr}_{\mu\nu\kappa}$ do not spoil this line of 
arguments and are suppressed by $\ord(1/Z^2)$ in complete analogy.  
This is indeed correct and can be shown in basically the same way, 
see Appendix \ref{hierarchyApp}.
More generally, we find that $\ell$-point correlations are suppressed 
as $\ord(1/Z^{\ell-1})$, i.e., 
\bea
\label{hierarchy}
\hat\rho_{\mu} &=& \ord\left(Z^0\right)
\nn
\hat\rho^{\rm corr}_{\mu\nu} &=& \ord\left(1/Z\right)
\nn
\hat\rho^{\rm corr}_{\mu\nu\kappa} &=& \ord\left(1/Z^2\right)
\nn
\hat\rho^{\rm corr}_{\mu\nu\kappa\lambda} &=& \ord\left(1/Z^3\right)
\,,
\ea
and so on, see Appendix \ref{hierarchyApp}. 
This hierarchy (\ref{hierarchy}) is related to 
the quantum de~Finetti theorem \cite{CKMR07}, 
the generalised cumulant expansion \cite{K62}, 
and the Bogoliubov-Born-Green-Kirkwood-Yvon (BBGKY)
hierarchy \cite{B75}, 
but we are considering lattice sites instead of particles. 
As an example for the four-point correlator, let us consider observables 
$\hat A_\mu$, $\hat B_\nu$, $\hat C_\kappa$, and $\hat D_\lambda$ at 
four different lattice sites, which have vanishing on-site expectation values 
$\langle\hat A_\mu\rangle=
\langle\hat B_\nu\rangle=
\langle\hat C_\kappa\rangle=
\langle\hat D_\lambda\rangle=0$. 
In this case, the hierarchy (\ref{hierarchy}) implies 
\bea
\langle\hat A_\mu\hat B_\nu\hat C_\kappa\hat D_\lambda\rangle 
&=&
\langle\hat A_\mu\hat B_\nu\rangle 
\langle\hat C_\kappa\hat D_\lambda\rangle 
+
\langle\hat A_\mu\hat C_\kappa\rangle 
\langle\hat B_\nu\hat D_\lambda\rangle 
+
\langle\hat A_\mu\hat D_\lambda\rangle 
\langle\hat B_\nu\hat C_\kappa\rangle 
\nn
&&
+
\ord\left(1/Z^3\right)
\,,
\ea
which resembles the Wick theorem in free quantum field theory
(even though the quantum system considered here is strongly interacting). 

%%%%%%%%%%%%%%%%%%%%%%%%%%%%%%%%%%%%%%%%%%%%%%%%%%%%%%%%%%%%%%%%%%%%%%%%%%%%%%%
\section{Mott Insulator State}\label{mottsection}
%%%%%%%%%%%%%%%%%%%%%%%%%%%%%%%%%%%%%%%%%%%%%%%%%%%%%%%%%%%%%%%%%%%%%%%%%%%%%%%

Now let us apply the hierarchy discussed above to the Bose-Hubbard model 
(\ref{Bose-Hubbard-Hamiltonian}).
To this end, we start with the factorising Mott state (\ref{Mott-state})
at zero hopping rate $J=0$ as our initial state 
\bea
\hat\rho^{\rm in}
=
\bigotimes\limits_\mu\hat\rho^{\rm in}_\mu
=
\bigotimes\limits_\mu\ket{1}_\mu\!\bra{1}
\,.
\ea
Then we slowly switch on the hopping rate $J(t)$ until we reach its 
final value. 
In view of the finite energy gap, the adiabatic theorem implies that 
we stay very close to the real ground state of the system if we do 
this slowly enough. 
Of course, we cannot cross the phase transition in this way 
(i.e., adiabatically) since the energy gap vanishes at the critical point, 
see Section \ref{superquench} below. 

Since we have $\langle\hat b_\mu\rangle=0$ in the Mott state, 
Eq.~(\ref{one-site-approx}) simplifies to 
\bea
\label{one-site-Mott}
i\partial_t\hat\rho_{\mu}
\approx
\frac{1}{Z}
\sum_{\kappa\neq\mu}\tr_{\kappa}\left\{
\li^S_{\mu \kappa}
\hat\rho_\mu \hat\rho_\kappa\right\}
+
\li_\mu\hat\rho_{\mu}
=0 
\,\leadsto\,
\hat\rho_{\mu}^0=\ket{1}_\mu\!\bra{1}
\,.
\eea
Thus, to zeroth order in $1/Z$ (i.e., on the Gutzwiller mean-field level),
the Mott insulator state $\hat\rho_{\mu}^0$ for finite $J$ has the same 
form as for $J=0$.
To obtain the first order in $1/Z$, we insert this result into 
(\ref{two-sites-approx}). 
Again using $\langle\hat b_\mu\rangle=0$, we find 
\bea
i \partial_t \hat\rho^{\rm corr}_{\mu \nu}
&=&
\left(\li_\mu+\li_\nu\right)\hat\rho^{\rm corr}_{\mu\nu}
+
\frac1Z\li_{\mu\nu}^S\hat\rho_\mu^0\hat\rho_\nu^0
\nn
&&+
\frac1Z
\sum_{\kappa\not=\mu,\nu} 
\tr_{\kappa}
\left\{
\li^S_{\mu\kappa}\hat\rho^{\rm corr}_{\nu\kappa}\hat\rho_{\mu}^0
+\li^S_{\nu\kappa}\hat\rho^{\rm corr}_{\mu\kappa}\hat\rho_{\nu}^0
\right\}
+\ord(1/Z^2)
\,.
\eea
Formally, this is an evolution equation for an infinite dimensional 
matrix $\hat\rho^{\rm corr}_{\mu \nu}$.
Fortunately, however, it suffices to consider a few elements only. 
If we introduce 
$\hat p_\mu=\ket{1}_\mu\!\bra{2}$ 
and 
$\hat h_\mu=\ket{0}_\mu\!\bra{1}$ 
as local particle and hole 
operators\footnote{These excitations are sometimes \cite{CBPE12,BPCK12,KJSW90} 
called doublons and holons.}, 
all the interesting physics can 
be captured by their correlation functions (for $\mu\neq\nu$)
\begin{eqnarray}
f_{\mu\nu}^{11}
&=&
\langle\hat{h}^\dagger_\mu\hat{h}_\nu^{} \rangle 
=
\tr\left\{\hat{\rho}\,\hat{h}^\dagger_\mu\hat{h}_\nu^{}\right\}
=
\tr_{\mu\nu}\left\{
\hat{\rho}^{\rm corr}_{\mu\nu}\hat{h}^\dagger_\mu\hat{h}_\nu^{}\right\}
\,,
\nn
f_{\mu\nu}^{12}
&=&
\langle\hat{h}^\dagger_\mu\hat{p}_\nu^{} \rangle 
=
\tr\left\{\hat{\rho}\,\hat{h}^\dagger_\mu\hat{p}_\nu^{}\right\}
=
\tr_{\mu\nu}\left\{
\hat\rho^{\rm corr}_{\mu\nu}\hat{h}^\dagger_\mu\hat{p}_\nu^{}\right\}
\,,
\nn
f_{\mu\nu}^{21}
&=&
\langle\hat{p}^\dagger_\mu\hat{h}_\nu^{} \rangle 
=
\tr\left\{\hat{\rho}\,\hat{p}^\dagger_\mu\hat{h}_\nu^{}\right\}
=
\tr_{\mu\nu}\left\{
\hat{\rho}^{\rm corr}_{\mu\nu}\hat{p}^\dagger_\mu\hat{h}_\nu^{}\right\}
\,,
\nn
f_{\mu\nu}^{22}
&=&
\langle\hat{p}^\dagger_\mu\hat{p}_\nu^{} \rangle 
=
\tr\left\{\hat{\rho}\,\hat{p}^\dagger_\mu\hat{p}_\nu^{}\right\}
=
\tr_{\mu\nu}\left\{
\hat{\rho}^{\rm corr}_{\mu\nu}\hat{p}^\dagger_\mu\hat{p}_\nu^{}\right\}
\,.
\end{eqnarray}
To first order in $1/Z$, these correlation functions form a closed set 
of equations 
\begin{eqnarray}
\label{f12-Mott}
i\partial_t f^{12}_{\mu\nu}
&=&
-\frac{J}{Z}\sum_{\kappa\neq \mu,\nu}
\left(T_{\mu\kappa}(f^{12}_{\kappa\nu}+\sqrt{2}f^{22}_{\kappa\nu})
+
\sqrt{2}T_{\nu\kappa}(f^{11}_{\mu\kappa}+\sqrt{2}f_{\mu\kappa}^{12})\right)
\nn
&&+Uf^{12}_{\mu\nu}-\frac{J\sqrt{2}}{Z}T_{\mu\nu}
\\
i\partial_t f^{21}_{\mu\nu}
&=&
+\frac{J}{Z}\sum_{\kappa\neq \mu,\nu}
\left(T_{\nu\kappa}(f^{21}_{\kappa\mu}+\sqrt{2}f^{11}_{\kappa\mu})
+
\sqrt{2}T_{\mu\kappa}(f^{22}_{\kappa\nu}+\sqrt{2}f_{\kappa\nu}^{12})\right)
\nn
&&-Uf^{21}_{\mu\nu}+\frac{J\sqrt{2}}{Z}T_{\mu\nu}
\\
\label{f11-Mott}
i\partial_t f^{11}_{\mu\nu}
&=&
i\partial_t f^{22}_{\mu\nu}
=
-\frac{\sqrt{2}J}{Z}\sum_{\kappa\neq\mu,\nu}\left(T_{\mu\kappa}
f_{\kappa\nu}^{21}-T_{\nu\kappa}f_{\mu\kappa}^{12}\right)
\,.
\end{eqnarray}
This truncation is due to the fact that the correlation functions 
$f^{mn}_{\mu\nu}$ involving higher occupation numbers $m\geq3$ or $n\geq3$
do not have any source terms of order $1/Z$ and hence do not contribute
at that level. 
Exploiting translational symmetry, we may simplify these equations by a 
spatial Fourier transformation with 
\begin{eqnarray}
\label{T_k}
T_{\mu\nu}
&=&
\frac{Z}{N}\sum_\mathbf{k}T_{\mathbf{k}}
e^{i\mathbf{k}\cdot(\mathbf{x}_\mu-\mathbf{x}_\nu)}
\\
\label{f_k}
f^{ab}_{\mu\nu}
&=&
\frac{1}{N}\sum_\mathbf{k}f^{ab}_{\mathbf{k}}
e^{i\mathbf{k}\cdot(\mathbf{x}_\mu-\mathbf{x}_\nu)}\,,
\end{eqnarray}
where $N$ denotes the number of lattice sites 
(which equals the number of particles in our case).
Formally, in order to Fourier transform equations 
(\ref{f12-Mott}-\ref{f11-Mott}), one should add the summands  
corresponding to $\kappa=\mu$ and $\kappa=\nu$.
Since these terms are of order $1/Z^2$, they do not spoil our first-order 
analysis.  
However, when going to second order $1/Z^2$ (see Section~\ref{Z2} below), 
they have to be taken into account.

After the Fourier transformation (\ref{T_k}) and (\ref{f_k}),
Eqs.~(\ref{f12-Mott}-\ref{f11-Mott}) become 
\begin{eqnarray}
(i\partial_t-U+3 J T_\mathbf{k})f_\mathbf{k}^{12}
&=&
-\sqrt{2}J T_\mathbf{k}(f_\mathbf{k}^{11}+f_\mathbf{k}^{22}+1)
\,,\label{diff0}
\\
(i\partial_t+U-3 J T_\mathbf{k})f_\mathbf{k}^{21}
&=&
+\sqrt{2}J T_\mathbf{k}(f_\mathbf{k}^{11}+f_\mathbf{k}^{22}+1)
\,,\label{diff1}
\\
i\partial_t f_\mathbf{k}^{11}=i\partial_t f_\mathbf{k}^{22}
&=&
\sqrt{2}JT_\mathbf{k}(f^{12}_\mathbf{k}-f^{21}_\mathbf{k})
\label{diff2}
\,.
\end{eqnarray}
From the last equation, we may infer an effective particle-hole symmetry 
$f_\mathbf{k}^{11}=f_\mathbf{k}^{22}$ valid to first order in $1/Z$.
With this symmetry, any stationary state (such as the ground state) 
with $\partial_tf_\mathbf{k}^{ab}=0$ must obey the condition 
\begin{eqnarray}
\label{stat}
f_\mathbf{k}^{12}=f_\mathbf{k}^{21}
=
\frac{\sqrt{2}JT_\mathbf{k}(2f_\mathbf{k}^{11}+1)}{U-3 JT_\mathbf{k}}
\,.
\end{eqnarray}
The remaining unknown quantity $f_\mathbf{k}^{11}$ can be obtained in the 
following way:
The evolution equations (\ref{diff0}-\ref{diff2}) leave the following 
bilinear quantity invariant 
\begin{eqnarray}\label{inv}
\partial_t
\left[
f_\mathbf{k}^{11}(f_\mathbf{k}^{11}+1)-f_\mathbf{k}^{12}f_\mathbf{k}^{21}
\right] 
=0
\,,
\end{eqnarray}
which remains valid even for time-dependent $J(t)$.
Thus, starting in the Mott state (\ref{Mott-state}) at zero hopping rate 
$J=0$ with vanishing correlations $f_\mathbf{k}^{ab}(t=0)=0$, we get the 
additional condition 
$f_\mathbf{k}^{11}(f_\mathbf{k}^{11}+1)=f_\mathbf{k}^{12}f_\mathbf{k}^{21}$
for all times $t>0$. 
Thus, to first order in $1/Z$, the ground state correlations read 
(for $\mu\neq\nu$)
\begin{eqnarray}
\langle\hat{h}^\dagger_\mu\hat{h}_\nu^{} \rangle_{\rm ground} 
=
\langle\hat{p}^\dagger_\mu\hat{p}_\nu^{} \rangle_{\rm ground} 
%=
%f^{11}_{\mu\nu,\mathrm{stat}}
&=&
\frac{1}{N}\sum_\mathbf{k}
\frac{U-3JT_\mathbf{k}-\omega_\mathbf{k}}{2\omega_\mathbf{k}}
e^{i\mathbf{k}\cdot(\mathbf{x}_\mu-\mathbf{x}_\nu)}
\label{ground1}
\\
\langle\hat{h}^\dagger_\mu\hat{p}_\nu^{} \rangle_{\rm ground} 
=
\langle\hat{p}^\dagger_\mu\hat{h}_\nu^{} \rangle_{\rm ground} 
%
%f^{12}_{\mu\nu,\mathrm{stat}}
&=&
%f^{21}_{\mu\nu,\mathrm{stat}}
%=
\frac{1}{N}\sum_\mathbf{k}\frac{\sqrt{2}JT_\mathbf{k}}{\omega_\mathbf{k}}
e^{i\mathbf{k}\cdot(\mathbf{x}_\mu-\mathbf{x}_\nu)}
\label{ground2}
\,.
\end{eqnarray}
Here we have used the abbreviation {\cite{KN11}}
\begin{eqnarray}
\label{eigen-frequency}
\omega_\mathbf{k}&=&\sqrt{U^2-6 J UT_\mathbf{k}+J^2 T_\mathbf{k}^2}
\,,
\end{eqnarray}
which is just the (non-zero) eigenfrequency of the evolution equations 
(\ref{diff0}-\ref{diff2}) for non-stationary states and will become 
important in the next Section. 

The above equations (\ref{ground1}) and (\ref{ground2}) describe the 
correlations and are valid for $\mu\neq\nu$ only. 
The correct on-site density matrix $\rho_\mu$ can be obtained from 
(\ref{one-site}) which shows that non-vanishing correlations lead 
to small deviations from the lowest-order result $\rho_\mu^0$.
As one would expect, the quantum ground-state fluctuations manifest 
themselves in a small depletion of the unit-filling state 
$\hat\rho_{\mu}^0=\ket{1}_\mu\!\bra{1}$ given by a small but finite 
probability for a particle 
$f_2=\tr\{\hat{\rho}_\mu|2\rangle_\mu\langle 2|\}$ 
or a hole 
$f_0=\tr\{\hat{\rho}_\mu|0\rangle_\mu\langle 0|\}$.
To first order in $1/Z$, we get from (\ref{one-site}) 
\begin{eqnarray}
\label{diff3}
i\partial_t f_0
=
i\partial_t f_2
=
\sum_\mathbf{k}\frac{\sqrt{2}J T_\mathbf{k}}{N}
(f_\mathbf{k}^{12}-f_\mathbf{k}^{21})
=
i \frac{1}{N}\sum_\mathbf{k} \partial_tf_\mathbf{k}^{11}
\,,
\end{eqnarray}
where we used equation (\ref{diff2}) in the last step.
This equation can be integrated easily and with the initial conditions 
$f_0(t=0)=f_2(t=0)=0$ we find the $1/Z$-corrections to the one-site 
density matrix
\begin{eqnarray}
\label{depletion}
\langle\hat{p}^\dagger_\mu\hat{p}_\mu^{}\rangle 
=
\langle\hat{h}_\mu^{}\hat{h}^\dagger_\mu\rangle 
=
f_0=f_2
=
\frac{1}{N}\sum_\mathbf{k}
\frac{U-3JT_\mathbf{k}-\omega_\mathbf{k}}{2\omega_\mathbf{k}}
%
%\frac{\sqrt{2}JT_\mathbf{k}}{\omega_\mathbf{k}}%
%
\,.
\end{eqnarray}
Note that, even though the right-hand side of the above equation looks 
like that of (\ref{ground1}) for $\mu=\nu$, one should be careful as they 
are derived from two different equations: (\ref{one-site}) and 
(\ref{two-sites}).
%

%%%%%%%%%%%%%%%%%%%%%%%%%%%%%%%%%%%%%%%%%%%%%%%%%%%%%%%%%%%%%%%%%%%%%%%%%%%%%%%
\section{Mott--Superfluid Quench}\label{superquench}
%%%%%%%%%%%%%%%%%%%%%%%%%%%%%%%%%%%%%%%%%%%%%%%%%%%%%%%%%%%%%%%%%%%%%%%%%%%%%%%

After having studied the ground state properties of the Mott phase, 
let us consider a quantum quench from the Mott state to the superfluid 
regime. 
This requires a time-dependent solution of the evolution equations 
(\ref{diff0}-\ref{diff2}) which crucially depends on the eigenfrequency 
(\ref{eigen-frequency}).
%
% \bea
% \omega_\mathbf{k}^2=U^2-6 J UT_\mathbf{k}+J^2 T_\mathbf{k}^2 
% \,.
% \ea
%
In view of the definition (\ref{T_k}), $T_{\mathbf{k}}$ adopts its 
maximum value $T_{\mathbf{k}=0}=1$ at $\mathbf{k}=0$.
Thus $\omega_{\mathbf{k}=0}=\Delta\mathcal E$ corresponds to the 
energy gap of the Mott state mentioned in 
Section~\ref{section-Bose-Hubbard-Model}. 
For $J=0$, we have a flat dispersion relation $\omega_\mathbf{k}=U$.
If we increase $J$, the dispersion relation $\omega_\mathbf{k}$ 
bends down and the minimum at $\mathbf{k}=0$ approaches the axis. 
Finally, at a critical value of the hopping rate \cite{NS10}
\bea
J_{\rm crit}=U(3-\sqrt{8})
\,,
\ea
the minimum $\omega_{\mathbf{k}=0}$ touches the axis and thus 
the energy gap vanishes $\Delta\mathcal E=0$.
This marks the transition to the superfluid regime and we cannot 
analytically or adiabatically continue beyond this point. 
However, nothing stops us from suddenly switching $J$ to a final 
value $J_{\rm out}>J_{\rm crit}$ beyond this point. 
Of course, this would not be adiabatic anymore and we would 
no longer be close to the ground state.
For hopping rates $J$ which are a bit larger than the critical 
value $J>J_{\rm crit}$, the dispersion relation dives below the 
axis and the $\omega_\mathbf{k}^2$ become negative for small $\mathbf{k}$.
Thus, the eigenfrequencies $\omega_\mathbf{k}$ become imaginary 
indicating an exponential growth of these modes, i.e., an instability. 
This is very natural since the quantum system ``feels'' that the Mott 
state is no longer the correct ground state. 

If we consider even larger $J$, we find that the original minimum 
of the dispersion relation $\omega_\mathbf{k}^2$ at $\mathbf{k}=0$
splits into degenerate minima at finite values of $\mathbf{k}$ 
when $J=3U$, while $\mathbf{k}=0$ becomes a local maximum. 
This local maximum even emerges $\omega_{\mathbf{k}=0}^2>0$ 
on the positive side again for $J>U(3+\sqrt{8})$.
Nevertheless, there are always 
unstable modes for some values of $\mathbf{k}$, 
see Fig.~\ref{fig-omega} and compare \cite{S11}.

\bigskip

\begin{figure}[h]
\begin{center}
\includegraphics[width=.49\columnwidth]{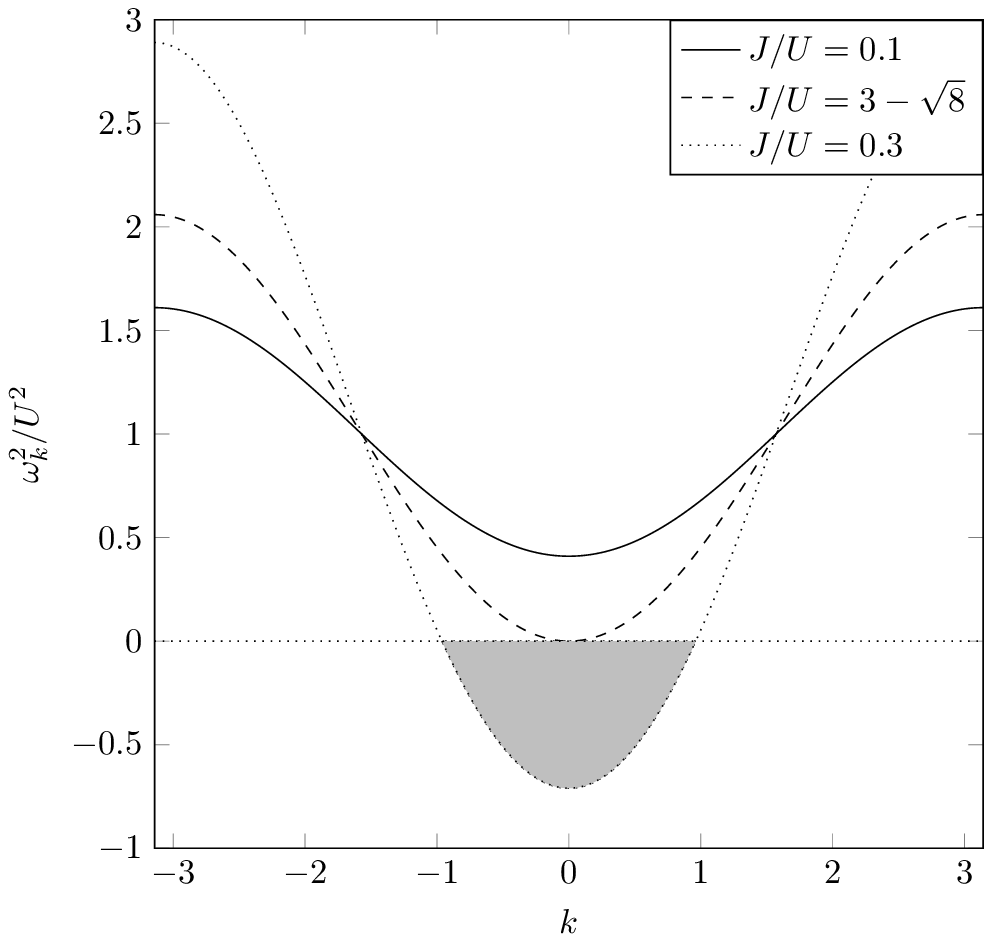}
\includegraphics[width=.49\columnwidth]{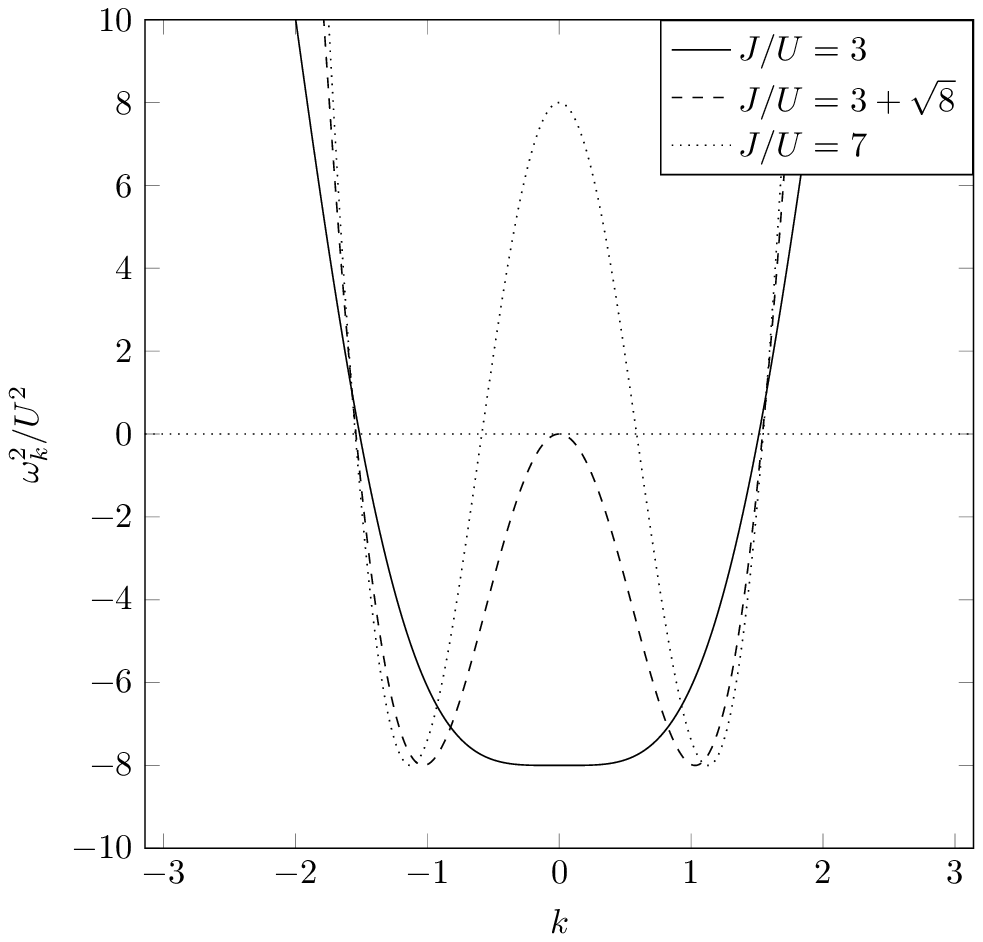}
\end{center}
\caption{Dispersion relation $\omega_k^2/U^2$ in one dimension for 
different values of $J/U$.}
\label{fig-omega}  
\end{figure}
\bigskip

After these preliminaries, let us study a quantum quench from the 
Mott state to the superfluid phase.
For simplicity, we consider a sudden change of $J(t)=J\Theta(t)$ 
from $J=0$ to the final value of $J$ 
(but the calculation can easily be generalised to other scenarios). 
Solving the evolution equations (\ref{diff0}-\ref{diff2}) for this case, 
we find 
\begin{eqnarray}
\label{quench-h+h}
\langle\hat{h}^\dagger_\mu\hat{h}_\nu^{} \rangle_{\rm quench} 
&=&
\langle\hat{p}^\dagger_\mu\hat{p}_\nu^{} \rangle_{\rm quench} 
%=
%f^{11}_{\mu\nu,\mathrm{Q}}
=
\frac{1}{N}\sum_\mathbf{k}
4J^2 T_\mathbf{k}^2\,
\frac{1-\cos(\omega_\mathbf{k}t)}{\omega^2_\mathbf{k}}\,
e^{i\mathbf{k}\cdot(\mathbf{x}_\mu-\mathbf{x}_\nu)}
\\
\label{quench-h+p}
\langle\hat{h}^\dagger_\mu\hat{p}_\nu^{} \rangle_{\rm quench} 
%=
%f^{12}_{\mu\nu,\mathrm{Q}}
&=&
\frac{1}{N}\sum_\mathbf{k}
\sqrt{2}JT_\mathbf{k}(U-3JT_\mathbf{k})
\frac{1-\cos(\omega_\mathbf{k}t)}{\omega_\mathbf{k}^2}\,
e^{i\mathbf{k}\cdot(\mathbf{x}_\mu-\mathbf{x}_\nu)}
\nn
&&+
\frac{i}{N}\sum_\mathbf{k}
\sqrt{2}JT_\mathbf{k}\,
\frac{\sin(\omega_\mathbf{k}t)}{\omega_\mathbf{k}}\,
e^{i\mathbf{k}\cdot(\mathbf{x}_\mu-\mathbf{x}_\nu)}
\,.
\end{eqnarray}
The remaining correlation can simply be obtained via 
$\langle\hat{p}^\dagger_\nu\hat{h}_\mu^{}\rangle=
(\langle\hat{h}^\dagger_\mu\hat{p}_\nu^{}\rangle)^*$. 
The correlator in terms of the original creation and 
annihilation operators $\hat b_\mu^\dagger$ and $\hat b_\nu$ 
is just a linear combination of these correlation functions 
\bea
\label{quench-b+b}
\langle\hat b_\mu^\dagger\hat b_\nu\rangle_{\rm quench} 
=
\frac{1}{N}\sum_\mathbf{k}4JUT_\mathbf{k}\,
\frac{1-\cos(\omega_\mathbf{k}t)}{\omega^2_\mathbf{k}}\,
e^{i\mathbf{k}\cdot(\mathbf{x}_\mu-\mathbf{x}_\nu)}
%\nn
%= 
%\sum_{\fk{k}} 
%e^{i\fk{k}\cdot(\fk{r}_\mu-\fk{r}_\nu)}
%\frac{4JUT_{\fk{k}}}{N}
%\frac{1-\cos(\omega_{\fk{k}}t)}{\omega^2_{\fk{k}}}
\,.
\eea 
Note that the momentum distribution 
\begin{eqnarray}
\label{momdist}
P(\mathbf{k})=\frac{1}{N^2}\sum_\mathbf{\mu\nu}
e^{i\mathbf{k}\cdot(\mathbf{x}_\mu-\mathbf{x}_\nu)}
\langle\hat b_\mu^\dagger\hat b_\nu\rangle
\end{eqnarray}
which is basically the Fourier transform of 
$\langle\hat b_\mu^\dagger\hat b_\nu\rangle$, 
can be measured by time-of-flight experiments \cite{G02,G01}
The quench $J(t)$ can be realised experimentally by decreasing the 
intensity of the laser field generating the optical lattice 
(which lowers the potential barrier for tunnelling and thus 
increases $J$). 
Thus the above prediction should be testable in experiments. 

As explained above, after a quench to the superfluid regime, 
the frequencies $\omega_\mathbf{k}$ become imaginary for some 
$\mathbf{k}$ and thus these modes grow exponentially. 
As a result, the expectation value will quickly be dominated by these 
fast growing modes and so most of the details of the initial state 
will become unimportant. 
Of course, this exponential growth cannot continue forever -- 
after some time, the $1/Z$-expansion breaks down since the quantum 
fluctuation are too strong and the growth will saturate. 

\begin{center}
\begin{figure}[h]
\includegraphics{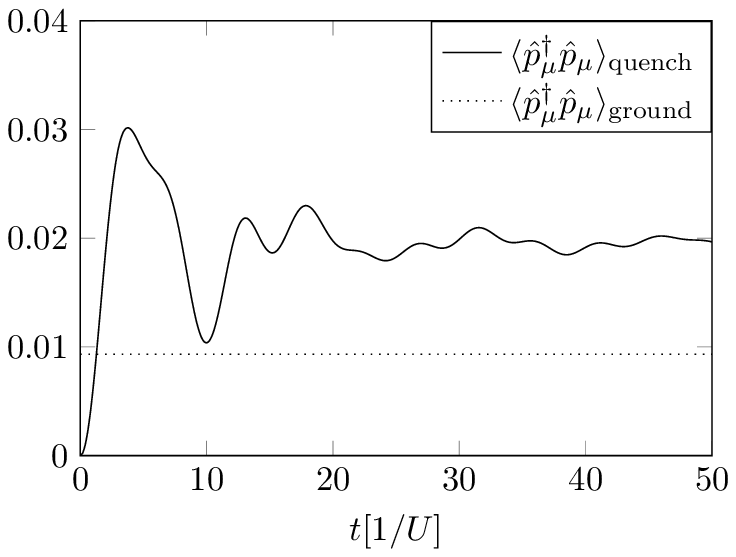}
\includegraphics{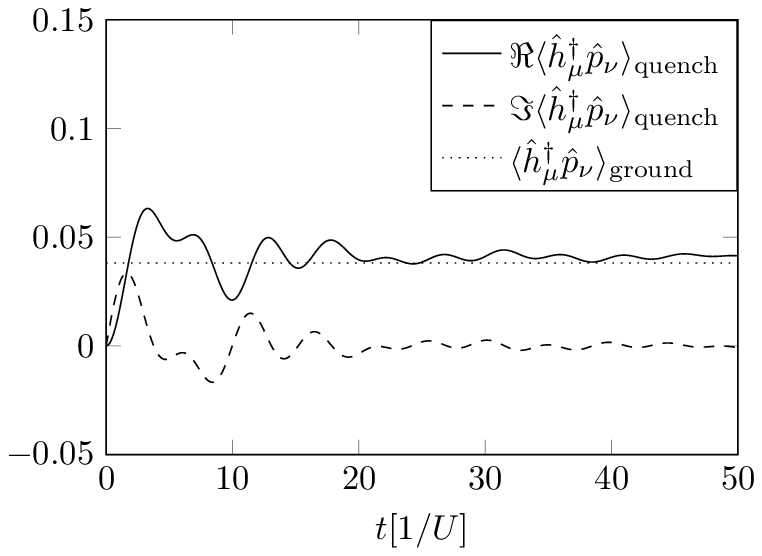}
\begin{center}
\includegraphics{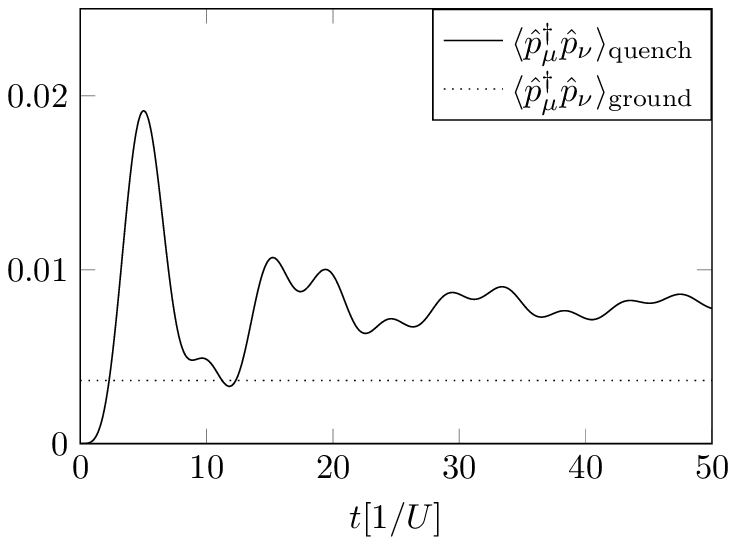}
\end{center}
\caption{Time-dependence of the depletion 
$\langle\hat{p}_{\mu}^\dagger\hat{p}_{\mu}\rangle$ 
and the nearest-neighbor correlations functions 
$\langle\hat{h}_{\mu}^\dagger\hat{p}_{\nu}\rangle$
and 
$\langle\hat{p}_{\mu}^\dagger\hat{p}_{\nu}\rangle$
in three dimensions after the quench within the Mott phase from 
$J/U=0$ to $J/U=0.14$ in comparison to their ground state values.  
After quasi-equilibration, 
$\langle\hat{p}_{\mu}^\dagger\hat{p}_{\nu}\rangle_\mathrm{quench}$ and 
$\langle\hat{p}_{\mu}^\dagger\hat{p}_{\nu}\rangle_\mathrm{ground}$
as well as 
$\langle\hat{p}_{\mu}^\dagger\hat{p}_{\mu}\rangle_\mathrm{quench}$ and 
$\langle\hat{p}_{\mu}^\dagger\hat{p}_{\mu}\rangle_\mathrm{ground}$ 
differ roughly by a factor of two.}
\end{figure}
\end{center}

%%%%%%%%%%%%%%%%%%%%%%%%%%%%%%%%%%%%%%%%%%%%%%%%%%%%%%%%%%%%%%%%%%%%%%%%%%%%%%%
\section{Equilibration versus Thermalisation}\label{equilibration}
%%%%%%%%%%%%%%%%%%%%%%%%%%%%%%%%%%%%%%%%%%%%%%%%%%%%%%%%%%%%%%%%%%%%%%%%%%%%%%%

Instead of a quench from the Mott to the superfluid phase, we can also study 
a quench within the Mott regime. 
Again, we consider a sudden change of $J$ from zero its final value for 
simplicity -- but now the final value $J$ lies below the critical point
$J<J_{\rm crit}$.
In this case, all frequencies are real $\omega_\mathbf{k}\in\mathbb R$ 
and thus there is no exponential growth -- all modes oscillate.
Apart from this point, we can use the same solution as in 
(\ref{quench-h+h}-\ref{quench-b+b}).  
For an infinite (or at least extremely large) lattice, the oscillations 
in (\ref{quench-h+h}-\ref{quench-b+b}) average out for sufficiently large 
times $t$ and thus these observables approach a quasi-equilibrium value 
\begin{eqnarray}
\label{equil-h+h}
\langle\hat{h}^\dagger_\mu\hat{h}_\nu^{} \rangle_{\rm equil} 
=
\langle\hat{p}^\dagger_\mu\hat{p}_\nu^{} \rangle_{\rm equil} 
%=
%f^{11}_{\mu\nu,\mathrm{Q,stat}}
&=&
\frac{1}{N}\sum_\mathbf{k}
\frac{4J^2 T_\mathbf{k}^2}{\omega^2_\mathbf{k}}\,
e^{i\mathbf{k}\cdot(\mathbf{x}_\mu-\mathbf{x}_\nu)}
\\
\label{equil-h+p}
\langle\hat{h}^\dagger_\mu\hat{p}_\nu^{} \rangle_{\rm equil} 
=
\langle\hat{p}^\dagger_\mu\hat{h}_\nu^{} \rangle_{\rm equil} 
%=
%f^{12}_{\mu\nu,\mathrm{Q,stat}}
&=&
\frac{1}{N}\sum_\mathbf{k}
\sqrt{2}JT_\mathbf{k}\,
\frac{U-3JT_\mathbf{k}}{\omega_\mathbf{k}^2}
e^{i\mathbf{k}\cdot(\mathbf{x}_\mu-\mathbf{x}_\nu)}
\,.
\end{eqnarray}
The quasi-equilibrium values for the local (on-site) particle or hole 
probability can be derived in complete analogy to the previous case 
\begin{eqnarray}\label{parthole}
\langle\hat{p}^\dagger_\mu\hat{p}_\mu^{}\rangle_{\rm equil}  
=
\langle\hat{h}_\mu^{}\hat{h}^\dagger_\mu\rangle_{\rm equil}  
=
\frac{1}{N}\sum_\mathbf{k}
\frac{4J^2 T_\mathbf{k}^2}{\omega^2_\mathbf{k}}
%
%\frac{U-3JT_\mathbf{k}-\omega_\mathbf{k}}{2\omega_\mathbf{k}}
%
%\frac{\sqrt{2}JT_\mathbf{k}}{\omega_\mathbf{k}}%
%
\,.
\end{eqnarray}
Having found that the observables considered above approach a 
quasi-equilibrium state, it is natural to ask the question of 
thermalisation.
This is one of the major unsolved questions 
(or rather a set of questions) in quantum many-body theory
\cite{KWE11,GME11,EHKKMWW,KISD11,BCH11}
In one version, this question can be posed in the following way:
Given an interacting quantum many-body system 
(for example the Bose-Hubbard model) on an infinite lattice 
in a appropriately excited state (such as after a quench), 
do all localised observables 
(e.g., $\langle\hat{p}^\dagger_\mu\hat{p}_\mu^{}\rangle$
and $\langle\hat{h}^\dagger_\mu\hat{h}_\nu^{}\rangle$) settle down 
to a value which is consistent with a thermal state described by 
a suitable temperature? 

Even though we cannot settle this question here, we can compare the 
quasi-equilibrium values obtained above with a thermal state.
To this end, we derive the thermal density matrix $\hat\rho_\beta$ 
corresponding to a given (inverse) temperature $k_{\rm B}T=1/\beta$. 
Using the grand canonical ensemble, the thermal density operator is 
given by
\begin{eqnarray}
\hat\rho_\beta
=
\frac{e^{-\beta(\hat{H}-\mu \hat{N})}}
{\tr\{e^{-\beta(\hat{H}-\mu \hat{N})}\}}
\,, 
\end{eqnarray}
where chemical potential $\mu$ will be chosen such that the filling 
is equal to unity.
Since we cannot calculate $\hat\rho_\beta$ exactly, we employ 
strong-coupling perturbation theory, i.e., an expansion in powers of $J$.
It is useful to introduce the operator \cite{CFMSE08}
\begin{eqnarray}
\hat R(\beta)
=
e^{\beta\hat{H}_0}e^{-\beta(\hat{H}_0+\hat{H}_1)}
\,,
\end{eqnarray}
where $\hat{H}_0$ is the diagonal on-site part of the grand canonical 
Hamiltonian $\hat{H}-\mu \hat{N}$ and $\hat{H}_1$ is the hopping term. 
This operator satisfies the differential equation
\begin{eqnarray}
\partial_\beta\hat{R}(\beta)=-\hat{H}_1(\beta)\hat{R}(\beta)
\,, 
\end{eqnarray}
where $\hat{H}_1(\beta)=e^{\beta\hat{H}_0} \hat{H}_1 e^{-\beta\hat{H}_0}$.
In analogy to time-dependent perturbation theory, the operator $\hat{R}$ 
can be calculated perturbatively by integrating this equation with respect 
to $\beta$. 
In first-order perturbation expansion, we have
\begin{eqnarray}
\label{thermal-perturbation}
\hat{\rho}_\beta
=
\frac{e^{-\beta\hat{H}_0}}{\tr\{e^{-\beta \hat{H}_0}\}}
\left(1+\frac{J}{Z}\sum_{\mu\nu}T_{\mu\nu}\,
\frac{e^{\beta U(\hat{n}_\mu-\hat{n}_\nu-1)}-1}
{U(\hat{n}_\mu-\hat{n}_\nu-1)}\,
\hat{a}^\dagger_\mu \hat{a}^{}_\nu
+\ord(J^2)
\right)
\,.
\end{eqnarray}
Obviously, the correction to first order in $J$ does not affect the 
on-site density matrix $\hat\rho_\mu$ but the two-point correlations.
Thus, we find that the quasi-equilibrium state of the on-site density 
matrix $\hat\rho_\mu$ can indeed be described by a thermal state
provided that we choose the chemical potential as $\mu=U/2$ which gives 
\bea\label{thermdens}
\hat\rho_\mu(\beta)
\approx
\frac{e^{-\beta U/2}}{2}\,\ket{0}_\mu\!\bra{0}
+
\left(1-e^{-\beta U/2}\right)\ket{1}_\mu\!\bra{1}
+
\frac{e^{-\beta U/2}}{2}\,\ket{2}_\mu\!\bra{2}
\,.
\ea
The particular value $\mu=U/2$ of the chemical potential ensures 
that (in first order thermal perturbation theory) we have on average one 
particle per lattice site and the particle-hole symmetry 
$\langle \hat{p}_\mu^\dagger \hat{p}_\mu\rangle=
\langle \hat{h}_\mu^\dagger \hat{h}_\mu\rangle$.
To obtain the correct probabilities, we have to select the temperature 
according to  
\bea
\label{temperature}
e^{-\beta U/2}=2\langle\hat{p}^\dagger_\mu\hat{p}_\mu^{}\rangle_{\rm equil}  
=
\frac{2}{N}\sum_\mathbf{k}
\frac{4J^2 T_\mathbf{k}^2}{\omega^2_\mathbf{k}}
=\ord(1/Z)
\,,
\ea
which can be deduced from (\ref{parthole}) and (\ref{thermdens}).
Since the depletion is small 
$\langle\hat{p}^\dagger_\mu\hat{p}_\mu^{}\rangle=\ord(1/Z)$,  
we obtain a low effective temperature which scales as $T=\ord(U/\ln Z)$. 
Accordingly, consistent with our $1/Z$-expansion, we can neglect 
higher Boltzmann factors such as $e^{-\beta U}$. 

Of course, the fact that the on-site density matrix $\hat\rho_\mu$ 
can be described (within our limits of accuracy) by a thermal state 
does not imply that the same is true for the correlations.
To study this point, let us calculate the thermal two-point correlator 
from (\ref{thermal-perturbation}). 
To first order in $J$ and $1/Z=\ord(e^{-\beta U}/2)$, we find 
\bea
\langle\hat{h}^\dagger_\mu\hat{p}_\nu^{} \rangle_\beta
=
\langle\hat{p}^\dagger_\mu\hat{h}_\nu^{} \rangle_\beta
=
\frac{\sqrt{2}JT_{\mu\nu}}{ZU}
+\ord(J^2)+\ord(1/Z^2)
\,,
\ea
while $\langle\hat{h}^\dagger_\mu\hat{h}_\nu^{} \rangle_\beta$
and $\langle\hat{p}^\dagger_\mu\hat{p}_\nu^{} \rangle_\beta$ 
vanish (to first order in $J$).
If we compare this to the quasi-equilibrium value 
$\langle\hat{h}^\dagger_\mu\hat{p}_\nu^{} \rangle_{\rm equil}$ in 
(\ref{equil-h+p}), we find that they coincide to first order in $J$ 
\bea
\langle\hat{h}^\dagger_\mu\hat{p}_\nu^{} \rangle_{\rm equil}
=
\langle\hat{p}^\dagger_\mu\hat{h}_\nu^{} \rangle_{\rm equil}
=
\frac{\sqrt{2}JT_{\mu\nu}}{ZU}
+\ord(J^2)+\ord(1/Z^2)
\,.
\ea
This is perhaps not too surprising since the same value can be obtained 
from the ground state fluctuations 
$\langle\hat{h}^\dagger_\mu\hat{p}_\nu^{} \rangle_{\rm ground}=
\langle\hat{p}^\dagger_\mu\hat{h}_\nu^{} \rangle_{\rm ground}$  
in (\ref{ground2}) after expanding them to first order in $J$. 
Due to the low effective temperature $T=\ord(U/\ln Z)$, the lowest 
Boltzmann factor is suppressed by $e^{-\beta U/2}=\ord(1/Z)$.
As a consequence, because the correlations are small $\ord(1/Z)$,
their finite-temperature corrections are even smaller $\ord(1/Z^2)$,
and thus can be neglected.  

The same is true for the other correlations 
$\langle\hat{h}^\dagger_\mu\hat{h}_\nu^{} \rangle=
\langle\hat{p}^\dagger_\mu\hat{p}_\nu^{} \rangle$.
All of them: the ground state correlators
$\langle\hat{h}^\dagger_\mu\hat{h}_\nu^{} \rangle_{\rm ground}=
\langle\hat{p}^\dagger_\mu\hat{p}_\nu^{} \rangle_{\rm ground}$ 
in (\ref{ground1}), the quasi-equilibrium correlators 
$\langle\hat{h}^\dagger_\mu\hat{h}_\nu^{} \rangle_{\rm equil}=
\langle\hat{p}^\dagger_\mu\hat{p}_\nu^{} \rangle_{\rm equil}$ 
in (\ref{equil-h+h}), as well as the thermal correlators 
$\langle\hat{h}^\dagger_\mu\hat{h}_\nu^{} \rangle_\beta$
and $\langle\hat{p}^\dagger_\mu\hat{p}_\nu^{} \rangle_\beta$ 
vanish to first order in $J$.
Therefore, to first order in $J$ and $1/Z$, the thermal state 
can describe the observable under consideration.
However, going to the next order in $J$, this description breaks down.
This failure can even be shown without explicitly calculating 
$\hat R(\beta)$ up to second order. 
If we compare the quasi-equilibrium correlators in (\ref{equil-h+h}) 
\bea
\langle\hat{h}^\dagger_\mu\hat{h}_\nu^{} \rangle_{\rm equil}
=
\langle\hat{p}^\dagger_\mu\hat{p}_\nu^{} \rangle_{\rm equil}
=
\frac{4J^2 }{U^2Z^2}\sum_\kappa T_{\mu\kappa}T_{\kappa\nu}
+\ord(J^3)+\ord(1/Z^2)
\,,
\ea
with the ground state correlations in (\ref{ground1}),
expanded to the same order in $J$  
\bea
\langle\hat{h}^\dagger_\mu\hat{h}_\nu^{} \rangle_{\rm ground}
=
\langle\hat{p}^\dagger_\mu\hat{p}_\nu^{} \rangle_{\rm ground}
=
\frac{2J^2 }{U^2Z^2}\sum_\kappa T_{\mu\kappa}T_{\kappa\nu}
+\ord(J^3)+\ord(1/Z^2)
\,,
\ea
we find a discrepancy by a factor of two.~ 
I.e., after the quench, these correlations settle down to a value 
which is twice as large as in the ground state.
This factor of two has already been found elsewhere in the context 
of standard time-dependent and time-independent perturbation theory, 
see also \cite{MK09}.
This is incompatible with the small Boltzmann factors 
$e^{-\beta U/2}=\ord(1/Z)$ and would require a comparably large 
effective temperature $T=\ord(U)$ instead of $T=\ord(U/\ln Z)$.
However, such a large effective temperature $T=\ord(U)$ is 
inconsistent with the small on-site depletion (\ref{temperature}).

In summary, the considered observables settle down to a quasi-equilibrium 
state -- but this state is not thermal. 
Thus, real thermalisation -- if it occurs at all -- requires much longer 
times scales.
This seems to be a generic feature and has been discussed for 
bosonic \cite{CFMSE08,FCMSE10,CDEO08} and fermionic systems
\cite{U09,MK08,EKW10,MK10,SGJ10,EKW09}
and is sometimes called ``pre-thermalisation''. 
This phenomenon can be visualised via the following intuitive picture:
The excited state generated by the quench can be viewed as a highly 
coherent superposition of correlated quasi-particles.
During the subsequent quantum evolution, these quasi-particles 
disperse and randomise their relative phases -- which results in 
a quasi-stationary state. 
However, the quasi-particles still retain their initial spectrum
(in energy and quasi-momentum), which could be approximately described 
by a generalised Gibbs ensemble (i.e., a momentum-dependent temperature). 
In this picture, thermalisation requires the exchange of energy and 
momentum between these quasi-particles due to multiple collisions, 
which changes the one-particle spectrum and takes much longer. 
Ergo, one would expect a separation of time scales -- i.e., first (quasi)
equilibration and only much later thermalisation -- for many systems 
in condensed matter, where the above quasi-particle picture applies.

%%%%%%%%%%%%%%%%%%%%%%%%%%%%%%%%%%%%%%%%%%%%%%%%%%%%%%%%%%%%%%%%%%%%%%%%%%%%%%%
\section{Tilted Mott Lattice}\label{tiltlattice}
%%%%%%%%%%%%%%%%%%%%%%%%%%%%%%%%%%%%%%%%%%%%%%%%%%%%%%%%%%%%%%%%%%%%%%%%%%%%%%%

In the following, we study the impact of a spatially constant but possibly 
time-dependent force on the particles, which could correspond to a tilt of 
the lattice, for example 
\cite{M01,M02,D96,QNS11,WWMK05,SSG02,W05,KK03,KK04,K04,K03,KB03,KKG09,CN99,M12,Si11}.
This scenario can be described by a generalisation of the 
Hamiltonian (\ref{Bose-Hubbard-Hamiltonian}) 
\bea
\label{Bose-Hubbard-Tilt}
\hat H
=
-\frac{J}{Z}\sum_{\mu\nu} T_{\mu\nu} \hat b^\dagger_\mu \hat b_\nu
+\frac{U}{2}\sum_{\mu} 
\hat n_{\mu}(\hat n_{\mu}-1)
+\sum_{\mu} V_\mu\hat n_{\mu}
\,.
\ea
The external potential $V_\mu(t)=\mathbf{x}_\mu\cdot\mathbf{E}(t)$
will be identified as an effective electric field $\mathbf{E}(t)$
and will be time-dependent in general. 
If we insert this modified Hamiltonian into (\ref{one-site-Mott}), 
we find that the potential $V_\mu$ has no effect to zeroth order $\ord(Z^0)$, 
i.e., the solution $\hat\rho_{\mu}^0=\ket{1}_\mu\!\bra{1}$ remains the 
same. 
In other words, the Gutzwiller mean field is not affected by the tilt
in the Mott state (in the superfluid phase, this would be different). 
However, the next-order $\ord(1/Z)$ quantum correlations can lead to 
the creation of particle-hole pairs via tunnelling over one or more 
lattice sites. 

In order to study this effect, let us generalise the evolution equations 
(\ref{f12-Mott}) and (\ref{f11-Mott}) in the presence of the potential
$V_\mu$ 
\begin{eqnarray}
\label{f12}
\left(i\partial_t+V_\mu-V_\nu-U\right)f^{12}_{\mu\nu}
&=&
-\frac{J}{Z}\sum_{\kappa\neq \mu,\nu}T_{\mu\kappa}
\left[3 f^{12}_{\kappa\nu}+\sqrt{2}f^{22}_{\kappa\nu}+
\sqrt{2}f^{11}_{\kappa\nu}\right]
\nn
& &
-\frac{J\sqrt{2}}{Z}T_{\mu\nu}
\,,
\\
\label{f11}
\left(i\partial_t+V_\mu-V_\nu\right) 
f^{11}_{\mu\nu}
%&=&
%\left(i\partial_t+V_\mu-V_\nu\right) %
%f^{22}_{\mu\nu}\nonumber\\
&=&
-\frac{\sqrt{2}J}{Z}\sum_{\kappa\neq\mu,\nu}T_{\mu\kappa}\left(
f_{\kappa\nu}^{21}-f_{\kappa\nu}^{12}\right)
\,,
\quad
\end{eqnarray}
and the same for $f^{22}_{\mu\nu}$, such that we again have an effective 
particle-hole operator symmetry $f^{11}_{\mu\nu}=f^{22}_{\mu\nu}$ to 
lowest order in $1/Z$. 
Here we have already used translational invariance.  
The tunnelling probability can now be obtained by solving the 
above equations.
However, instead of solving them directly, we can simplify the problem 
by effectively factorising these equations: 
If we introduce the {\em effective} differential equations for 
$\hat{h}_\mu$ and $\hat{p}_\mu$,
\begin{eqnarray}
\label{effective}
\left[i\partial_t-V_\mu-\frac{U}{2}\right]\hat{p}_\mu
&=&
-\frac{J}{Z}\sum_\nu T_{\mu\nu}
\left[\frac32\,\hat{p}_\nu+\sqrt{2}\,\hat{h}_\nu\right]
\,,
\nonumber\\
\left[i\partial_t-V_\mu+\frac{U}{2}\right]\hat{h}_\mu
&=&
\frac{J}{Z}\sum_\nu T_{\mu\nu}
\left[\frac32\,\hat{h}_\nu+\sqrt{2}\,\hat{p}_\nu\right]
\,,
\end{eqnarray}
and exploit the initial conditions 
$\langle\hat h_\mu^\dagger\hat h_\nu\rangle_0=\delta_{\mu\nu}$
and 
$\langle\hat h_\mu^\dagger\hat p_\nu\rangle_0=
\langle\hat p_\mu^\dagger\hat h_\nu\rangle_0=
\langle\hat p_\mu^\dagger\hat p_\nu\rangle_0=0$
valid in the Mott state, we exactly recover 
(\ref{f12}) and (\ref{f11}) to first order in $1/Z$. 

For potentials of the form $V_\mu(t)=\mathbf{x}_\mu\cdot\mathbf{E}(t)$
it is possible to apply the Peierls transformation and absorb the 
potential in a phase.
After the Fourier transformations
\begin{eqnarray}
\hat{h}_\mu(t)
&=&
\exp\left\{-i\int_0^t dt' V_\mu(t')\right\}
\sum_\mathbf{k}\hat{h}_\mathbf{k}(t)\exp\{i\mathbf{k\cdot x}_\mu\}
\,,
\\
\hat{p}_\mu(t)
&=&
\exp\left\{-i\int_0^t dt' V_\mu(t')\right\}
\sum_\mathbf{k}\hat{p}_\mathbf{k}(t)\exp\{i\mathbf{k\cdot x}_\mu\}
\,,
\\
T_{\mu\nu}(t)
&=&
\frac{Z}{N}\sum_\mathbf{k}T_\mathbf{k}(t)
\exp\left\{i\mathbf{k}\cdot(\mathbf{x}_\mu-\mathbf{x}_\nu)
+i\int_0^t dt'[V_\mu(t')-V_\nu(t')]\right\}
\label{Peierls}
\,,
\end{eqnarray}
the effective evolution equations (\ref{effective}) become
\begin{eqnarray}
i\partial_t \hat{h}_\mathbf{k}
&=&
+\frac{1}{2}\left[3JT_\mathbf{k}(t)-U\right]\hat{h}_\mathbf{k}+
\sqrt{2}J T_\mathbf{k}(t)\hat{p}_\mathbf{k}
\,,
\label{holedgl}
\\
i\partial_t \hat{p}_\mathbf{k}
&=&
-\frac{1}{2}\left[3JT_\mathbf{k}(t)-U\right]\hat{p}_\mathbf{k}-
\sqrt{2}J T_\mathbf{k}(t)\hat{h}_\mathbf{k}
\label{partdgl}
\,.
\end{eqnarray}
Note that $T_\mathbf{k}(t)$ explicitly depends on time here and this 
time-dependence encodes the potential $V_\mu(t)$.
Introducing the effective vector potential $\mathbf{A}(t)$ which generates
the effective electric field $\mathbf{E}(t)$ in 
$V_\mu(t)=\mathbf{x}_\mu\cdot\mathbf{E}(t)$
via $\mathbf{E}(t)=\partial_t\mathbf{A}(t)$, this is equivalent to the 
substitution $\mathbf{k}\to\mathbf{k}+\mathbf{A}(t)$
in complete analogy to the minimal coupling procedure 
$T_\mathbf{k}(t)=T_{\mathbf{k}+\mathbf{A}(t)}$ known from
electrodynamics.  

The most general solution of (\ref{holedgl}) and (\ref{partdgl}) 
can be written as
\begin{eqnarray}
\hat{h}_\mathbf{k}
&=&
f_\mathbf{k}^+(t)\hat{A}_\mathbf{k}+
f_\mathbf{k}^-(t)\hat{B}_\mathbf{k}
\,,
\\
\hat{p}_\mathbf{k}
&=&
g_\mathbf{k}^+(t)\hat{A}_\mathbf{k}+
g_\mathbf{k}^-(t)\hat{B}_\mathbf{k}
\,,
\end{eqnarray}
where $\hat{A}_\mathbf{k}$ and $\hat{B}_\mathbf{k}$ are time-independent 
operators while $f_\mathbf{k}^\pm$ and $g_\mathbf{k}^\pm$ are time-dependent 
c-number functions. 
In analogy to the previous case, we assume that we start in the 
Mott state (\ref{Mott-state}) with $J=V_\mu=0$.
In this case, the equations (\ref{holedgl}) and (\ref{partdgl}) decouple 
and we may choose $\hat{A}_\mathbf{k}=\hat{h}_\mathbf{k}^{\rm in}$ and 
$\hat{B}_\mathbf{k}=\hat{p}_\mathbf{k}^{\rm in}$ which imply 
$\langle \hat{A}_\mathbf{k}^\dagger \hat{A}_\mathbf{p}\rangle_0=
\delta_{\mathbf{k},\mathbf{p}}$ and
$\langle \hat{B}_\mathbf{k}^\dagger \hat{B}_\mathbf{p}\rangle_0=0$,
as well as, $f_\mathbf{k}^+(t)=\exp(i U t/2)$ and 
$g_\mathbf{k}^-(t)=\exp(-i U t/2)$ while the other two vanish. 

Now we imagine the following sequence:
First we switch on $J$ adiabatically, then we apply the potential $V_\mu(t)$
for a finite period of time, and finally we switch off $J$ adiabatically.
Thus, at the very end, the equations (\ref{holedgl}) and (\ref{partdgl}) 
decouple again and the final particle operator $\hat{p}_\mathbf{k}^{\rm out}$
oscillates with positive frequencies $\exp(-i U t/2)$ while the final 
hole operator $\hat{h}_\mathbf{k}^{\rm out}$
oscillates with negative frequencies $\exp(+i U t/2)$.
However, during the time-evolution, positive and negative frequencies will
get mixed in general by the time-dependence of 
$T_\mathbf{k}(t)=T_{\mathbf{k}+\mathbf{A}(t)}$, 
i.e., the potential $V_\mu(t)$.
Thus, the initial and final particle/hole-operators will be connected by 
a Bogoliubov transformation 
\bea
\label{Bogoliubov}
\hat{p}_\mathbf{k}^{\rm out} 
&=& 
\alpha_\mathbf{k}\hat{p}_\mathbf{k}^{\rm in}+ 
\beta_\mathbf{k}\hat{h}_\mathbf{k}^{\rm in}
\,,
\nn 
\hat{h}_\mathbf{k}^{\rm out} 
&=& 
\alpha^*_\mathbf{k}\hat{h}_\mathbf{k}^{\rm in}+ 
\beta^*_\mathbf{k}\hat{p}_\mathbf{k}^{\rm in}
\,,
\ea
where the Bogoliubov coefficients $\alpha_\mathbf{k}$ and $\beta_\mathbf{k}$
satisfy the relation $|\alpha_\mathbf{k}|^2-|\beta_\mathbf{k}|^2=1$.
%see Appendix {\red ...}
%
In view of the initial conditions 
$\langle \hat{A}_\mathbf{k}^\dagger \hat{A}_\mathbf{p}\rangle_0=
\delta_{\mathbf{k},\mathbf{p}}$ and
$\langle \hat{B}_\mathbf{k}^\dagger \hat{B}_\mathbf{p}\rangle_0=0$,
we find 
\begin{eqnarray}
\langle\hat{p}^\dagger_\mathbf{k}\hat{p}_\mathbf{k}\rangle_{\rm out}
=
|\beta_\mathbf{k}|^2
\,,
\end{eqnarray}
which gives the probability to create a particle in the mode $\mathbf{k}$.
Since particles (i.e., doubly occupied lattice sites) and holes  
(i.e., empty lattice sites) are always created in pairs, we get the 
same probability for the holes. 
Note that $\mathbf{k}$ is the canonical and not necessarily the mechanical 
momentum due to the substitution $\mathbf{k}\to\mathbf{k}+\mathbf{A}(t)$
mentioned above. 

%%%%%%%%%%%%%%%%%%%%%%%%%%%%%%%%%%%%%%%%%%%%%%%%%%%%%%%%%%%%%%%%%%%%%%%%%%%%%%%
\section{\label{section-Analogue}Analogue of Sauter-Schwinger Tunnelling}
%%%%%%%%%%%%%%%%%%%%%%%%%%%%%%%%%%%%%%%%%%%%%%%%%%%%%%%%%%%%%%%%%%%%%%%%%%%%%%%

The precise amount of mixing which determines the Bogoliubov coefficients 
$\alpha_\mathbf{k}$ and $\beta_\mathbf{k}$ can be derived from the evolution
equations (\ref{holedgl}) and (\ref{partdgl}).
Turning these two first-order differential equations into one second-order 
equation, we get for $g^+_\mathbf{k}$ and $f^+_\mathbf{k}$,
\begin{eqnarray}
\partial_t^2 f^+_\mathbf{k}-
\frac{\dot{T}_\mathbf{k}}{T_\mathbf{k}}\partial_t f^+_\mathbf{k}
+\left(
\frac{\omega^2_\mathbf{k}}{4}+i \frac{U\dot{T}_\mathbf{k}}{2T_\mathbf{k}}
\right)f^+_\mathbf{k}
&=&0
\,,
\\
\partial_t^2 g^+_\mathbf{k}-
\frac{\dot{T}_\mathbf{k}}{T_\mathbf{k}}\partial_t g^+_\mathbf{k}
+\left(
\frac{\omega^2_\mathbf{k}}{4}-i \frac{U\dot{T}_\mathbf{k}}{2T_\mathbf{k}}
\right)g^+_\mathbf{k}
&=&0
\,.
\end{eqnarray}
With the substitutions $f^+_\mathbf{k}=\sqrt{T_\mathbf{k}}u_\mathbf{k}$ 
and $g^+_\mathbf{k}=\sqrt{T_\mathbf{k}}v_\mathbf{k}$, we may eliminate 
the first-order terms and arrive at 
\begin{eqnarray}
\partial_t^2 u_\mathbf{k}+
\left(
\frac{\omega^2_\mathbf{k}}{4}+i \frac{U\dot{T}_\mathbf{k}}{2T_\mathbf{k}}+
\frac{\ddot{T}_\mathbf{k}}{2T_\mathbf{k}}-
\frac{3\dot{T}_\mathbf{k}^2}{4T_\mathbf{k}^2}
\right)u_\mathbf{k}&=&0
\,,
\label{dglexakt}
\nn
\partial_t^2 v_\mathbf{k}+
\left(
\frac{\omega^2_\mathbf{k}}{4}-i \frac{U\dot{T}_\mathbf{k}}{2T_\mathbf{k}}+
\frac{\ddot{T}_\mathbf{k}}{2T_\mathbf{k}}-
\frac{3\dot{T}_\mathbf{k}^2}{4T_\mathbf{k}^2}
\right)v_\mathbf{k}&=&0
\,.
\end{eqnarray}
Now we consider a small tilt of the lattice, corresponding to a weak 
electric field $|\mathbf{E}|\ll U$. 
In this case, we may approximate the above equations by neglecting the
terms with $\dot{T}_\mathbf{k}$ and $\ddot{T}_\mathbf{k}$ since 
$\dot{T}_\mathbf{k}=\ord(\mathbf{E})$. 
Furthermore, for a weak tilt, the particles have to tunnel across many 
lattice sites in order to gain enough energy and to be able to overcome
the energy gap before a particle-hole pair can be created. 
Thus, we may consider large length scales, corresponding 
to small wavenumbers $\mathbf{k}$ and Taylor expand the 
${T}_\mathbf{k}(t)$
%
%In this limit, we may Taylor expand the ${T}_\mathbf{k}(t)$ and we find 
%
\bea
T_\mathbf{k}(t)=T_{\mathbf{k}+\mathbf{A}(t)}=
1-\xi[\mathbf{k}+\mathbf{A}(t)]^2+\ord(\mathbf{k}^4)
\,,
\ea
where $\xi$ is the stiffness.
With these approximations, we find that (\ref{dglexakt}) simplify to 
\begin{eqnarray}
\label{klein-gordon}
\partial_t^2 \phi_\mathbf{k}+
\left(
m_{\rm eff}^2 c^4_{\rm eff}+
c^2_{\rm eff}[\mathbf{k}+\mathbf{A}(t)]^2
\right)
\phi_\mathbf{k}=0
\,,
\end{eqnarray}
which is just the Klein-Fock-Gordon equation describing charged scalar 
particles in an external electromagnetic field, provided that we identify
the effective speed of light 
\begin{eqnarray}
\label{c_eff}
c^2_{\rm eff}= \frac{\xi}{2}\,J(3U-J)
\,,
\end{eqnarray}
while the effective mass is given by half the energy gap $\Delta{\cal E}$ 
\begin{eqnarray}
\label{m_eff}
m_{\rm eff}^2 c^4_{\rm eff}=\frac{1}{4}(U^2-6 J U+J^2)
\,.
\end{eqnarray}
Consequently, there is a quantitative analogy between the tilted 
Bose-Hubbard lattice and the Sauter-Schwinger effect, i.e., 
electron-positron pair creation out of the quantum vacuum due to 
an external electric field, sketched in Fig.~\ref{sauter-analogy}   
and the following table: 

\bigskip

\begin{center}
\begin{tabular}{|c|c|}
\hline
Sauter-Schwinger effect & \;Bose-Hubbard model\; \\ 
\hline
electrons \& positrons & particles \& holes \\ 
Dirac sea & Mott state \\ 
\;mass of electron/positron\; & energy gap $\Delta{\cal E}$ \\
electric field $\mathbf{E}$ & lattice tilt $V_\mu$ \\ 
speed of light $c$ & velocity $c_{\rm eff}$ \\ 
\hline
\end{tabular}
\end{center}

\bigskip 

\begin{figure}[h]
\begin{center}
\includegraphics[width=.8\columnwidth]{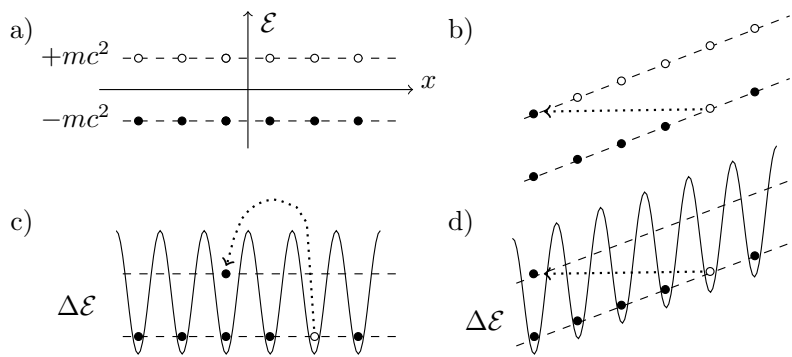}
\end{center}
\psfrag{test}{$\Delta\cal E$}
\caption{Sketch of the analogy: 
a) Dirac sea for $E=0$, 
b) Sauter-Schwinger tunneling for $E\neq0$, 
c) Mott state with energy gap $\Delta\cal E$, 
d) tunneling in tilted lattice.}
\label{sauter-analogy}
\end{figure}

We can now use this analogy to apply our knowledge of the 
Sauter-Schwinger effect \cite{S31,S51,K65,BI70,D09,SGD08} 
to particle-hole creation in the 
tilted Bose-Hubbard model -- as Richard Feynman said: 
{\em The same equations have the same solutions.}
For example, consider a purely time-dependent electric field of the 
following form
\begin{eqnarray}
\label{sauterpulse}
\mathbf{E}=\frac{E_0\mathbf{e}_z}{\cosh^2\left(t/\tau\right)}
\,.
\end{eqnarray}
Such a profile is called Sauter pulse since F.~Sauter was the first to 
realise (already in 1931) that the Dirac equation and the Klein-Fock-Gordon 
equation in the presence of such a field can be solved exactly 
in terms of hypergeometric functions 
(although he considered the form with $t$ and $x$ interchanged). 
From the exact solution of the scalar field case, one obtains 
\cite{KLY08,GG96}
\begin{eqnarray}
\label{sauter}
|\beta_\mathbf{k}|^2
=
%\frac{1}{\delta_\mathbf{0}^2}
\frac{\cosh\left(\pi\tau[\omega_+-\omega_-]\right)+
\cosh\left(\pi\sqrt{4 E_0^2 c^2\tau^4-1}\right)}
{2\sinh(\pi \tau \omega_+)\sinh(\pi \tau \omega_-)}
\,,
\end{eqnarray}
with the abbreviations 
\begin{eqnarray}
\omega_\pm=\sqrt{c^2(k_z\mp E_0\tau)^2+m_e^2c^4+k_\perp^2 c^2}
\,.
\end{eqnarray}
Here $k_\perp$ denotes the momentum perpendicular to the electric field 
and $m_e$ is the mass of the electron. 
Via the analogy established above, expression~(\ref{sauter}) yields the 
momentum dependent particle-hole pair creation probability in a Mott 
state subject to a time-dependent tilt according to Eq.~(\ref{sauterpulse}). 
For various pulse lengths $\tau$, this result plotted in 
Fig.~\ref{Pexc-N_12-L_12} and compared to numerical results for a 
one-dimensional Bose-Hubbard lattice.

In the static limit $\tau\rightarrow\infty$, Eq.~(\ref{sauter}) reproduces 
the well-known expression 
\begin{eqnarray}
\label{sauter-inf}
|\beta_\mathbf{k}|^2
=
%\frac{1}{\delta_\mathbf{0}^2}
\exp\left(-\pi\frac{m_e^2 c^4+k_\perp^2 c^2}{E_0 c}\right)
\,.
\end{eqnarray}
As we see, the electron-positron pair creation probability is 
exponentially suppressed for small electric fields $E_0$. 
Inserting the translation formula (\ref{c_eff}) and (\ref{m_eff}),
we get the same exponential suppression for the particle-hole pair 
creation probability via tunnelling in tilted Mott lattices. 
Thus, in order to actually verify this prediction experimentally,
the tilt should not be too small.
In this case, the terms with $\dot{T}_\mathbf{k}$ and 
$\ddot{T}_\mathbf{k}$ we have neglected earlier due to 
$\dot{T}_\mathbf{k}=\ord(\mathbf{E})$ might start to play a role. 
Thus, let us estimate the impact of these contributions. 
Including the terms involving $\dot{T}_\mathbf{k}$ and $\ddot{T}_\mathbf{k}$, 
we find
\begin{eqnarray}
\partial_t^2 u_\mathbf{k}
+
\left[
%\partial_t^2+
m_{\rm eff}^2 c^4_{\rm eff}+k_\perp^2 c^2_{\rm eff}
+c^2_{\rm eff}(k_z-E_0 t)^2 
\right]u_\mathbf{k}
&&
\nn
+\xi
\left[-E_0^2+i E_0 U(k_z-E_0 t)\right]
u_\mathbf{k}
&=&0\,,
\end{eqnarray}
where we have assumed a constant field ($\tau\rightarrow\infty$)
for simplicity.
The above differential equation can be solved in terms of parabolic 
cylindrical functions from which the pair creation probability 
is determined to be
%
% \begin{eqnarray}
% |\beta_\mathbf{k}|^2
% &\approx&
% %\frac{1}{\delta^2_\mathbf{0}}
% \exp\Bigg[
% -\frac{\pi}{E_0c_{\rm eff}}
% \Bigg(
% m_{\rm eff}^2 c^4_{\rm eff}+c^2_{\rm eff} k_\perp^2
% \nonumber\\
% & &\hspace{3cm}-\frac{2c_{\rm eff}^2 E_0^2}{J(3U-J)} 
% +\frac{c_{\rm eff}^2 E_0^2 U^2}{J^2(3U-J)^2}
% \Bigg)
% \Bigg]\,.
% \end{eqnarray}
% %
% {\red mit $\xi$}
\begin{eqnarray}\label{staticfield}
|\beta_\mathbf{k}|^2
\approx
%\frac{1}{\delta^2_\mathbf{0}}
\exp\left[
-\frac{\pi}{E_0c_{\rm eff}}
\left(
m_{\rm eff}^2 c^4_{\rm eff}+c^2_{\rm eff} k_\perp^2
-\xi E_0^2
+\xi^2\frac{E_0^2 U^2}{4c^2_{\rm eff}}
\right)
\right]\,.
\end{eqnarray}
In Fig.~\ref{static}, we depicted the dependence of the 
particle-hole creation probability 
$\langle\hat{p}^\dagger_\mu\hat{p}_\mu\rangle=
\sum_\mathbf{k}|\beta_\mathbf{k}|^2/N$ on the potential gradient.

\begin{figure}[h]
\begin{center}
\includegraphics{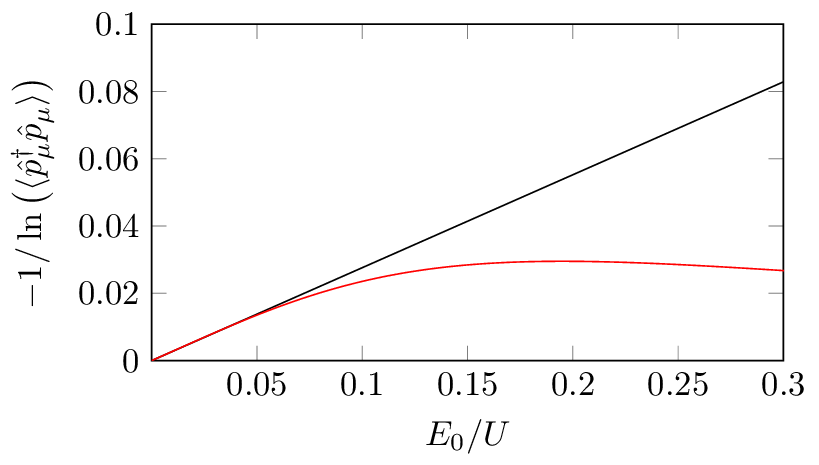}
\caption{Dependence of $-\ln(\langle\hat{p}^\dagger_\mu\hat{p}_\mu\rangle)$ 
on the lattice tilt for $J/U=0.1$.
The black line represents the standard result (\ref{sauter-inf}) 
for the static Sauter-Schwinger effect while the red curve deviates 
due to perturbative corrections in $E_0$ given by the lattice structure, 
see equation (\ref{staticfield}).}
\label{static}
\end{center}
\end{figure}

%%%%%%%%%%%%%%%%%%%%%%%%%%%%%%%%%%%%%%%%%%%%%%%%%%%%%%%%%%%%%%%%%%%%%%%%%%%%%%%
\section{Second Order in $1/Z$}\label{Z2}
%%%%%%%%%%%%%%%%%%%%%%%%%%%%%%%%%%%%%%%%%%%%%%%%%%%%%%%%%%%%%%%%%%%%%%%%%%%%%%%

So far, we have only considered the first order in $1/Z$.
Now let us discuss the effect of higher orders by means of few examples.
Let us go back to the derivation from (\ref{two-sites}) to 
(\ref{two-sites-approx}) and include $1/Z^2$ corrections. 
To achieve this level of accuracy, we should not replace the exact 
on-site density matrix $\hat\rho_\mu$ by it lowest-order approximation 
$\hat\rho_\mu^0$ but include its first-order corrections in 
(\ref{depletion}), i.e., the quantum depletion $f_0=\ord(1/Z)$ of the unit 
filling (Mott) state.  
This results in a renormalisation of the eigenfrequency 
\begin{eqnarray}\label{renomega}
\omega_\mathbf{k}^{\rm ren}
=
\sqrt{U^2-6JT_\mathbf{k}(1-3f_0)+J^2T_\mathbf{k}^2(1-3f_0)^2}
\,,
\end{eqnarray}
which indicates a shift of the Mott-superfluid transition to slightly 
higher values of $J$,
\begin{eqnarray}
J_{\rm crit}^{\rm ren}
=
U\,\frac{3-2\sqrt{2}}{1-3f_0}>U(3-2\sqrt{2})
\,,
\end{eqnarray}
see Appendix \ref{secondorder}.
Since the net effect can roughly be understood as a reduction of the 
effective hopping rate $J^{\rm ren}=J(1-3f_0)$, it is easy to visualise 
that this implies also a decrease of the effective propagation velocity.

There are also other $1/Z^2$ corrections in (\ref{two-sites-approx}) such as
the three-point correlator $\hat\rho^{\rm corr}_{\mu\nu\kappa}$ but they act
as source terms and do not affect the eigenfrequency (at second order).
However, there are other quantities where these source terms are crucial. 
In particular we consider correlation functions which are of the form
\begin{equation}
F_{\cal O}(\mu,\nu)
=
\langle
    \hat{\cal O}_\mu
    \hat{\cal O}_\nu
\rangle
-
\langle
    \hat{\cal O}_\mu
\rangle
\langle
    \hat{\cal O}_\nu
\rangle
\;,
\end{equation}
and vanish to first order in $1/Z$, in contrast to the off-diagonal 
long-range order 
$\langle\hat{a}_\mu^\dagger \hat{a}_\nu\rangle$ discussed above.
One important example is the particle-number correlation, i.e.,  
$\langle \hat{n}_\mu \hat{n}_\nu\rangle-1$. 
After a somewhat lengthy calculation, we find for the ground state 
correlations (see Appendix \ref{secondorder})
\begin{eqnarray}\label{number}
F_{n}(\mu,\nu)
&=&
\frac{2}{N^2}\sum_{\mathbf{p},\mathbf{q}}
e^{i(\mathbf{p}+\mathbf{q})\cdot(\mathbf{x}_\mu-\mathbf{x}_\nu)}
\left(f^{11}_\mathbf{p}f^{11}_\mathbf{q}-f^{12}_\mathbf{p}f^{21}_\mathbf{q}
\right)
\,,
\end{eqnarray}
where $f^{12}_\mathbf{p},f^{21}_\mathbf{p}$ and 
$f^{11}_\mathbf{p}$ are given through equations (\ref{stat}) and (\ref{inv}).
It is also possible to calculate this quantity via a perturbation expansion
into powers of $J/U$. 
Note, however, that the above result is not perturbative in $J/U$, see, 
for example, the non-polynomial dependence of $\omega_\mathbf{k}$ on $J$. 

These predictions could be tested experimentally by site-resolved imaging,
i.e., measurements on single lattice sites \cite{B10,W09,GZHC09,S10}. 
In some of these experiments, the particle number per site is not directly
measured, but only the parity -- i.e., whether the number of particles on
a given lattice site is even or odd \cite{CBPE12}.
The parity correlator reads
\begin{eqnarray}
\label{parity}
F_{(-1)^n}(\mu,\nu)
&=&
\frac{8}{N^2}\sum_{\mathbf{p},\mathbf{q}}
e^{i(\mathbf{p}+\mathbf{q})\cdot(\mathbf{x}_\mu-\mathbf{x}_\nu)}
%\times
%\nn
%&&
%\times
\left(f^{11}_\mathbf{p}f^{11}_\mathbf{q}+f^{12}_\mathbf{p}f^{21}_\mathbf{q}
\right)
%\nn
%&&
%-\frac{2}{N^2}\sum_{\mathbf{p},\mathbf{q}}
%\frac{
%3 J(T_\mathbf{q} (U-\omega_\mathbf{p})
%+T_\mathbf{p}(U-\omega_\mathbf{q}))
%}{\omega_\mathbf{p}\omega_\mathbf{q}}
%e^{i(\mathbf{p}+\mathbf{q})(\mathbf{x}_\mu-\mathbf{x}_\nu)}
%\nn
%&&
%+\frac{2}{N^2}\sum_{\mathbf{p},\mathbf{q}}
%\frac{
%17 J^2 T_\mathbf{p} T_\mathbf{q}+
%(U-\omega_\mathbf{p})(U-\omega_\mathbf{q})
%}{\omega_\mathbf{p}\omega_\mathbf{q}}
%e^{i(\mathbf{p}+\mathbf{q})(\mathbf{x}_\mu-\mathbf{x}_\nu)}
\,.
\end{eqnarray}
Again, this result can be compared with a perturbative expansion into
powers of $J/U$.
Assuming a hyper-cubic lattice ${\mathbb Z}^D$ in $D$ dimensions with 
nearest-neighbour hopping (i.e., $Z=2D$), we obtain up to quartic order 
in $J$ 
\begin{eqnarray}
F_{(-1)^n}(\mu,\nu)
&=&
\left(\frac{J}{ZU}\right)^2 8n(n+1)+
\nonumber\\
&&
+\left(\frac{J}{ZU}\right)^4\frac{2  n (1 + n)}{3} 
\left[n(n+1)(70 - 208 Z + 48 Z^2)+
\right.
\nonumber \\
&&
\quad
\left.
%+ n (70 - 208 Z + 48 Z^2)+ 
+4 - 22 Z + 9 Z^2 \right]
+\ord(J^6)
\,,
\end{eqnarray}
where $\mu$ and $\nu$ are nearest neighbours and 
$n=\langle\hat{n}_\mu\rangle$ is an arbitrary (integer) filling. 
Inserting $n=1$ and keeping only the lowest-order $1/Z^2$ terms, we may 
compare this result with (\ref{parity}), after an expansion into powers 
of $J$, and find perfect agreement. 

However, there is an interesting observation regarding the above equation: 
In one spatial dimension with $Z=2$ nearest neighbours, the $J^4$ 
contribution in the above equation is negative. 
This suggests that the parity correlator assumes its maximum at a finite 
value of $J$ (in the Mott phase), which can indeed be confirmed by numerical 
simulations, see, e.g., \cite{E11} and Section \ref{numerical}.  
In two or more spatial dimensions, the situation is different. 
Even though the parity correlator should still assume its maximum at some 
finite value of $J$, this value is quite close to the phase transition or 
already in the superfluid regime.
Thus, this maximum is not visible in our $1/Z$-expansion starting in the 
Mott state, which predicts a monotonously increasing parity correlation 
in its region of applicability. 

In analogy to Sections \ref{superquench} and \ref{equilibration}, 
we can also study the correlations after a quantum quench with 
$J(t)=J\Theta(t)$.
Again, there are no contributions to the particle-number and parity 
correlations in first order $1/Z$ -- but, to second order $1/Z$, 
we find formally the same expressions as in the static case 
(\ref{parity}) and (\ref{number})
\begin{eqnarray}
\label{numberquench}
F_{n}(\mu,\nu)
&=&
\frac{2}{N^2}\sum_{\mathbf{p},\mathbf{q}}
e^{i(\mathbf{p}+\mathbf{q})\cdot(\mathbf{x}_\mu-\mathbf{x}_\nu)}
\left(
f^{11}_\mathbf{p}(t)f^{11}_\mathbf{q}(t)-
f^{12}_\mathbf{p}(t)f^{21}_\mathbf{q}(t)
\right)
\,,
\end{eqnarray}
and
\begin{eqnarray}
\label{parityquench}
F_{(-1)^n}(\mu,\nu)
&=&
\frac{8}{N^2}\sum_{\mathbf{p},\mathbf{q}}
e^{i(\mathbf{p}+\mathbf{q})\cdot(\mathbf{x}_\mu-\mathbf{x}_\nu)}
%\times
%\nn
%&&
%\times
\left(
f^{11}_\mathbf{p}(t)f^{11}_\mathbf{q}(t)
+f^{12}_\mathbf{p}(t)f^{21}_\mathbf{q}(t)
\right)
\,,
\end{eqnarray}
where $f^{12}_\mathbf{p}(t)$,$f^{21}_\mathbf{p}(t)$ and $f^{11}_\mathbf{p}(t)$
are now given by equations (\ref{quench-h+h}) and (\ref{quench-h+p}).
The parity correlations after a quench have been experimentally observed
in a one-dimensional setup \cite{CBPE12}.
Although the hierarchical expansion relies on a large coordination number,
we find qualitative agreement between the theoretical prediction 
(\ref{parityquench}) for $Z=2$ and the results from \cite{CBPE12}.
For large times $t$ and distances $\mathbf{x}_\mu-\mathbf{x}_\nu$, 
we may estimate the integrals over $\mathbf{p}$ and $\mathbf{q}$ in 
the expressions (\ref{numberquench}) and (\ref{parityquench}) via the 
stationary-phase or saddle-point approximation.
The dominant contributions stem from the momenta satisfying the 
saddle-point condition
\begin{eqnarray}
\label{statphase}
\nabla_\mathbf{k}
\left[
\mathbf{k}\cdot(\mathbf{x}_\mu-\mathbf{x}_\nu)\pm\omega_\mathbf{k}t
\right]
=0
\,.
\end{eqnarray}
Thus their structure is determined by the group velocity 
$\mathbf{v}_\mathbf{k}=\nabla_\mathbf{k}\omega_\mathbf{k}$. 
If the equation $\mathbf{x}_\mu-\mathbf{x}_\nu=\pm\mathbf{v}_\mathbf{k}t$ 
has a real solution $\mathbf{k}$, i.e., if the distance 
$\mathbf{x}_\mu-\mathbf{x}_\nu$ can be covered in the time $t$ with the 
group velocity $\mathbf{v}_\mathbf{k}$, then we get a stationary-phase
solution -- otherwise the integral will be exponentially suppressed 
(i.e., the saddle point $\mathbf{k}$ becomes complex). 
For small values of $J$, the maximum group velocity is given by 
$\mathbf{v}_\mathbf{k}^{\rm max}\approx3J$, which determines the 
maximum propagation speed of the correlations, i.e., the effective 
light-cone.

\section{Numerical Simulations for the Bose-Hubbard model}\label{numerical}

In the following we analyze the Bose-Hubbard system 
(\ref{Bose-Hubbard-Hamiltonian}) 
numerically in one and two dimensions.
All calculations are carried out on a finite lattice with $L$
lattice sites and $N$ bosons.
%\subsection{Ground state and excitations in 1D}
%------------------------------------------------------------------------

%------------------------------------------------------------------------
\subsection{General formalism for the one-dimensional Hubbard model}
%------------------------------------------------------------------------

The eigenstates of lattice systems are calculated by means of
exact numerical diagonalisation of the Hamiltonian
matrix~\cite{EKO,KPS,KPPS,KS,RB,RB04,KT08,HSTR,ZD}
which can be obtained from the Hamiltonian (\ref{Bose-Hubbard-Hamiltonian}) 
using the basis of Fock states
\begin{equation}
\label{Fock}
|{\bf n}_\Gamma\rangle
=
\bigotimes_{\mu=1}^L
|n_{\Gamma\mu}\rangle
\;,
\quad
\Gamma=1,\dots,{\cal D}
\;,\quad
{\cal D}=\frac{(N+L-1)!}{N!(L-1)!}
\;,
\end{equation}
where $\Gamma$ labels the configuration of the bosons and the occupation 
numbers 
of individual lattice sites $n_{\Gamma\mu}$ satisfy the condition
\begin{equation}
N
=
\sum_{\mu=1}^L
n_{\Gamma\mu}
\;.
\end{equation}
The matrix dimension can be reduced by factor $L$ for homogeneous lattices 
with periodic boundary conditions ($\hat b_{L+1}\equiv\hat b_{1}$).
In this case the Hamiltonian commutes with the unitary translation operator
$\hat{\cal T}$ which acts through cyclic permutation on the lattice bosons.
Due to the periodic boundary conditions, the operator satisfies 
$\hat{{\cal T}}^L=1$.
As a basis one can use linear combinations of the Fock states~(\ref{Fock}) 
in the form
\begin{equation}
\label{basis}
|{\bf n}_{K\Gamma}\rangle
=
{\cal N}_\Gamma
\sum_{\mu=1}^{L}
\left(
    \frac{\hat{\cal T}}{\tau_K}
\right)^{\mu-1}
 |{\bf n}_\Gamma\rangle
\;,
\end{equation}
which are eigenstates of the operator $\hat{{\cal T}}$ for the eigenvalue
$
\tau_K
=
\exp
\left(
    i K
\right)
$.
${\cal N}_\Gamma$ are normalisation constants chosen such that
$\langle{\bf n}_{K\Gamma}|{\bf n}_{K'\Gamma'}\rangle =
\delta_{\Gamma\Gamma'}\delta_{KK'}$.
Here the state $|{\bf n}_\Gamma\rangle$ cannot be obtained by cyclic 
permutations of
$|{\bf n}_{\Gamma'}\rangle$ with $\Gamma'\ne\Gamma$.
The eigenstates of the Hamiltonian have the following form
\begin{equation}
|K\Omega\rangle
=
\sum_{\Gamma=1}^{{\cal D}_K}
C_{K\Omega\Gamma}
|{\bf n}_{K\Gamma}\rangle
\;,\quad
\Omega=1,\dots,{\cal D}_K
\;,\quad
\sum_{K=0}^{L-1}
{\cal D}_K
=
{\cal D}
\;,
\end{equation}
and the corresponding eigenenergies are denoted by $E_{K\Omega}$.

If the complete eigenvalue problem is solved, one can work out
expectation value of any operator $\hat{\cal O}$
at arbitrary temperature in a canonical ensemble as
\begin{equation}
\langle
    \hat{\cal O}
\rangle_T
=
\frac{1}{{\cal Z}(T)}
\sum_{K\Omega}
\langle
    K\Omega|\hat{\cal O}|K\Omega
\rangle
\exp
\left(
    -\frac{E_{K\Omega}}{ T}
\right)
\end{equation}
with the partition function
\begin{equation}
{\cal Z}(T)
=
\sum_{K\Omega}
\exp
\left(
    -\frac{E_{K\Omega}}{T}
\right)
\;.
\end{equation}
Using the set of the basis states (\ref{basis}) we were able to solve 
numerically the complete eigenvalue problem for $N=L=9$.

In order to study zero-temperature properties of the system, 
it is sufficient to calculate the ground state. 
This can be done exactly with the aid
of iterative numerical solvers for sparse matrices of large dimensions
along the lines of \cite{RB}.
We were able to do this up to $N=L=14$.

%------------------------------------------------------------------------
\subsection{Energy spectrum}\label{energyspectrum}
%------------------------------------------------------------------------

In the limit of vanishing hopping, $J=0$, the basis states~(\ref{basis}) are
the eigenstates of the Hamiltonian (\ref{Bose-Hubbard-Hamiltonian}) which are,
apart from the ground state, degenerate.
The ground state has equal occupation numbers at each lattice site, that is
$n_{\Gamma\mu}\equiv n$.
It exists only at $K=0$ and has the energy 
\begin{equation}
 E_{01}=L\frac{U}{2}n(n-1)\,.
\end{equation}
The energy eigenvalues of the degenerated excited states are given through
\begin{equation}
E_{K\Gamma}
=
\sum_{\mu=1}^L
\frac{U}{2}
n_{\Gamma\mu}
\left(
    n_{\Gamma\mu}-1
\right)
\end{equation}
and do not depend on $K$, corresponding to flat energy bands.
The lowest band contains $L(L-1)$ degenerate eigenstates with the energies
$E_{K\Gamma}=E_{01}+U$.
These states correspond to bosonic configurations with the same 
occupation numbers $n$ at any site except two, one of which contains 
$n-1$ boson and the other one $n+1$.
The highest band contains $L$ degenerate states with all atoms sitting at one
lattice site.
These states have the energy $E_{K\Gamma}=UN(N-1)/2$.
A finite hopping rate $J$ lifts the degeneracy,
the bands aquire finite widths and can even overlap 
if the tunneling parameter is large enough.

The full energy spectrum calculated for $N=L=9$ and $J/U=0.1$,
which corresponds to the Mott-insulator state, 
is shown in Fig.~\ref{sp_01}(a).
The lowest dot at $K=0$ is the ground state energy $E_{01}$, 
see also  Fig.~\ref{sp_01}(b).
Also the lowest excited state is located at $K=0$ and has the energy $E_{02}$.
Together with the energies $E_{K2}$, where $K\ne0$, they form the lowest 
excitation branch shown by the black lines in Figs.~\ref{sp_01}(b,c).
An increase of the system size leads to more dense distribution of the 
points and the solid line becomes smoother.
We see that, at small momenta $K$, the lowest excitation branch 
can be approximated by a pseudo-relativistic form
%
%Assuming that at small momenta the lowest excitation branch has 
%relativistic form
%
\begin{equation}
\label{rel}
\omega_{\bf K}^2
\equiv
(E_{K2}-E_{01})^2
=
(\Delta{\cal E})^2
+
K^2
v_{\rm eff}^2
\;. 
\end{equation}
Thus we can estimate the effective velocity $v_{\rm eff}$ using the 
numerically calculated values of $E_{01}$, $E_{02}$ and $E_{11}$. 
The results for different system sizes are shown in Fig.~\ref{sp_01}(d).
With the increase of $J/U$ the energy bands become broader and
the gap in the excitation spectrum $E_{02}-E_{01}$ becomes smaller.

The energy eigenvalues in the lowest band in Fig.~\ref{sp_01} 
correspond to particle-hole excitations of the form 
$\hat{p}^\dagger_{k_p}\hat{h}_{k_h}|\Psi_\mathrm{Mott}\rangle$
with the total momentum $k_p+k_h=K$.
When discussing the ground-state properties or the dynamics after a quench 
in the previous Sections, it was sufficient to consider translationally
invariant states with $K=0$, i.e., $k_p=-k_h=k$, where $k$ corresponds to 
the relative momentum. 
In the discussion of the Sauter-Schwinger analogue, we considered a 
spatially constant potential gradient and absorbed it into the 
Fourier coefficients $T_\mathbf{k}(t)$ via a Peierls transformation, 
finally arriving at the evolution equations (\ref{holedgl}) and 
(\ref{partdgl}) for particle and hole operators with $k_p=-k_h=k$.
However, for arbitrary potentials $V_\mu$, this is no longer possible 
in this simple form. 
In order to satisfy the equations of motion (\ref{f12-Mott}-\ref{f11-Mott})
for the correlation functions for an arbitrary potential $V_\mu$, 
one should employ the generalised evolution equations
\begin{eqnarray}
\left(i\partial_t-V_\mu+\frac{U}{2}\right)\hat{h}_\mu
&=&
\frac{J}{Z}\sum_\kappa T_{\mu\kappa}
\left(\hat{h}_\kappa+\sqrt{2}\hat{p}_\kappa\right)\,,
\label{holegen}\\
\left(i\partial_t-V_\mu-\frac{U}{2}\right)\hat{p}_\mu
&=&
-\frac{\sqrt{2}J}{Z}\sum_\kappa T_{\mu\kappa}
\left(\hat{h}_\kappa+\sqrt{2}\hat{p}_\kappa\right)
\label{partgen}\,.
\end{eqnarray}
%
%When the potential $V_\mu$ is periodic on a lattice, it makes sense to 
%define a translation operator with discete eigenvalues $\exp(iK)$ and 
%
Here the particle and hole-operators are fixed up to a $k_{h,p}$-independent 
phase.
For the limiting case $V_\mu\to0$ we find from Eqs.~(\ref{holegen}) and 
(\ref{partgen}) the following eigenfrequencies 
\begin{eqnarray}
\omega^h_{k_h}&=&-\frac{1}{2}(JT_{k_h}+\omega_{k_h})\,,
\\
\omega^p_{k_p}&=&-\frac{1}{2}(JT_{k_p}-\omega_{k_p})\,,
\end{eqnarray}
where $\omega_{k}$ are the eigenfrequencies defined in 
(\ref{eigen-frequency}).
These excitations define the lowest band with the energies 
\begin{eqnarray}
\label{lowest-band}
E_{K\Omega}=\omega^p_{k_p}-\omega^h_{k_h}\,,
\end{eqnarray}
where $k_h+k_p=K$.
The spectrum is depicted in Fig.~\ref{partholeband} and the effective 
velocity reads
\begin{eqnarray}
v_\mathrm{eff}=\frac{1}{2}\sqrt{J(3 U-J)}
\end{eqnarray}
which has for $J/U=0.1$
the numerical value $v_\mathrm{eff}/U=0.269$.
Including the second order corrections in $1/Z^2$ we have a slightly 
lower value $v_\mathrm{eff}/U=0.264$, compare Fig.~\ref{sp_01}(d).

%------------------------------------------------------------------------
\begin{figure}[t]
 \includegraphics[width=15cm]{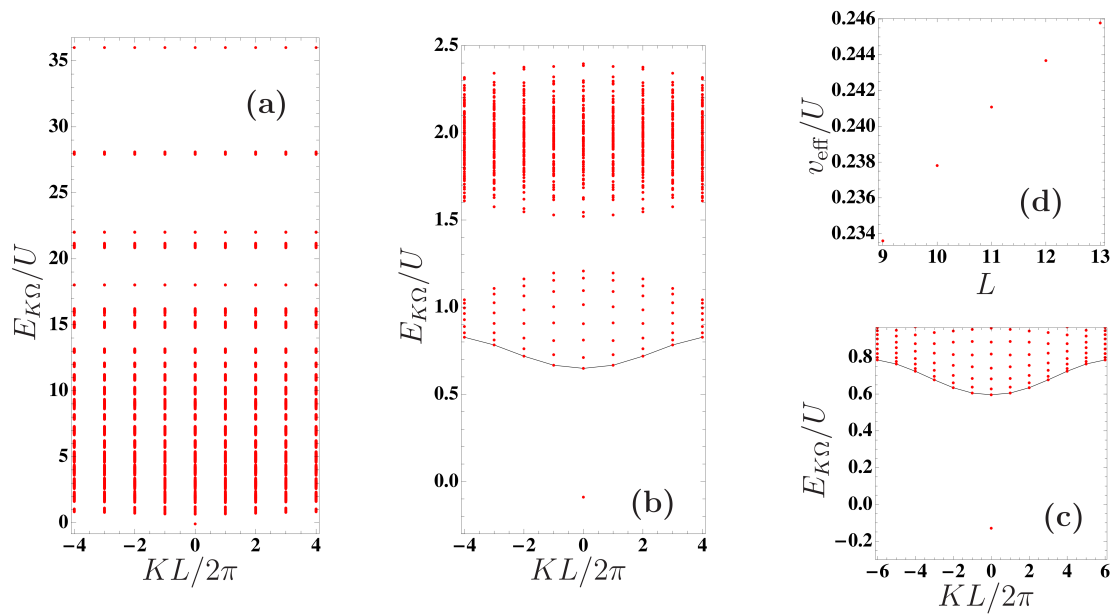}
%\centering
% \begin{minipage}[c]{9.5cm}
% \psfrag{k}[c]{$KL/2\pi$}
% \psfrag{E}[b]{$E_{K\Omega}/U$}
% 
% \psfrag{a}[c]{\bf(a)}
% \includegraphics[width=4cm]{spectrum-J_0.05-N_9-L_9.eps}
% \hspace{1cm}
% \psfrag{b}[c]{\bf(b)}
% \includegraphics[width=4cm]{spectrum_low-J_0.05-N_9-L_9.eps}
% \end{minipage}
% \hspace{1cm}
% \begin{minipage}[c]{4cm}
% \psfrag{L}[c]{$L$}
% \psfrag{c}[b]{$v_{\rm eff}/U$}
% 
% \psfrag{d}[c]{\bf(d)}
% \includegraphics[width=4cm]{ceff_L-J_0.05-n0_1-small.eps}
% 
% \vspace{0.5cm}
% 
% \psfrag{k}[c]{$KL/2\pi$}
% \psfrag{E}[b]{$E_{K\Omega}/U$}
% \psfrag{c}[c]{\bf(c)}
% \includegraphics[width=4cm]{spectrum_low-J_0.05-N_13-L_13.eps}
% \end{minipage}
\caption{
         {\bf(a)} Full energy spectrum and {\bf(b)} its lowest part for $N=L=9$.
         {\bf(c)} Lowest part of the spectrum for $N=L=13$.
         {\bf(d)} Effective velocity in a one-dimensional lattice with $n=1$ 
atom per site and 
         $J/U=0.1$.
        }
\label{sp_01}
\end{figure}
\begin{figure}
\begin{center}
\includegraphics[width=10cm]{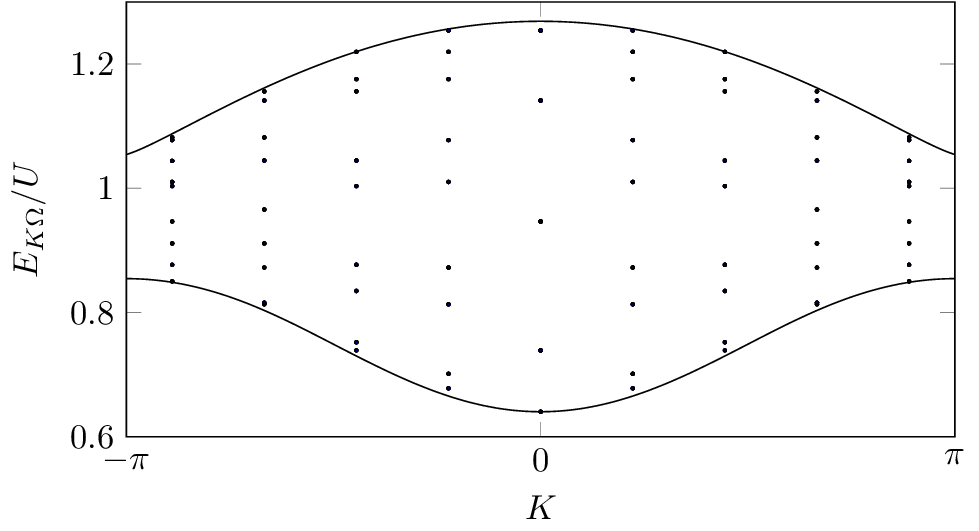}
\caption{Boundaries of the lowest energy excitation band in the continuum limit (solid lines) and 
energy excitations for $L=9$ (points) from Eq.~(\ref{lowest-band}) in a one-dimensional lattice
with $n=1$ and $J/U=0.1$.}
\label{partholeband}
\end{center}
\end{figure}
%------------------------------------------------------------------------

%------------------------------------------------------------------------
\subsection{Probability distribution of the occupation numbers}
%------------------------------------------------------------------------

We calculate the ground state and then the probabilities $p(n_\mu)$
to have $n_\mu$ atoms at a lattice site which satisfy the normalisation 
condition
\begin{equation}
\sum_{n_\mu=0}^N
p(n_\mu)
=1
\;.
\end{equation}
As in the previous Section, we consider the system with $N=L$.
From Eqs. (\ref{Mott-state}) and (\ref{superfluid-state}) it follows that
in the limit $J=0$ we have $p(n_\mu)=\delta_{n_\mu,1}$,
and in the opposite limit $U=0$ the probabilities are given by the binomial distribution
\begin{equation}
p(n_\mu)
=
\frac{N!}{(N-n_\mu)! n_\mu!}\left(\frac{1}{N}\right)^{n_\mu}\left(1-\frac{1}{N}\right)^{N-n_\mu}
\;.
\end{equation}
The result for $N=14$ at arbitrary $J/U$ and at zero temperature
is shown in Fig.~\ref{p_n}.
One can clearly see that the probability to have three particles or more at one
lattice site is very small which is consistent with the approximations
used in the $1/Z$ expansion.

The particle-number distribution at finite temperature is shown in 
Fig.~\ref{p_n-T}(b).
Comparison with the zero-temperature result for the same system size
[Fig.~\ref{p_n-T}(a)]
indicates that temperature has stronger influence at smaller values of $J/U$.

%------------------------------------------------------------------------

\begin{figure}[t]
%\centering
\psfrag{J}[c]{$J/U$}
\psfrag{p}[b]{$p(n_\mu)$}
\begin{center}
\includegraphics[width=7cm]{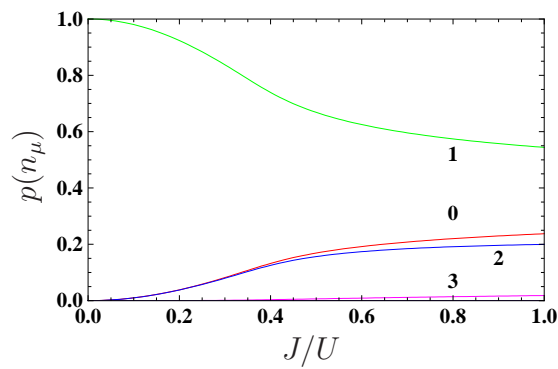}
\end{center}
\caption{Probabilities to have $n_\mu=0$ (red), $1$ (green), $2$ (blue), 
$3$ (magenta)
         atoms at a lattice site in a one-dimensional lattice
         of $L=14$ sites with $n=1$ atom per site at zero temperature.
        }
\label{p_n}
\end{figure}

%------------------------------------------------------------------------

%------------------------------------------------------------------------
\begin{figure}[t]
%\centering

\psfrag{J}[c]{$J/U$}
\psfrag{p}[b]{$p(n_\mu)$}

\psfrag{a}[c]{\bf(a)}
\includegraphics[width=7cm]{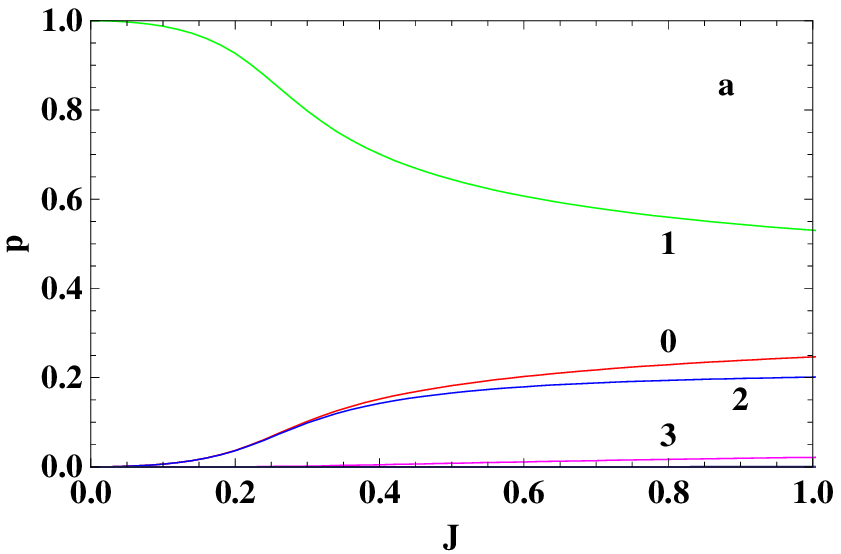}
\hspace{1cm}
\psfrag{b}[c]{\bf(b)}
\includegraphics[width=7cm]{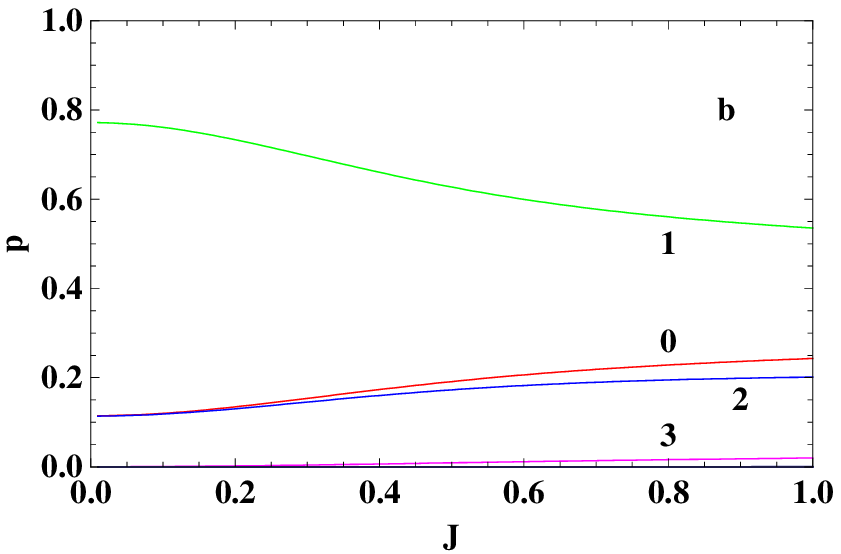}

\psfrag{a}[c]{\bf(c)}
\includegraphics[width=7cm]{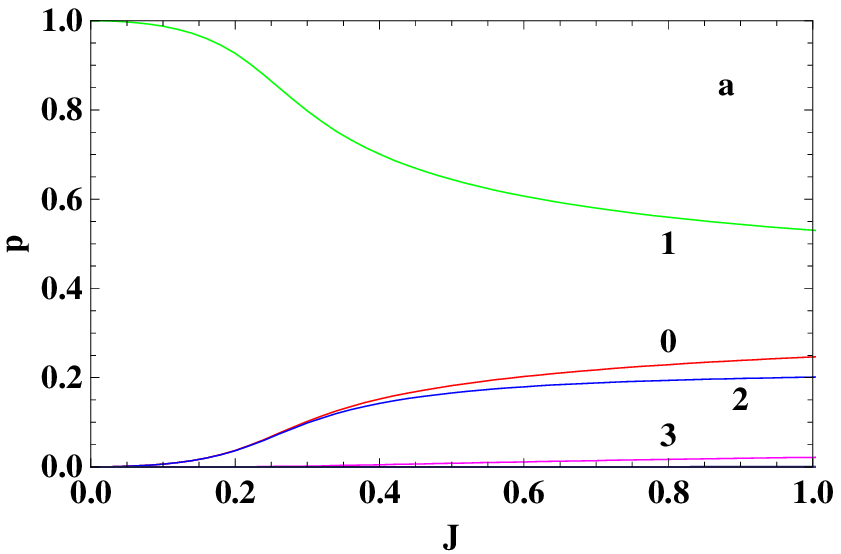}
\hspace{1cm}
\psfrag{b}[c]{\bf(d)}
\includegraphics[width=7cm]{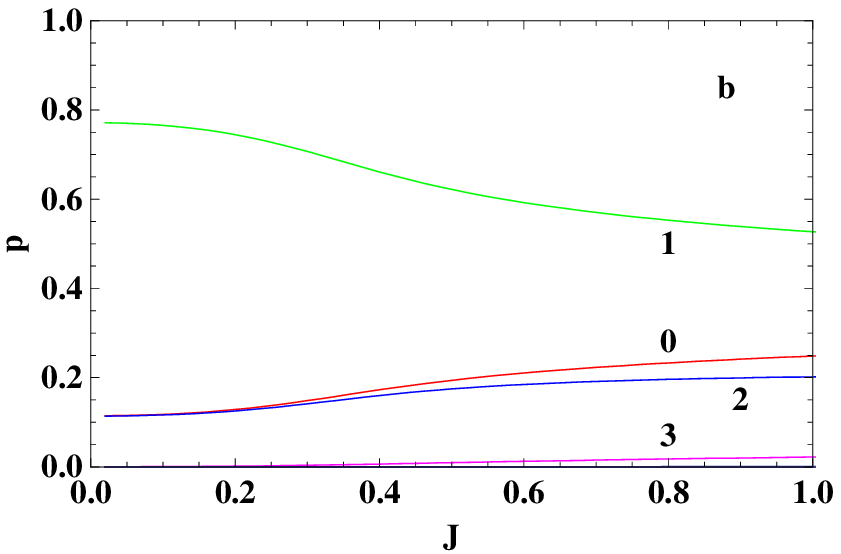}

\caption{Probabilities to have $n_\mu=0$ (red), $1$ (green), $2$ (blue), 
$3$ (magenta)
         atoms at a site in a lattice with $n=1$ atom per site.
         {\bf(a)} and {\bf(b)}: one-dimensional lattice of $L=9$ sites;
         {\bf(c)} and {\bf(d)}: two-dimensional lattice of $3\times 3$ sites;
         {\bf(a)} and {\bf(c)}: zero temperature;
         {\bf(b)} and {\bf(d)}: $T/U=0.3$.
        }
\label{p_n-T}
\end{figure}
%------------------------------------------------------------------------

%------------------------------------------------------------------------
\subsection{Two-point correlation functions}
%------------------------------------------------------------------------

In this Subsection, we consider two-point correlation functions 
which have been discussed in Section~\ref{Z2}.
%
% \begin{equation}
% F_{\cal O}(\mu,\nu)
% =
% \langle
%     \hat{\cal O}_\mu
%     \hat{\cal O}_\nu
% \rangle
% -
% \langle
%     \hat{\cal O}_\mu
% \rangle
% \langle
%     \hat{\cal O}_\nu
% \rangle
% \;,
% \end{equation}
% where $\hat{\cal O}_\mu$ is a local operator.
%
Due to the translational invariance, the correlation functions depend
on the distance $s=|x_\mu-x_\nu|$ and have the property
$F_{\cal O}(\mu,\nu)=F_{\cal O}(s)=F_{\cal O}(L-s)$ 
in view of the periodic boundary conditions.

First we consider the parity correlation function 
$F_{(-1)^n}(s)$.
% defined as
% \begin{equation}
% C_s
% =
% \langle(-1)^{\hat n_\mu + \hat n_\nu}\rangle
% -
% \langle(-1)^{\hat n_\mu}\rangle
% \langle(-1)^{\hat n_\nu}\rangle
% \;,\quad
% s=|x_\mu-x_\nu|
% \;.
% \end{equation}
Its dependence on $J/U$ as well as on the distance $s$ is shown
in Fig.~\ref{par-dd}.
This result is in a very good quantitative agreement with
the DMRG-calculations~\cite{E11}.
Note that our definition of $J$ differs by a factor $Z$ from that 
used in \cite{E11}.
Fig.~\ref{par-dd} shows the number correlation function 
$F_{n}(s)$.

%------------------------------------------------------------------------
\begin{figure}[t]
%\centering

\psfrag{J}[c]{$J/U$}
\psfrag{par}[b]{$F_{(-1)^n}(s)$}

\psfrag{a}[c]{\bf(a)}
\includegraphics[width=7cm]{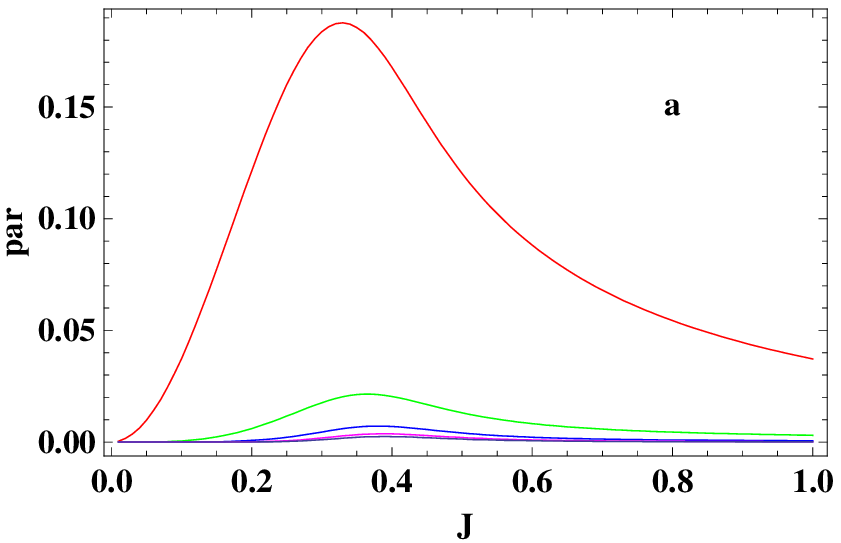}
\hspace{1cm}
\psfrag{n}[c]{$s$}
\psfrag{b}[c]{\bf(b)}
\includegraphics[width=7cm]{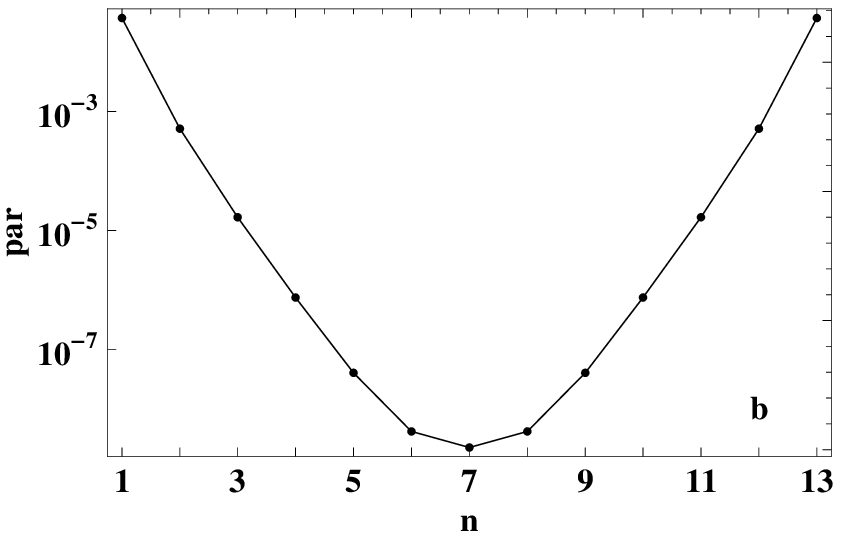}

\psfrag{J}[c]{$J/U$}
\psfrag{dd}[b]{$F_{n}(s)$}

\psfrag{a}[c]{\bf(c)}
\includegraphics[width=7cm]{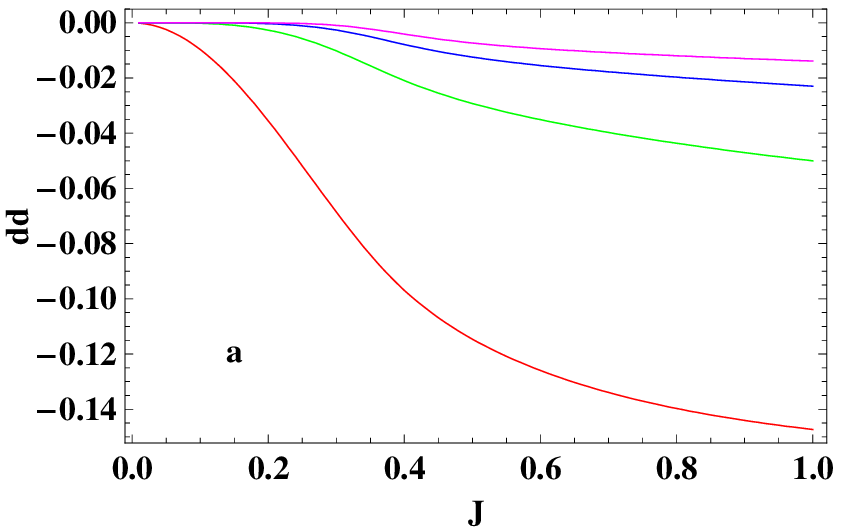}
\hspace{1cm}
\psfrag{n}[c]{$s$}
\psfrag{ddm1}[b]{$\left|F_{n}(s)\right|$}
\psfrag{b}[c]{\bf(d)}
\includegraphics[width=7cm]{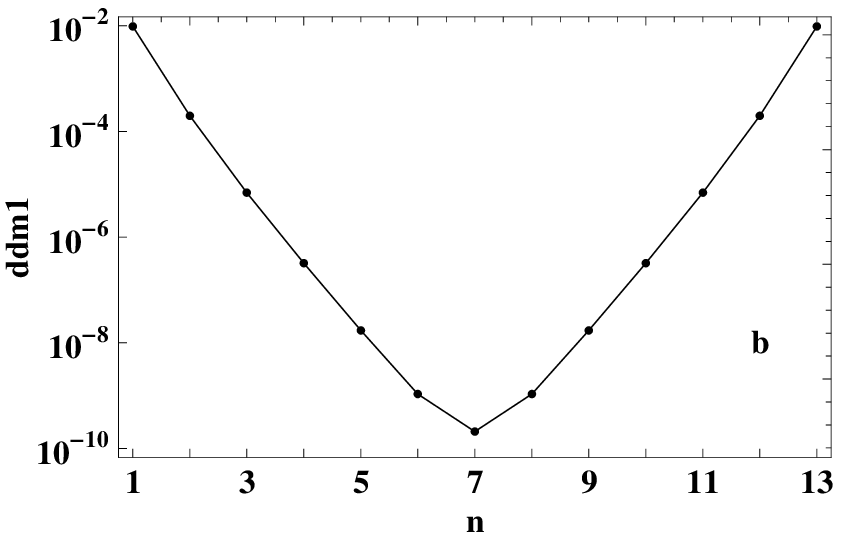}

\caption{
         Parity correlation [{\bf (a)} and {\bf (b)}]
         and density-density correlation [{\bf (c)} and {\bf (d)}]
         in a one-dimensional lattice of $L=14$ sites with $n=1$
         atom per site at zero temperature.
         {\bf (a)} and {\bf (c)}: dependence on $J/U$ for
         $s=1$ (red), $2$ (green), $3$ (blue), $4$ (magenta);
         {\bf (b)} and {\bf (d)}: Dependence on $s$ for $J/U=0.1$.
         Due to the periodic boundary conditions
         correlation functions increase starting from $s=7$.
        }
\label{par-dd}
\end{figure}
%------------------------------------------------------------------------

Fig.~\ref{obdm} shows matrix elements of the one-body density matrix
$\langle \hat b_\mu^\dagger \hat b_{\mu+s} \rangle$
as well as its momentum distribution (\ref{momdist}).
% \begin{equation}
% P(k)
% =
% \frac{1}{N}
% \sum_{\mu,\nu}
% e^{ik(x_\mu-x_\nu)}
% \langle \hat b_\mu^\dagger \hat b_\nu \rangle
% \end{equation}
% which is basically the Fourier transform of
% $\langle \hat b_\mu^\dagger \hat b_{\mu+s} \rangle$.
In a finite lattice the quasi-momentum takes discrete values which are integer
multiples of $2\pi/L$. These allowed values are marked in Fig.~\ref{obdm}(a) 
by dots.
The momentum distribution calculated in the first order of $1/Z$ is also 
shown for comparison.
We observe good quantitative agreement.

%------------------------------------------------------------------------
\begin{figure}[t]
%\centering

\psfrag{k}[c]{$k$}
\psfrag{pdis}[b]{$P(k)$}
\psfrag{b}[c]{\bf(b)}
\psfrag{a}[c]{\bf(a)}
\includegraphics[width=7cm]{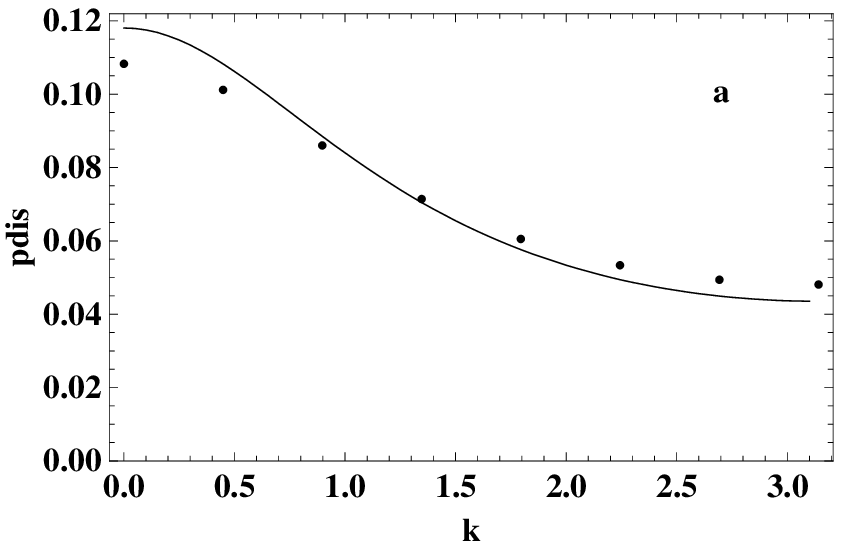}
\hspace{1cm}
\psfrag{dm}[b]{$\langle \hat b_\mu^\dagger \hat b_{\mu+s} \rangle$}
\psfrag{n}[c]{$s$}
\includegraphics[width=7cm]{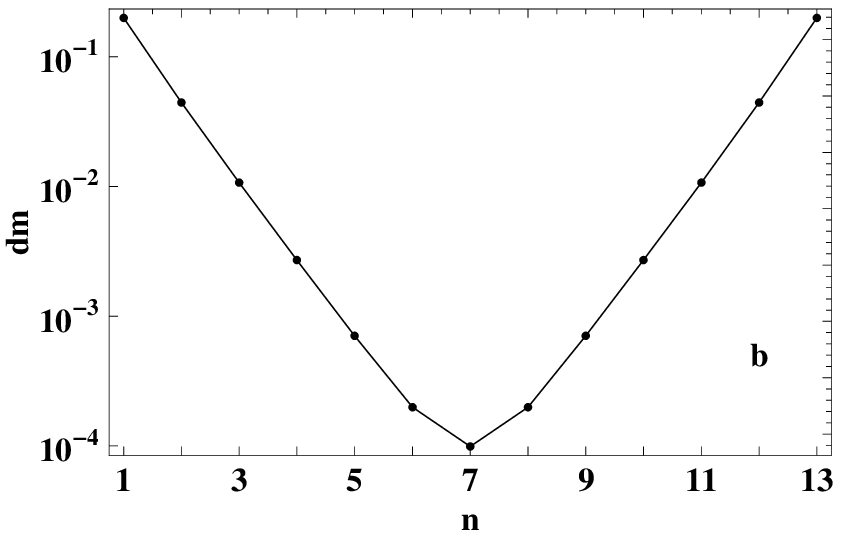}

\caption{
         {\bf(a)}: Momentum distribution (\ref{momdist}) in a one-dimensional lattice
	 with $n=1$ atom per site at $J/U=0.1$ and zero temperature.
	 The result obtained by exact diagonalisation for a finite 
	 lattice of $L=14$ is shown by dots. The solid line is a calculation 
	 in the first order of $1/Z$ for an infinite lattice.
	 {\bf(b)}:
         Correlation function $\langle\hat b_\mu^\dagger\hat b_{\mu+s}\rangle$ 
	 calculated by exact diagonalisation for a finite lattice with the 
	 same values of parameters as in {\bf(a)}.
        }
\label{obdm}
\end{figure}
%------------------------------------------------------------------------

%------------------------------------------------------------------------
\subsection{Particle-number distribution and correlation functions in 2D}
%------------------------------------------------------------------------

The whole procedure of exact numerical diagonalisation described for 
one-dimensional
systems can be generalised to higher dimensions.
We did numerical simulations for two-dimensional square lattices of $3\times3$
lattice sites with periodic boundary conditions.
Exact calculations for square lattices of the size $4\times4$ and larger 
were not
possible due to the problem with computer memory.
Due to the fact that the size of the two-dimensional system is very small
one can expect strong finite-size effects. However, as we will see later
numerical calculations give qualitatively correct predictions valid for 
large systems.

The probability distribution of the occupation numbers is shown
in Figs.~\ref{p_n-T}(c) and (d).
It is very similar to the one-dimensional case.
In a lattice of $3\times 3$ sites with periodic boundary conditions
the distance $s$ takes only three values $s=0,1,\sqrt{2}$ which makes the study
of the long-range behavior of the two-point correlation functions practically
impossible. Nevertheless some useful information can be obtained in the 
Mott-insulator
phase where the correlations decay exponentially.
Fig.~\ref{par-dd-obdm-2D}(a)
shows the dependence of the parity correlation function on $J/U$.
As in the one-dimensional case, $F_{(-1)^n}(s)$ has a maximum
at a finite value of $J$, which is, however, 
not in the Mott phase.
The results in Fig.~\ref{par-dd-obdm-2D} are in a good agreement with those 
obtained in \cite{E11} by DMRG and MPS calculations for large systems.
Correlations $\langle \hat n_\mu \hat n_\nu \rangle$ %{\color{red}($F_n$)}
and $\langle \hat b_\mu^\dagger \hat b_\nu \rangle$ %{\color{red}($F_a$)}
are shown in Figs.~\ref{par-dd-obdm-2D}(b),~\ref{par-dd-obdm-2D}(c), 
respectively.
They become stronger for increasing values of $J/U$.

%------------------------------------------------------------------------
\begin{figure}[t]
%\centering

\begin{center}
\psfrag{J}[c]{$J/U$}
\psfrag{par}[b]{$F_{(-1)^n}(s)$}
\psfrag{a}[c]{\bf(a)}
\includegraphics[width=7cm]{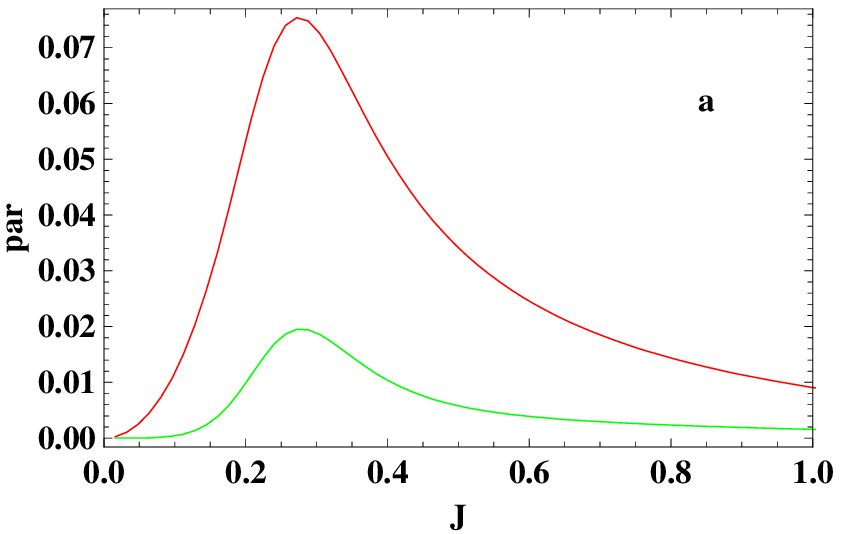}
\psfrag{J}[c]{$J/U$}
\psfrag{dd}[b]{$F_{n}(s)$}
\psfrag{b}[c]{\bf(b)}
\hspace{0.4cm}\includegraphics[width=7cm]{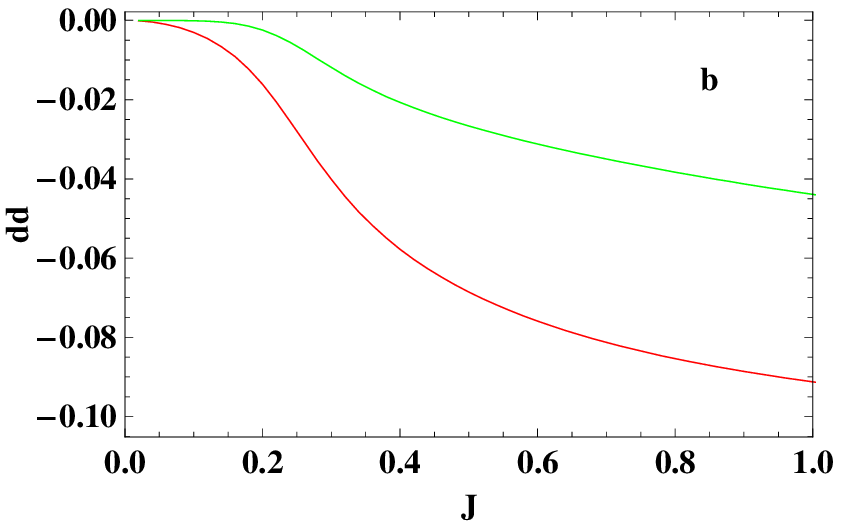}
 \end{center}

\psfrag{J}[c]{$J/U$}
\psfrag{dm}[b]{$\langle \hat b_\mu^\dagger \hat b_\nu \rangle$}
\psfrag{c}[c]{\bf(c)}
\begin{center}
\includegraphics[width=7cm]{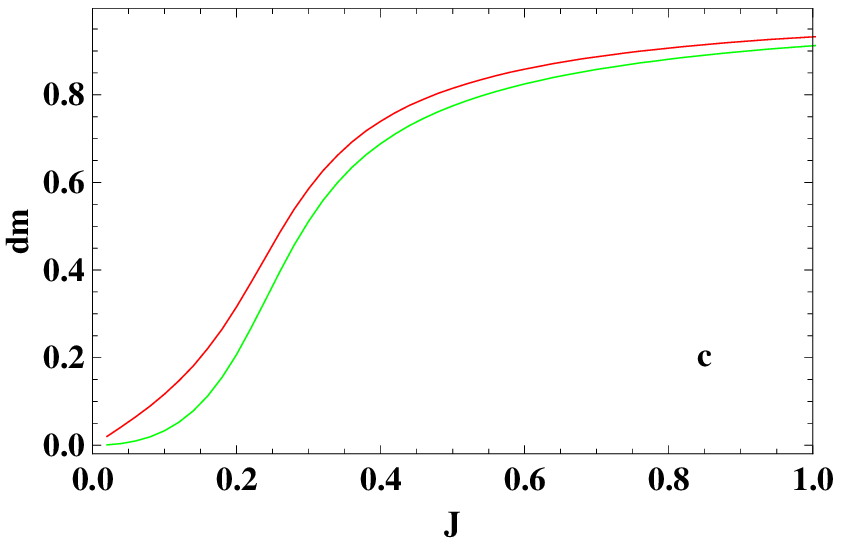}
 \end{center}

\caption{Parity correlation (a), density-density correlation (b)
         and elements of the one-body density matrix (c)
         in a two-dimensional lattice of $3\times 3$ sites with $n=1$
         atom per site at zero temperature.
         $s=1$ (red), $\sqrt{2}$ (green).
        }
\label{par-dd-obdm-2D}
\end{figure}
%------------------------------------------------------------------------

%------------------------------------------------------------------------
\subsection{Time evolution after quench}
%------------------------------------------------------------------------

If the complete set of the eigenvalues and eigenstates of the Hamiltonian 
is known, the time evolution of an arbitrary initial state $|\psi(0)\rangle$
can be calculated exactly without numerical integration,
provided that the Hamiltonian does not depend explicitly on time.
The initial state can be decomposed into the eigenstates of the Hamiltonian as
\begin{equation}
|\psi(0)\rangle
=
\sum_K
\sum_{\Omega=1}^{{\cal D}_K}
c_{K\Omega}
|K\Omega\rangle
\;,\quad
c_{K\Omega}
=
\langle
    \psi(0)|K\Omega
\rangle
\;.
\end{equation}
If the parameters of the Hamiltonian do not depend on time,
the evolution of the initial state is given by
\begin{equation}
|\psi(t)\rangle
=
\sum_K
\sum_{\Gamma=1}^{{\cal D}_K}
c_{K \Gamma}(t)
|{\bf n}_{K\Gamma}\rangle
\end{equation}
with
\begin{equation}
c_{K\Gamma}(t)
=
\sum_{\Omega=1}^{{\cal D}_K}
c_{K\Omega}
C_{K\Omega\Gamma}
\exp
\left(
    -i E_{K\Omega} t
\right)
\;.
\end{equation}
The time dependence of the expectation value of any operator 
$\hat{\cal O}$ can be
calculated as
\begin{equation}
\langle
    \hat{\cal O}
\rangle
(t)
=
\sum_{K\Gamma}
\sum_{K'\Gamma'}
\langle
    {\bf n}_{K\Gamma} |\hat{\cal O}| {\bf n}_{K'\Gamma'}
\rangle
c_{K\Gamma}^*(t)
c_{K'\Gamma'}(t)
\;.
\end{equation}
We will be dealing with operators $\hat{\cal O}$ which have the property
\begin{equation}
\langle
    {\bf n}_{K\Gamma} |\hat{\cal O}| {\bf n}_{K'\Gamma'}
\rangle
=
\langle
    {\bf n}_{K\Gamma} |\hat{\cal O}| {\bf n}_{K\Gamma'}
\rangle
\delta_{KK'}
\;.
\end{equation}
Then the expectation value averaged over the evolution time is given by
\begin{eqnarray}
\overline
{
\langle
    \hat{\cal O}
\rangle
}
&=&
\lim_{t\to\infty}
\frac{1}{t}
\int_0^t
\langle
    \hat{\cal O}
\rangle(t')
dt'\nonumber\\
& =&
\sum_{K}
\sum_{\Gamma}
\sum_{\Gamma'}
\langle
    {\bf n}_{K\Gamma} |\hat{\cal O}| {\bf n}_{K\Gamma'}
\rangle
\sum_{\Omega=1}^{{\cal D}_K}
\left|
    c_{K\Omega}
\right|^2
C^*_{K\Omega\Gamma}
C_{K\Omega\Gamma'}
\;.
\end{eqnarray}
We study time evolution of the initial state with exactly one atom at each
lattice site which is the ground state with $K=0$ of the Bose-Hubbard 
Hamiltonian
in the limit $J=0$. Since the Hamiltonian after quench preserves the 
translational
invariance, the time evolution involves only states with $K=0$.
In Figs.~\ref{pn-1D-t} and \ref{pn_2D-t}
we present numerical results for the particle-number distribution 
$p(n_\mu)$ and correlation function
$\langle \hat b_\mu^\dagger \hat b_\nu \rangle$ in one and two dimensions.
The purpose of this study is to address the question of (quasi) equilibration 
versus thermalisation in closed quantum systems.
Time evolution of the probabilities $p(n_\mu)$ for one- and two-dimensional 
systems is shown
in Figs.~\ref{pn-1D-t}(a),~\ref{pn_2D-t}(a), respectively.
On large time scales they oscillate around the averaged values shown by 
straight horizontal lines.
For the chosen value of $J/U=0.1$, the behavior $p(0)$ is almost 
indistinguishable from that of $p(2)$.
Figs.~\ref{pn-1D-t}(b),~\ref{pn_2D-t}(b) show the dependence of the 
probabilities on the temperature.
Averaged values of the probabilities correspond to the effective 
temperature which is slightly less than $0.15~U$.

%------------------------------------------------------------------------
\begin{figure}[t]
%\centering

\psfrag{t}[c]{$tU$}
\psfrag{Pn}[b]{$p(n_\mu)$}
\psfrag{a}[c]{\bf(a)}

\includegraphics[width=7cm]{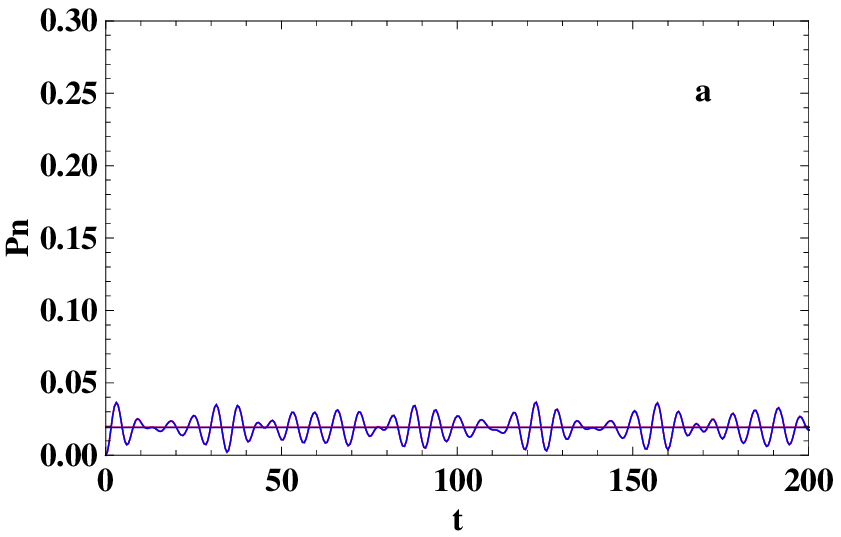}
\hspace{1cm}
\psfrag{T}[c]{$T/U$}
\psfrag{b}[c]{\bf(b)}
\includegraphics[width=7cm]{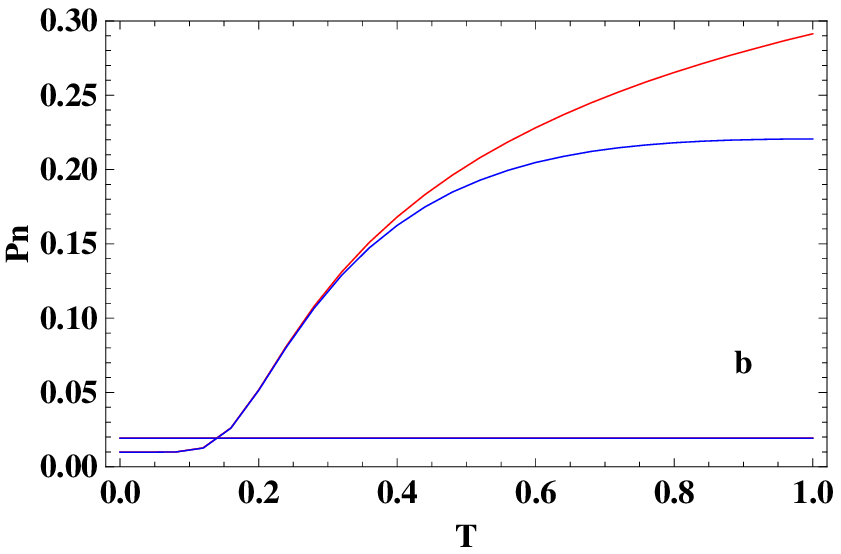}

\psfrag{OBDM}[b]{$\langle \hat b_\mu^\dagger \hat b_{\mu+s} \rangle$}

\psfrag{t}[c]{$tU$}
\psfrag{a}[c]{\bf(c)}
\includegraphics[width=7cm]{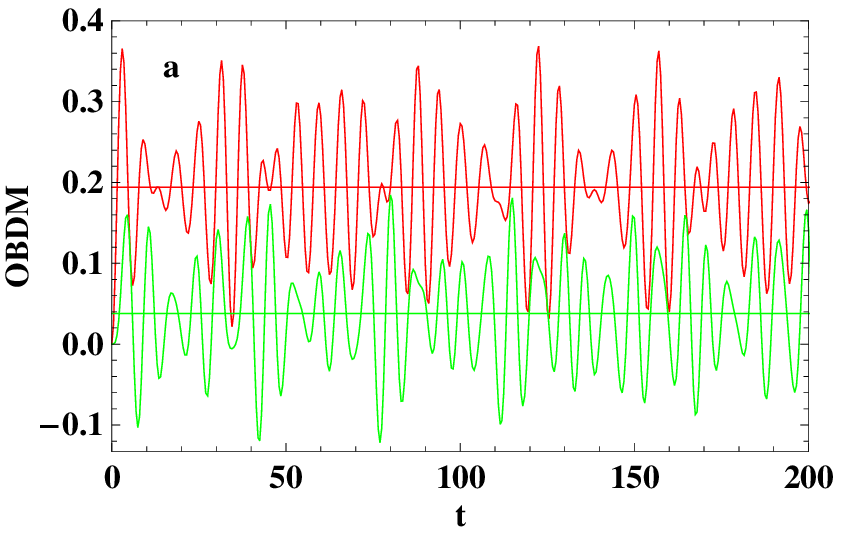}
\hspace{1cm}
\psfrag{T}[c]{$T/U$}
\psfrag{b}[c]{\bf(d)}
\includegraphics[width=7cm]{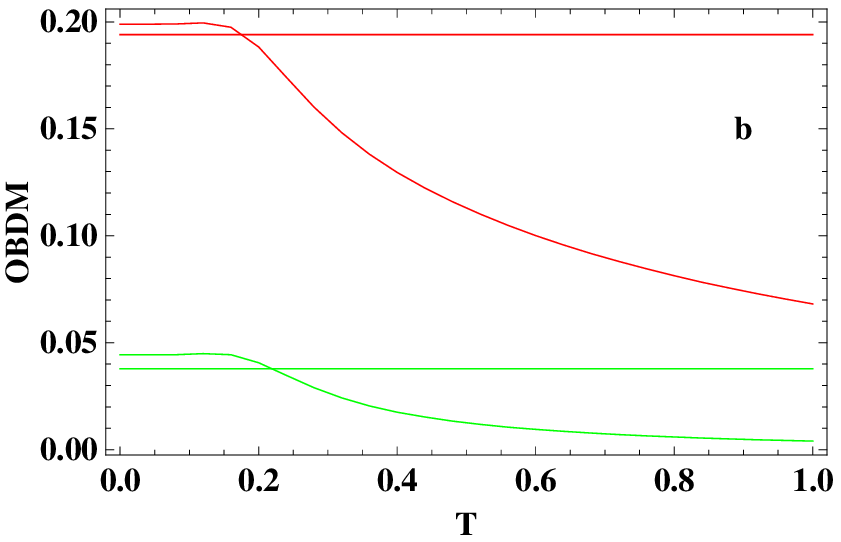}

\caption{Quench and (quasi) equilibration in a one-dimensional lattice 
of $L=9$ sites
         with $n=1$ atom per site.
         {\bf (a)}
         Time evolution of the
         probabilities to have $n_\mu=0$ (red), $2$ (blue)
         atoms at a lattice site after quench $J/U = 0 \to 0.1$.
         {\bf (b)}
         Probabilities to have $n_\mu=0$ (red), $2$ (blue)
         atoms at a lattice site for $J/U=0.1$ as a function of temperature.
         Straight horizontal lines in both panels show the values of probabilities
         averaged over the infinite evolution time.
         {\bf (c)}
         Elements of the one-body density matrix with $s=1$ (red), $2$ (green),
         after quench $J/U = 0 \to 0.1$.
         {\bf (d)}
         Elements of the one-body density matrix with $s=1$ (red), $2$ (green),
         for $J/U = 0.1$ as a function of temperature.
        }
\label{pn-1D-t}
\end{figure}
%------------------------------------------------------------------------

The time dependence of the correlation functions
$\langle \hat b_\mu^\dagger \hat b_\nu \rangle$
presented in Figs.~\ref{pn-1D-t}(c),~\ref{pn_2D-t}(c)
displays the same oscillating character.
In the one-dimensional system, the effective temperature corresponding 
to the averaged values of $\langle \hat b_\mu^\dagger \hat b_\nu \rangle$ 
can be defined but appears to be higher than that of the probabilities 
$p(n_\mu)$.
In contrast, in the two-dimensional case, the correlation functions 
$\langle \hat b_\mu^\dagger \hat b_\nu \rangle$ cannot be described by 
a thermal state, see Fig.~\ref{pn_2D-t}(d). 
The absence of effective temperature in the two-dimensional system is 
consistent with the result obtained within the $1/Z$ expansion 
in Section \ref{equilibration}.
%However, the $1/Z$ expansion fails to describe the thermalisation 
%in one dimension.

% it is only possible
%to define an effective temperature for $p(n_\mu)$.
%For the correlation functions 
%$\langle \hat b_\mu^\dagger \hat b_\nu \rangle$ instead,
%we find no effective temperature for $J/U=0.2$, see Fig.~\ref{pn_2D-t}(d).
%However, for sufficiently large values of $J/U$ it is possible to 
%define an effective temperature for 
%$\langle \hat b_\mu^\dagger \hat b_\nu \rangle$, 
%since the curves representing the thermal
%and the time-averaged expectation values intersect.

%------------------------------------------------------------------------
\begin{figure}[t]
%\centering

\psfrag{t}[c]{$tU$}
\psfrag{Pn}[b]{$p(n_\mu)$}
\psfrag{a}[c]{\bf (a)}

\includegraphics[width=7cm]{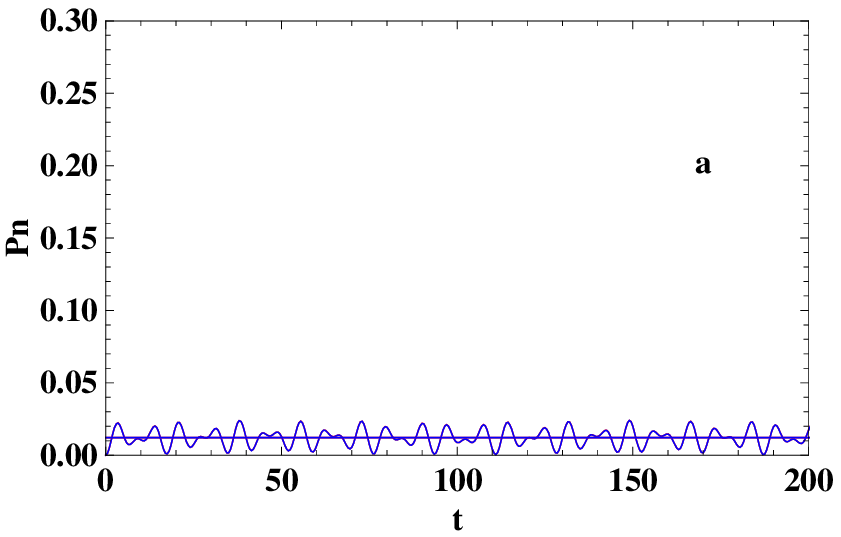}
\hspace{1cm}
\psfrag{T}[c]{$ T/U$}
\psfrag{b}[c]{\bf (b)}
\includegraphics[width=7cm]{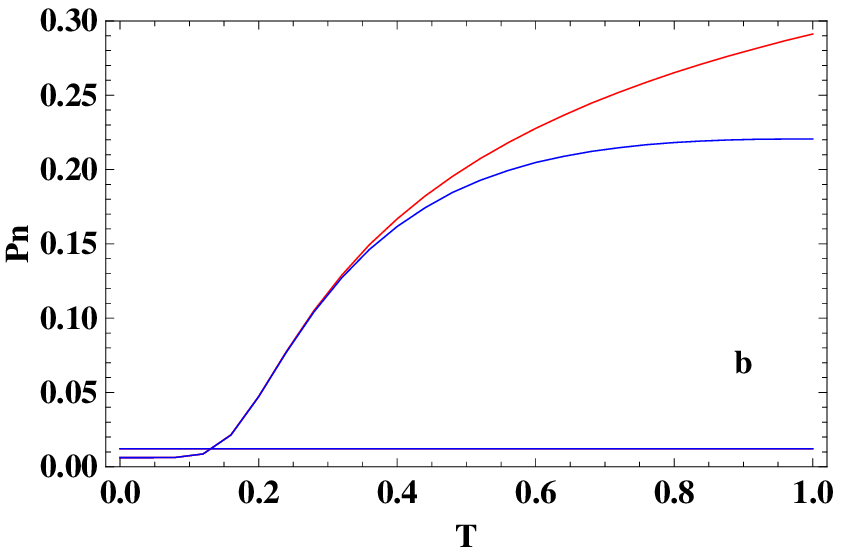}

\psfrag{OBDM}[b]{$\langle \hat b_\mu^\dagger \hat b_\nu \rangle$}

\psfrag{t}[c]{$tU$}
\psfrag{a}[c]{\bf (c)}
\includegraphics[width=7cm]{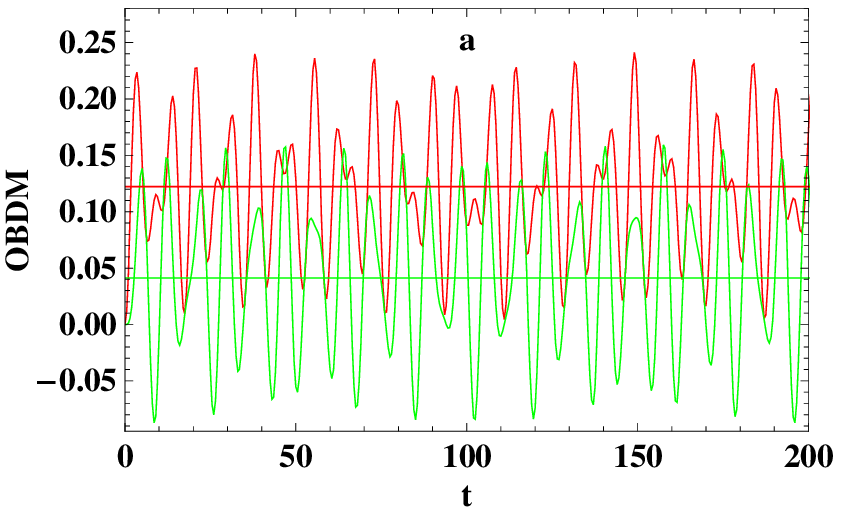}
\hspace{1cm}
\psfrag{T}[c]{$T/U$}
\psfrag{b}[c]{\bf (d)}
\includegraphics[width=7cm]{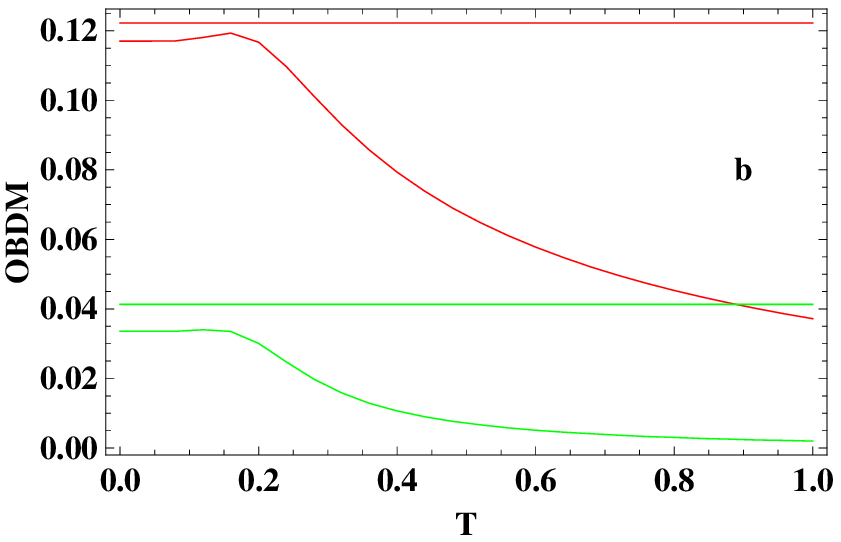}

\caption{Quench and (quasi) equilibration in a two-dimensional lattice
         of $3\times 3$ sites with $n=1$ atom per site.
         {\bf (a)}
         Probabilities to have $n_\mu=0$ (red), $2$ (blue)
         atoms at a lattice site after quench $J/U = 0 \to 0.1$.
         {\bf (b)}
         Probabilities to have $n_1=0$ (red), $2$ (blue)
         atoms at a lattice site for $J/U=0.1$ as a function of temperature.
         {\bf (c)}
         Elements of the one-body density matrix with 
$s=1$ (red), $\sqrt{2}$ (green)
         after quench $J/U = 0 \to 0.1$.
         {\bf (d)}
         Elements of the one-body density matrix 
with $s=1$ (red), $\sqrt{2}$ (green)
         for $J/U=0.1$ as a function of temperature.
        }
\label{pn_2D-t}
\end{figure}
%------------------------------------------------------------------------

%------------------------------------------------------------------------
\subsection{Tilt in one dimension}
%------------------------------------------------------------------------

In this Section, we calculate the probability to create a particle-hole 
excitation
due to the time-dependent tilt. Initial state of the system $|\psi(0)\rangle$
is the ground state of the Hamiltonian (\ref{Bose-Hubbard-Hamiltonian})
for finite value of $J/U$ in the Mott-insulator phase.
During the time evolution the system is described by the Hamiltonian 
(\ref{Bose-Hubbard-Tilt})
with the on-site energies
\begin{equation}
V_\mu
=
E_0
f(t/\tau)
x_\mu
\;,
\end{equation}
where function $f(s)$ has similar form as (\ref{sauterpulse})
\begin{equation}
\label{fs}
f(s)
=
\left\{
    \cosh^{-2}
    \left(
        s-\frac{5}{2}
    \right)
    -
    \cosh^{-2}
    \left(
        \frac{5}{2}
    \right)
\right\}
\left[
    1
    -
    \cosh^{-2}
    \left(
        \frac{5}{2}
    \right)
\right]^{-1}
\,,
\end{equation}
with $0<s<5$,
such that $f(0)=f(5)=0$ and $f(5/2)=1$. 

%------------------------------------------------------------------------
\begin{figure}[t]
%\centering
% \psfrag{a}[c]{\bf (a)}
% \psfrag{eps}[c]{$E_0/U$}
% \psfrag{pe}[b]{$P_{\rm exc}$}
% %\psfrag{p0}[b]{$-\Delta\epsilon\,\ln P_{exc}$}
% 
% \includegraphics[width=7cm]{Pexc-J_0.05-N_12-L_12.eps}
% \hspace{1cm}
% %\includegraphics[width=7cm]{Pexc_log-N_12-L_12.eps}
% \psfrag{eps}[c]{$E_0/U$}
% \psfrag{b}[c]{\bf (b)}
% \psfrag{p0}[b]{$P_{\rm exc}$}
% \includegraphics[width=7cm]{Pexc_log_log-J_0.05-N_12-L_12.eps}
% 
\begin{center}
\includegraphics[width=.45\textwidth]{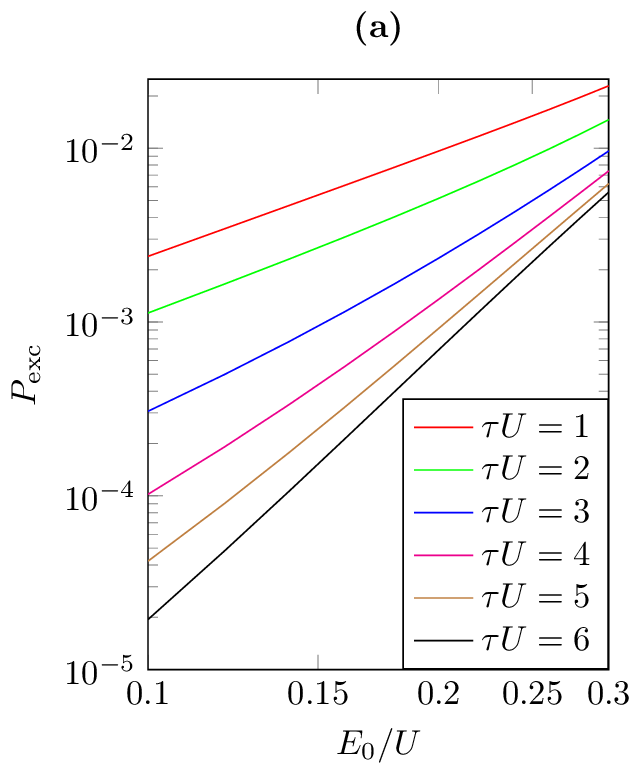} 
\includegraphics[width=.45\textwidth]{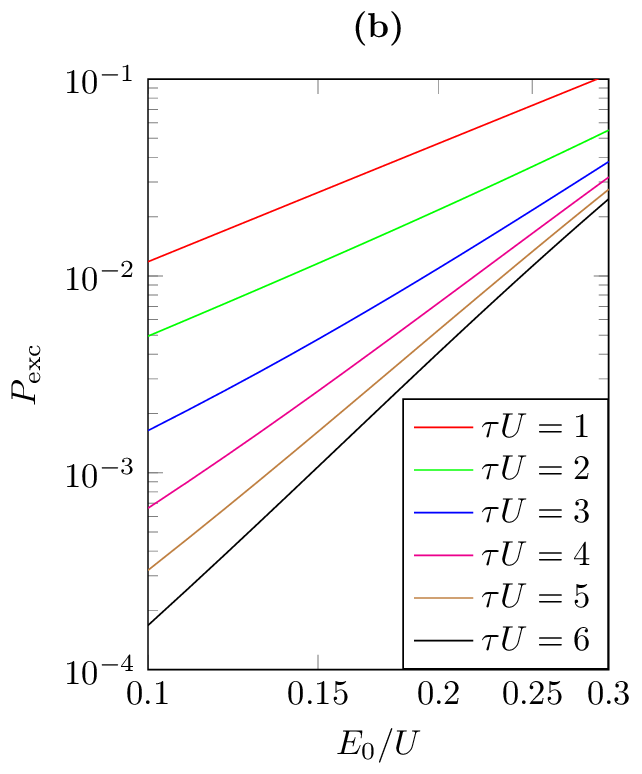} 
\caption{{\bf a)} Numerical results for the excitation probability per unit time in logarithmic scales 
for $N=L=12$ and $J/U=0.1$.
{\bf b)} Analytical results (\ref{sauter}) for the excitation probability per unit time where the 
expressions for the effective mass 
(\ref{m_eff}) and the effective velocity (\ref{c_eff}) have been used.}
\label{Pexc-N_12-L_12}
\end{center}
%\end{figure}
% 
%\begin{figure}[h]
% \begin{center}
% \includegraphics[width=.445\textwidth]{Decaysauter2.ps}
% \includegraphics[width=.45\textwidth]{Decaysauterloglog2.ps} 
% \caption{Analytical result (\ref{sauter}) for the 
% $\langle \hat{p}^\dagger_\mu\hat{p}_\mu\rangle/\tau$ after a Sauter pulse
% for $J/U=0.1$ and $N=12$.% where the expressions for the effective mass 
% %(\ref{m_eff}) and the effective velocity (\ref{c_eff}) have been used.
% }
% \label{analyticsauter}
% \end{center}
%\label{Pexc-sauter}
\end{figure}

% \begin{figure}[t]
% \begin{center}
% \includegraphics[width=.445\textwidth]{Decaysauter3.ps}
% \includegraphics[width=.45\textwidth]{Decaysauterloglog3.ps} 
% \caption
% {
% Analytical result (\ref{sauter}) for the 
% $\langle \hat{p}^\dagger_\mu\hat{p}_\mu\rangle/\tau$ after a Sauter pulse.
% }
% \end{center}
% \end{figure}

In contrast to all the previous numerical calculation we do not impose anymore
periodic boundary conditions. Instead of that we consider the case when the 
particles
cannot tunnel between the lattice sites $\mu=1$ and $\mu=L$ and
perform calculations using the basis of Fock states (\ref{Fock}).
Numerical integration of the Schr\"odinger equation
is done by means of the fourth-order Runge-Kutta method.
The accuracy of the numerical integration is controlled by the conservation 
of norm of the state $|\psi(t)\rangle$.

The results for the excitation probability per unit time
\begin{equation}
\label{Pexc}
P_{\rm exc}
=
\frac
{1-\left|\langle\psi(0)|\psi(5\tau)\rangle\right|^2}
{\tau}
\;,
\end{equation}
where $5\tau$ is the total evolution time,
are shown in Fig.~\ref{Pexc-N_12-L_12}(a).
At short evolution times, the excitation probability $P_{\rm exc}$
has a power-law dependence
on $\Delta\epsilon$ which corresponds to the perturbative regime 
of the pair production.
One should keep in mind that 
finite-size effects start to play an important role
if the evolution time exceeds $L/v_{\rm eff}$, where $v_{\rm eff}$ 
is an effective
velocity
for the propagation of excitations discussed in Section \ref{energyspectrum}.
For $L=N=12$ and $J/U=0.1$ this leads to the requirement $\tau U < 7$.
At much longer times $\tau$, the dynamics of a finite-size system will 
be adiabatic
and the excitation probability will tend to zero in contrast to the 
infinite system,
where the excitation probability remains finite in the limit 
$\tau\rightarrow\infty$ as determined by Eq.~(\ref{sauter-inf}).

In Fig.~\ref{Pexc-N_12-L_12}(b) we show the results of the same calculations 
obtained
in Section~\ref{section-Analogue} in the first order of the $1/Z$-expansion 
where 
corrections due to time-derivatives of $T_\mathbf{k}$ have been 
neglected, see Eq.~(\ref{dglexakt}).
The excitation probability~(\ref{Pexc}) and the particle-hole creation rate are related
via
\begin{eqnarray}
P_{\rm exc}
=
\frac{1-\langle\Psi_\mathrm{Mott}|\hat{\rho}(\infty)|\Psi_\mathrm{Mott}\rangle}{\tau}
\approx
\frac{1- (1-2\langle \hat{p}^\dagger_\mu\hat{p}_\mu\rangle)^N}{\tau}
\approx
2N\frac{\langle \hat{p}^\dagger_\mu\hat{p}_\mu\rangle}{\tau}
\;,
\end{eqnarray}
where
$
\langle \hat{p}^\dagger_\mu\hat{p}_\mu\rangle
=
\sum_{\bf k}
\left|
    \beta_{\bf k}
\right|^2/N
$ and $\beta_\mathbf{k}$ is the Bogoliubov coefficient defined in equation 
(\ref{sauter}).
We observe a very good qualitative agreement with the results of exact 
numerical calculations, although the latter give somewhat
smaller values of $P_{\rm exc}$.

%%%%%%%%%%%%%%%%%%%%%%%%%%%%%%%%%%%%%%%%%%%%%%%%%%%%%%%%%%%%%%%%%%%%%%%%%%%%%%%
\section{Fermi-Hubbard Model}\label{Fermi-Hubbard Model}
%%%%%%%%%%%%%%%%%%%%%%%%%%%%%%%%%%%%%%%%%%%%%%%%%%%%%%%%%%%%%%%%%%%%%%%%%%%%%%%

Now, after having studied the bosonic case, let us investigate the 
Fermi-Hubbard model \cite{H63,EFGKK05,F91}.
We shall find many similarities to the Bose-Hubbard model -- but also crucial 
differences.
The Hamiltonian is given by 
\begin{eqnarray}
\label{Fermi-Hubbard-Hamiltonian}
\hat H
=
-\frac{J}{Z}\sum_{\mu\nu,s}T_{\mu\nu}\hat{c}_{\mu,s}^\dagger\hat{c}_{\nu,s}
+U\sum_{\mu}\hat{n}_\mu^\uparrow\hat{n}_\mu^\downarrow
\,.
\end{eqnarray}
The nomenclature is the same as in the bosonic case 
(\ref{Bose-Hubbard-Hamiltonian}) but with an additional spin label $s$ 
which can assume two values $s=\uparrow$ or $s=\downarrow$.
In the following, we consider the case of half-filling 
$\langle\hat{n}_\mu^\uparrow+\hat{n}_\mu^\downarrow\rangle=1$
where half the particles are in the $s=\uparrow$ state and the other 
have $s=\downarrow$.
Note that the total particle numbers  
$\hat N^\uparrow=\sum_\mu \hat n_\mu^\uparrow$ 
and 
$\hat N^\downarrow=\sum_\mu \hat n_\mu^\downarrow$
for each spin species are conserved separately 
$[\hat H,\hat N^\uparrow]=[\hat H,\hat N^\downarrow]=0$. 
The creation and annihilation operators satisfy the fermionic 
anti-commutation relations 
\bea
\label{fermionic-commutation}
\left\{\hat c_{\nu,a},\hat c_{\mu,b}^\dagger\right\}
=\delta_{\mu\nu}\delta_{ab}
\;,\,
\left\{\hat c_{\nu,a},\hat c_{\mu,b}\right\}
=
\left\{\hat c_{\nu,a}^\dagger,\hat c_{\mu,b}^\dagger\right\}
=0
\,.
\ea
The fermionic nature of the particles has important consequences.
For example, let us estimate the expectation value of the hopping 
Hamiltonian $\hat H_J$. 
Introducing the ``coarse-grained'' operator 
\bea
\label{c-sigma}
\hat c_{\mu,s}^\Sigma
=
\frac{1}{\sqrt{Z}}\sum_{\nu}T_{\mu\nu}\hat{c}_{\nu,s}
\,,
\ea
we may write the expectation value of the tunnelling energy $\hat H_J$ 
per lattice site for one spin species $s$ as 
$-J\langle\hat c_{\mu,s}^\dagger\hat c_{\mu,s}^\Sigma\rangle/\sqrt{Z}$. 
This expectation value can be interpreted as a scalar product of the 
two vectors $\hat c_{\mu,s}\ket{\Psi}$ and $\hat c_{\mu,s}^\Sigma\ket{\Psi}$
and hence it is bounded by  
\bea
\left|\bra{\Psi}\hat c_{\mu,s}^\dagger\hat c_{\mu,s}^\Sigma\ket{\Psi}\right|
\leq
||\hat c_{\mu,s}\ket{\Psi}||\cdot||\hat c_{\mu,s}^\Sigma\ket{\Psi}||
\,.
\ea
Inserting $||\hat c_{\mu,s}\ket{\Psi}||^2=
\bra{\Psi}\hat c_{\mu,s}^\dagger\hat c_{\mu,s}\ket{\Psi}=
\bra{\Psi}\hat n_{\mu,s}\ket{\Psi}$, we get the expectation value of the 
number operator $\hat n_{\mu,s}$. 
In contrast to the bosonic case, this operator is bounded and thus we find 
$||\hat c_{\mu,s}\ket{\Psi}||\leq1$. 
Furthermore, the operator $\hat c_{\mu,s}^\Sigma$ in (\ref{c-sigma}) obeys 
the same anti-commutation relations (\ref{fermionic-commutation}) and thus 
we find $||\hat c_{\mu,s}^\Sigma\ket{\Psi}||\leq1$ in complete analogy.
Consequently, the absolute value of the tunnelling energy per lattice site 
is below $2J/\sqrt{Z}$, i.e., decreases for large $Z$. 

The above result implies that the interaction term $\propto U$ always 
dominates (except in the trivial case $U=0$) in the limit $Z\to\infty$ 
under consideration.
Hence, we are in the strongly interacting Mott regime and do not find 
anything analogous to the Mott--superfluid transition as in the bosonic 
case. 
Note that often \cite{MV89,F99} a different $Z$-scaling is considered, 
where the hopping term scales with $J/\sqrt{Z}$ instead of $J/Z$ as in 
(\ref{Fermi-Hubbard-Hamiltonian}).
Using this $J/\sqrt{Z}$ scaling, one can study the transition from the 
Mott state to a metallic state which is supposed to occur at a critical 
value of $J$ where -- roughly speaking -- the hopping term starts to 
dominate over the interaction term.
However, this transition is not as well understood as the Mott--superfluid 
transition in the bosonic case. 
With our $J/Z$-scaling in (\ref{Fermi-Hubbard-Hamiltonian}), we study a 
different corner of the phase space where we can address question such 
as tunnelling in tilted lattices and equilibration vs thermalisation etc. 

\subsection{Symmetries and Degeneracy}

In addition to the usual invariances already known from the bosonic case, 
the Fermi-Hubbard model has some more symmetries.
For example, the particle-hole symmetry 
$\hat c_{\mu,s}^\dagger\leftrightarrow\hat c_{\mu,s}$ and thus 
$\hat n_{\mu,s}=\hat c_{\mu,s}^\dagger\hat c_{\mu,s}
\leftrightarrow
\hat{\bar n}_{\mu,s}=\hat c_{\mu,s}\hat c_{\mu,s}^\dagger
=1-\hat n_{\mu,s}$ 
is no longer an effective approximate symmetry, but becomes exact 
(for the case of half-filling considered here). 

Furthermore, there is an effective $SU(2)$-symmetry corresponding to the
spin degrees of freedom. 
To specify this, let us introduce the effective spin operators
\bea
\label{spin-operators}
\hat S_\mu^z
=
\frac12
\sum\limits_{ab}
\hat c_{\mu,a}^\dagger\,
\sigma^z_{ab}\,
\hat c_{\mu,b}
=
\frac12
\left(\hat{n}_\mu^\uparrow-\hat{n}_\mu^\downarrow\right)
\,,
\ea
and analogously 
$\hat S_\mu^x=\sum_{ab}\hat c_{\mu,a}^\dagger\sigma^x_{ab}\hat c_{\mu,b}/2$
as well as 
$\hat S_\mu^y=\sum_{ab}\hat c_{\mu,a}^\dagger\sigma^x_{ab}\hat c_{\mu,b}/2$
where $\sigma^{x,y,z}_{ab}$ are the usual Pauli spin matrices.
These operators satisfy the usual spin, i.e., $SU(2)$, commutation relations 
and the Fermi-Hubbard Hamiltonian (\ref{Fermi-Hubbard-Hamiltonian})
is invariant under global $SU(2)$ rotations generated by the total spin 
operators $\hat{\mathbf S}_{\rm tot}=\sum_\mu\hat{\mathbf S}_\mu$.

In the case of zero hopping $J=0$, this global $SU(2)$ invariance even
becomes a local symmetry, i.e., we may perform a spin rotation at each 
site without changing the energy.
As a result, the ground state (at half filling) is highly degenerate for
$J=0$ in contrast to the Bose-Hubbard model (at integer filling). 
This degeneracy can be lifted by an additional staggered magnetic field 
(see Appendix \ref{staggered}) and is related to the spin modes which become 
arbitrarily soft for small $J$. 
In this limit $J\ll U$, their dynamics can be described by an effective 
Hamiltonian, which is basically the Heisenberg model 
\begin{eqnarray}
\label{Heisenberg-Hamiltonian}
\hat H
=
\frac{2J^2}{Z^2U}
\sum_{\mu\nu}T_{\mu\nu}\,\hat{\mathbf S}_\mu\cdot\hat{\mathbf S}_\nu
\,,
\end{eqnarray}
with an effective anti-ferromagnetic coupling constant
of order $1/Z^2$. 
This effective Hamiltonian describes the Fermi-Hubbard Hamiltonian 
(\ref{Fermi-Hubbard-Hamiltonian}) for half-filling in the low-energy 
sub-space where we have one particle per site, but with a variable spin
$\hat{\mathbf S}_\mu$. 

In order to avoid complications such as frustration for the 
anti-ferromagnetic Heisenberg model (\ref{Heisenberg-Hamiltonian}), 
we assume a bipartite lattice -- i.e., we can divide the total lattice
into two sub-lattices $\cal A$ and $\cal B$ such that, for each site in 
$\mu\in\cal A$, all the neighbouring sites $\nu$ belong to $\cal B$ 
and {\em vice versa}.
In this case, the ground state of the Heisenberg model 
(\ref{Heisenberg-Hamiltonian}) approaches the N\'eel state for large $Z$
\begin{eqnarray}
\label{Neel}
\hat{\rho}_{\rm Neel}
=
\bigotimes_{\mu\in\cal A}
\bigotimes_{\nu\in\cal B}
\hat{n}_\mu^\downarrow\,
\hat{\bar{n}}_\mu^\uparrow\,
\hat{n}_\nu^\uparrow\,
\hat{\bar{n}}_\nu^\downarrow
\,,
\end{eqnarray}
which is just the state with exactly one particle per site, but in 
alternating spin states, i.e., $s=\downarrow$ for $\mu\in\cal A$ and 
$s=\uparrow$ for $\nu\in\cal B$. 
Note that $\hat{n}_\mu^\downarrow$ is the projector on the 
$\ket{1}_\mu^\downarrow$ state 
$\hat{n}_\mu^\downarrow=\ket{1^\downarrow}_\mu\bra{1^\downarrow}$
while $\hat{\bar{n}}_\mu^\uparrow$
projects on the $\ket{0}_\mu^\uparrow$ state etc. 
As usual, this state (\ref{Neel}) breaks the original symmetry group of 
the Hamiltonian (\ref{Fermi-Hubbard-Hamiltonian}) containing particle-hole 
symmetry, $SU(2)$ invariance, and translational symmetry, down to a 
sub-group, which includes invariance under a combined spin-flip and 
particle-hole exchange etc. 

Let us stress that the N\'eel state (\ref{Neel}) is only the lowest-order 
approximation of the real ground state of the Heisenberg model 
(\ref{Heisenberg-Hamiltonian}), there are quantum spin fluctuations
of order $\ord(1/Z)$.  
These quantum spin fluctuations do not vanish in the limit $J\to 0$ since
$J$ only appears in the overall pre-factor in front of the 
Heisenberg Hamiltonian (\ref{Heisenberg-Hamiltonian}) while the internal 
structure remains the same. 
Only after adding a suitable staggered magnetic field 
(see Appendix \ref{staggered}), the N\'eel state (\ref{Neel}) 
is the exact unique ground state (for $J\to 0$). 
Either way, in analogy to the bosonic case, we can now use this fully 
factorising state (\ref{Neel}) as the starting point for our $1/Z$-expansion. 

%%%%%%%%%%%%%%%%%%%%%%%%%%%%%%%%%%%%%%%%%%%%%%%%%%%%%%%%%%%%%%%%%%%%%%%%%%%%%%%
\section{Charge Modes}\label{chargemodes}
%%%%%%%%%%%%%%%%%%%%%%%%%%%%%%%%%%%%%%%%%%%%%%%%%%%%%%%%%%%%%%%%%%%%%%%%%%%%%%%

Starting with the N\'eel state (\ref{Neel}) as the zeroth order in $1/Z$,
let us now derive the first-order corrections. 
To this end, let us consider the Heisenberg equations of motion 
\begin{eqnarray}
i\partial_t \hat{c}_{\mu s}
&=&
-\frac{J}{Z}\sum_{\kappa\neq \mu} T_{\mu\kappa}\hat{c}_{\kappa s}
+U\hat{c}_{\mu s} \hat{n}_{\mu \bar{s}}
%-A_{\mu s}\hat{c}_{\mu s}
\label{annihilation-operator}\\
i\partial_t \hat{c}_{\mu s}^\dagger
&=&
+\frac{J}{Z}\sum_{\kappa\neq \mu} T_{\mu\kappa}\hat{c}^\dagger_{\kappa s}
-U\hat{c}^\dagger_{\mu s} \hat{n}_{\mu \bar{s}}
%+A_{\mu s}\hat{c}^\dagger_{\mu s}
\label{creation-operator}\\
i\partial_t\hat{n}_{\mu s}
&=&
-i\partial_t\hat{\bar{n}}_{\mu s}
=
\frac{J}{Z}\sum_{\kappa\neq \mu}T_{\mu\kappa}
\left(
\hat{c}_{\kappa s}^\dagger \hat{c}_{\mu s}-
\hat{c}_{\mu s}^\dagger \hat{c}_{\kappa s}
\right)
\label{number-operator}
\,,
\end{eqnarray}
where $\bar s$ denotes the spin label opposite to $s$, i.e., either 
$(s,\bar{s})=(\uparrow,\downarrow)$ or $(s,\bar{s})=(\downarrow,\uparrow)$.
If we now insert these evolution equations into the correlation functions
$\langle\hat{c}_{\mu a}^\dagger\hat{c}_{\nu b}
\hat{n}_{\mu\bar{a}}\hat{n}_{\nu\bar{b}}\rangle$, 
$\langle\hat{c}_{\mu a}^\dagger\hat{c}_{\nu b}
\hat{\bar n}_{\mu\bar{a}}\hat{n}_{\nu\bar{b}}\rangle$, 
$\langle\hat{c}_{\mu a}^\dagger\hat{c}_{\nu b}
\hat{n}_{\mu\bar{a}}\hat{\bar n}_{\nu\bar{b}}\rangle$, 
and 
$\langle\hat{c}_{\mu a}^\dagger\hat{c}_{\nu b}
\hat{\bar n}_{\mu\bar{a}}\hat{\bar n}_{\nu\bar{b}}\rangle$, 
we find that they form a closed set of equations to first order in $1/Z$, 
where we can neglect three-point correlations 
\begin{eqnarray}
i\partial_t 
\langle \hat{c}_{\mu a}^\dagger\hat{c}_{\nu b}
\hat{n}_{\mu\bar{a}}\hat{n}_{\nu\bar{b}}\rangle
&=&
+\frac{J}{Z}\langle\hat{n}_{\mu \bar{a}}\rangle_0
\sum_{\kappa\neq \mu,\nu}T_{\mu\kappa}
\langle\hat{c}_{\kappa a}^\dagger\hat{c}_{\nu b}
(\hat{n}_{\kappa\bar{a}}+\hat{\bar{n}}_{\kappa\bar{a}})
\hat{n}_{\nu\bar{b}}\rangle
\nonumber\\
& &
-\frac{J}{Z}\langle\hat{n}_{\nu \bar{b}}\rangle_0
\sum_{\kappa\neq \mu,\nu}T_{\nu\kappa}
\langle\hat{c}_{\mu a}^\dagger\hat{c}_{\kappa b}
\hat{n}_{\mu\bar{a}}(\hat{n}_{\kappa\bar{b}}+\hat{\bar{n}}_{\kappa\bar{b}})
\rangle
\nonumber\\
%& &
%+(A_{\mu a}-A_{\nu b})\langle \hat{c}_{\mu a}^\dagger\hat{c}_{\nu b}
%\hat{n}_{\mu\bar{a}}\hat{n}_{\nu\bar{b}}\rangle
%\nonumber\\
& &
+\frac{J}{Z}T_{\mu\nu}
\langle \hat{c}_{\nu a}^\dagger\hat{c}_{\nu b}
\hat{n}_{\mu\bar{a}}\hat{n}_{\nu\bar{b}}\rangle_0
-\frac{J}{Z}T_{\mu\nu}
\langle \hat{c}_{\mu a}^\dagger\hat{c}_{\mu b}
\hat{n}_{\mu\bar{a}}\hat{n}_{\nu\bar{b}}\rangle_0\label{corr1}
\,,
\end{eqnarray}
where the expectation values $\langle\hat{n}_{\mu \bar{a}}\rangle_0$ and 
$\langle\hat{n}_{\nu \bar{b}}\rangle_0$ as well as those in the last line 
are taken in the zeroth-order N\'eel state (\ref{Neel}).
In complete analogy, we obtain for the remaining three correlators 
\begin{eqnarray}
i\partial_t 
\langle \hat{c}_{\mu a}^\dagger\hat{c}_{\nu b}
\hat{n}_{\mu\bar{a}}\hat{\bar{n}}_{\nu\bar{b}}\rangle
&=&
+\frac{J}{Z}\langle\hat{n}_{\mu \bar{a}}\rangle_0
\sum_{\kappa\neq \mu,\nu}T_{\mu\kappa}
\langle\hat{c}_{\kappa a}^\dagger\hat{c}_{\nu b}
(\hat{n}_{\kappa\bar{a}}+\hat{\bar{n}}_{\kappa\bar{a}})
\hat{\bar{n}}_{\nu\bar{b}}\rangle
\nonumber\\
& &
-\frac{J}{Z}\langle\hat{\bar{n}}_{\nu \bar{b}}\rangle_0
\sum_{\kappa\neq \mu,\nu}T_{\nu\kappa}
\langle\hat{c}_{\mu a}^\dagger\hat{c}_{\kappa b}
\hat{n}_{\mu\bar{a}}(\hat{n}_{\kappa\bar{b}}+\hat{\bar{n}}_{\kappa\bar{b}})
\rangle
\nonumber\\
& &
-U%(U-A_{\mu a}+A_{\nu b})
\langle \hat{c}_{\mu a}^\dagger\hat{c}_{\nu b}
\hat{n}_{\mu \bar{a}}\hat{\bar{n}}_{\nu \bar{b}}\rangle
\nonumber\\
& &
+\frac{J}{Z}T_{\mu\nu}
\langle \hat{c}_{\nu a}^\dagger\hat{c}_{\nu b}
\hat{n}_{\mu\bar{a}}\hat{\bar{n}}_{\nu\bar{b}}\rangle_0
-\frac{J}{Z}T_{\mu\nu}
\langle \hat{c}_{\mu a}^\dagger\hat{c}_{\mu b}
\hat{n}_{\mu\bar{a}}\hat{\bar{n}}_{\nu\bar{b}}\rangle_0
\,,
\end{eqnarray}
as well as 
\begin{eqnarray}
i\partial_t 
\langle \hat{c}_{\mu a}^\dagger\hat{c}_{\nu b}
\hat{\bar{n}}_{\mu\bar{a}}\hat{n}_{\nu\bar{b}}\rangle
&=&
+\frac{J}{Z}\langle\hat{\bar{n}}_{\mu \bar{a}}\rangle_0
\sum_{\kappa\neq \mu,\nu}T_{\mu\kappa}
\langle\hat{c}_{\kappa a}^\dagger\hat{c}_{\nu b}
(\hat{n}_{\kappa\bar{a}}+\hat{\bar{n}}_{\kappa\bar{a}})
\hat{n}_{\nu\bar{b}}\rangle
\nonumber\\
& &
-\frac{J}{Z}\langle\hat{n}_{\nu \bar{b}}\rangle_0
\sum_{\kappa\neq \mu,\nu}T_{\nu\kappa}
\langle\hat{c}_{\mu a}^\dagger\hat{c}_{\kappa b}
\hat{\bar{n}}_{\mu\bar{a}}
(\hat{n}_{\kappa\bar{b}}+\hat{\bar{n}}_{\kappa\bar{b}})\rangle
\nonumber\\
& &
+U%(U+A_{\mu a}-A_{\nu b})
\langle \hat{c}_{\mu a}^\dagger\hat{c}_{\nu b}
\hat{\bar{n}}_{\mu \bar{a}}\hat{n}_{\nu \bar{b}}\rangle
\nonumber\\
& &
+\frac{J}{Z}T_{\mu\nu}
\langle \hat{c}_{\nu a}^\dagger\hat{c}_{\nu b}
\hat{\bar{n}}_{\mu\bar{a}}\hat{n}_{\nu\bar{b}}\rangle_0
-\frac{J}{Z}T_{\mu\nu}
\langle \hat{c}_{\mu a}^\dagger\hat{c}_{\mu b}
\hat{\bar{n}}_{\mu\bar{a}}\hat{n}_{\nu\bar{b}}\rangle_0
\,,
\end{eqnarray}
and finally 
\begin{eqnarray}
i\partial_t 
\langle \hat{c}_{\mu a}^\dagger\hat{c}_{\nu b}
\hat{\bar{n}}_{\mu\bar{a}}\hat{\bar{n}}_{\nu\bar{b}}\rangle
&=&
+\frac{J}{Z}\langle\hat{\bar{n}}_{\mu \bar{a}}\rangle_0
\sum_{\kappa\neq \mu,\nu}T_{\mu\kappa}
\langle\hat{c}_{\kappa a}^\dagger\hat{c}_{\nu b}
(\hat{n}_{\kappa\bar{a}}+\hat{\bar{n}}_{\kappa\bar{a}})
\hat{\bar{n}}_{\nu\bar{b}}\rangle
\nonumber\\
& &
-\frac{J}{Z}\langle\hat{\bar{n}}_{\nu \bar{b}}\rangle_0
\sum_{\kappa\neq \mu,\nu}T_{\nu\kappa}
\langle\hat{c}_{\mu a}^\dagger\hat{c}_{\kappa b}
\hat{\bar{n}}_{\mu\bar{a}}
(\hat{n}_{\kappa\bar{b}}+\hat{\bar{n}}_{\kappa\bar{b}})\rangle
\nonumber\\
%& &
%+(A_{\mu a}-A_{\nu b})
%\langle \hat{c}_{\mu a}^\dagger\hat{c}_{\nu b}
%\hat{\bar{n}}_{\mu\bar{a}}\hat{\bar{n}}_{\nu\bar{b}}\rangle
%\nonumber\\
& &
+\frac{J}{Z}T_{\mu\nu}
\langle \hat{c}_{\nu a}^\dagger\hat{c}_{\nu b}
\hat{\bar{n}}_{\mu\bar{a}}\hat{\bar{n}}_{\nu\bar{b}}\rangle_0
-\frac{J}{Z}T_{\mu\nu}
\langle \hat{c}_{\mu a}^\dagger\hat{c}_{\mu b}
\hat{\bar{n}}_{\mu\bar{a}}\hat{\bar{n}}_{\nu\bar{b}}\rangle_0
\label{sectors}
\,.
\end{eqnarray}
We observe that the spin structure is conserved in these equations, 
i.e., the four correlators containing 
$\hat{c}_{\mu\uparrow}^\dagger\hat{c}_{\nu\uparrow}$ decouple from 
those with 
$\hat{c}_{\mu\uparrow}^\dagger\hat{c}_{\nu\downarrow}$ etc. 
Thus we can treat the four sectors separately.  
Let us focus on the correlators containing 
$\hat{c}_{\mu\downarrow}^\dagger\hat{c}_{\nu\downarrow}$ and introduce
the following short-hand notation:
If ${\mu}\in\cal A$ and ${\nu}\in\cal B$, we denote the correlations by 
$\langle \hat{c}_{\mu \downarrow}^\dagger\hat{c}_{\nu \downarrow}
\hat{n}_{\mu\uparrow}\hat{n}_{\nu\uparrow}\rangle=f_{\mu\nu}^{1_A1_B}$,
and 
$\langle \hat{c}_{\mu \downarrow}^\dagger\hat{c}_{\nu \downarrow}
\hat{\bar{n}}_{\mu\uparrow}\hat{n}_{\nu\uparrow}\rangle=f_{\mu\nu}^{0_A1_B}$,
etc.
Inserting the zeroth-order N\'eel state (\ref{Neel}), 
we find four trivial equations which fully decouple
\begin{eqnarray}
i\partial_t f^{1_A0_B}_{\mu\nu}
&=&
-U f^{1_A0_B}_{\mu\nu}
\,,
\nn
i\partial_t f^{0_B1_A}_{\mu\nu}
&=&
+U f^{0_B1_A}_{\mu\nu}
\,,
\nn
i\partial_t f^{0_B0_B}_{\mu\nu}
&=&0
\,,
\nn
i\partial_t f^{1_A1_A}_{\mu\nu}
&=&0
\label{hom1}
\,.
\end{eqnarray}
Thus, if these correlations vanish initially, they remain zero 
(to first order in $1/Z$).
Setting these correlations (\ref{hom1}) to zero, 
we get four pairs of coupled equations 
\begin{eqnarray}
i\partial_t f^{0_A0_B}_{\mu\nu}
&=&
+\frac{J}{Z}\sum_{\kappa\neq{\mu,\nu}}T_{\mu\kappa}
f^{1_B0_B}_{\kappa\nu}
%\left(f^{1_B0_B}_{\kappa\nu}+f^{0_B0_B}_{\kappa\nu}\right)
%+a f^{0_A0_B}_{\mu\nu}
\,,
\nn
i\partial_t f^{1_B0_B}_{\mu\nu}
&=&
+\frac{J}{Z}\sum_{\kappa\neq{\mu,\nu}}T_{\mu\kappa}
f^{0_A0_B}_{\kappa\nu}
%\left(f^{0_A0_B}_{\kappa\nu}+f^{1_A0_B}_{\kappa\nu}\right)
-Uf^{1_B0_B}_{\mu\nu}
\,,
\label{hom2}\\
i\partial_t f^{0_B0_A}_{\mu\nu}
&=&
-\frac{J}{Z}\sum_{\kappa\neq{\mu,\nu}}T_{\kappa\nu}
f^{0_B1_B}_{\mu\kappa}
%\left(f^{0_B1_B}_{\mu\kappa}+f^{0_B0_B}_{\mu\kappa}\right)
%-af^{0_B0_A}_{\mu\nu}
\nn
i\partial_t f^{0_B1_B}_{\mu\nu}
&=&
-\frac{J}{Z}\sum_{\kappa\neq{\mu,\nu}}T_{\kappa\nu}
f^{0_B0_A}_{\mu\kappa}
%\left(f^{0_B0_A}_{\mu\kappa}+f^{0_B1_A}_{\mu\kappa}\right)
+Uf^{0_B1_B}_{\mu\nu}
\,,
\\
i\partial_t f^{1_B1_A}_{\mu\nu}
&=&
+\frac{J}{Z}\sum_{\kappa\neq{\mu,\nu}}T_{\mu\kappa}
f^{0_A1_A}_{\kappa\nu}
%\left(f^{0_A1_A}_{\kappa\nu}+f^{1_A1_A}_{\kappa\nu}\right)
%-af^{1_B1_A}_{\mu\nu}
\nn
i\partial_t f^{0_A1_A}_{\mu\nu}
&=&
+\frac{J}{Z}\sum_{\kappa\neq{\mu,\nu}}T_{\mu\kappa}
f^{1_B1_A}_{\kappa\nu}
%\left(f^{0_B1_A}_{\kappa\nu}+f^{1_B1_A}_{\kappa\nu}\right)
+Uf^{0_A1_A}_{\mu\nu}
\,,
\\
i\partial_t f^{1_A1_B}_{\mu\nu}
&=&
-\frac{J}{Z}\sum_{\kappa\neq{\mu,\nu}}T_{\kappa\nu}
f^{1_A0_A}_{\mu\kappa}
%\left(f^{1_A0_A}_{\mu\kappa}+f^{1_A1_A}_{\mu\kappa}\right)
%+af^{1_A1_B}_{\mu\nu}
\nn
i\partial_t f^{1_A0_A}_{\mu\nu}
&=&
-\frac{J}{Z}\sum_{\kappa\neq{\mu,\nu}}T_{\kappa\nu}
f^{1_A1_B}_{\mu\kappa}
%\left(f^{1_A0_B}_{\mu\kappa}+f^{1_A1_B}_{\mu\kappa}\right)
-Uf^{1_A0_A}_{\mu\nu}
\,.
\label{hom12}
\end{eqnarray}
Again, since these equations do not have any non-vanishing source terms 
(to first order in $1/Z$), they can be set to zero if we start in an 
initially uncorrelated state (see Appendix \ref{staggered}). 
Note that they would acquire non-zero source terms if we go away from 
half-filling. 
The positive and negative eigenfrequencies of these modes behave as 
\bea\label{eigenmodes}
\omega_\mathbf{k}^\pm=\frac{U\pm\sqrt{U^2+4 J^2T_\mathbf{k}^2}}{2}
\,.
\ea
Thus we have soft modes which scale as 
$\omega_\mathbf{k}^-\sim J^2/U$ for small $J$ and hard modes 
$\omega_\mathbf{k}^+\approx U$.
These modes are important for making contact to the $t$-$J$ model 
\cite{A94} which describes the low-energy excitations of the 
Fermi-Hubbard Hamiltonian (\ref{Fermi-Hubbard-Hamiltonian}) for small $J$
away from half-filling.
However, at half-filling, we can set them to zero. 
After doing this, we are left with four coupled equations, which do have 
non-vanishing source terms 
\begin{eqnarray}
i\partial_t f^{0_A0_A}_{\mu\nu}
&=&
\frac{J}{Z}\sum_{\kappa\neq{\mu,\nu}}
\left\{
T_{\mu\kappa}
f^{1_B0_A}_{\kappa\nu}
%\left(f^{0_B0_A}_{\kappa\nu}+f^{1_B0_A}_{\kappa\nu}\right)
-T_{\kappa\nu}
f^{0_A1_B}_{\mu\kappa}
%\left(f^{0_A0_B}_{\mu\kappa}+f^{0_A1_B}_{\mu\kappa}\right)
\right\}
\label{charge1}
\,,
\\
i\partial_t f^{0_A1_B}_{\mu\nu}
&=&
\frac{J}{Z}\sum_{\kappa\neq{\mu,\nu}}
\left\{
T_{\mu\kappa}
f^{1_B1_B}_{\kappa\nu}
%\left(f^{1_B1_B}_{\kappa\nu}+f^{0_B1_B}_{\kappa\nu}\right)
-T_{\kappa\nu}
f^{0_A0_A}_{\mu\kappa}
%\left(f^{0_A0_A}_{\mu\kappa}+f^{0_A1_A}_{\mu\kappa}\right)
\right\}
%\nonumber\\
%& &
+U f^{0_A1_B}_{\mu\nu}-\frac{J}{Z}T_{\mu\nu}
\,,
\\
i\partial_t f^{1_B0_A}_{\mu\nu}
&=&
\frac{J}{Z}\sum_{\kappa\neq{\mu,\nu}}
\left\{
T_{\mu\kappa}
f^{0_A0_A}_{\kappa\nu}
%\left(f^{0_A0_A}_{\kappa\nu}+f^{1_A0_A}_{\kappa\nu}\right)
-T_{\kappa\nu}
f^{1_B1_B}_{\mu\kappa}
%\left(f^{1_B1_B}_{\mu\kappa}+f^{1_B0_B}_{\mu\kappa}\right)
\right\}
%\nonumber\\
%& &
-U f^{1_B0_A}_{\mu\nu}+\frac{J}{Z}T_{\mu\nu}
\,,
\\
i\partial_t f^{1_B1_B}_{\mu\nu}
&=&
\frac{J}{Z}\sum_{\kappa\neq{\mu,\nu}}
\left\{
T_{\mu\kappa}
f^{0_A1_B}_{\kappa\nu}
%\left(f^{0_A1_B}_{\kappa\nu}+f^{1_A1_B}_{\kappa\nu}\right)
-T_{\kappa\nu}
f^{1_B0_A}_{\mu\kappa}
%\left(f^{1_B0_A}_{\mu\kappa}+f^{1_B1_A}_{\mu\kappa}\right)
\right\}
\,.
\label{charge4}
\end{eqnarray}
Due to the source terms $JT_{\mu\nu}/Z$, these modes will develop 
correlations if we slowly (or suddenly) switch on the hopping rate $J$,
even if there are no correlations initially. 
The eigenfrequencies of these modes behave as 
\begin{eqnarray}
\label{omega-fermi}
\omega_\mathbf{k}=\sqrt{U^2+4 J^2T_\mathbf{k}^2}
\,.
\end{eqnarray}
A similar dispersion relation can be derived from a mean-field 
approach \cite{F91}. 
In contrast to the bosonic case, the origin of the Brillouin zone at 
$\mathbf{k}=0$ does not have minimum but actually maximum excitation 
energy $\omega_\mathbf{k}$.
The minimum is not a point but a hyper-surface where $T_\mathbf{k}=0$
(or, more generally, $T_\mathbf{k}^2$ assumes its minimum).
After Fourier transformation of (\ref{charge1})-(\ref{charge4}) we find 
that the equations of motion conserve a bilinear quantity, that is
\begin{eqnarray}\label{invfermi}
\partial_t\left[
\left(f^{1_B1_B}_\mathbf{k}-1\right)f^{1_B1_B}_\mathbf{k}+
f^{0_A1_B}_\mathbf{k}f^{1_B0_A}_\mathbf{k}
\right]=0\,.
\end{eqnarray}
This relation holds, as in the bosonic case, also for time-dependent $J(t)$.

%%%%%%%%%%%%%%%%%%%%%%%%%%%%%%%%%%%%%%%%%%%%%%%%%%%%%%%%%%%%%%%%%%%%%%%%%%%%%%%
\section{Mott State}\label{Fermi-Mott}
%%%%%%%%%%%%%%%%%%%%%%%%%%%%%%%%%%%%%%%%%%%%%%%%%%%%%%%%%%%%%%%%%%%%%%%%%%%%%%%

\subsection{Ground state correlations}

In complete analogy to the bosonic case, we now imagine switching $J$ 
adiabatically from zero (where all the charge fluctuations vanish) to 
a finite value.
Thus we find the following non-zero ground-state correlations 
\begin{eqnarray}
\label{ground-11}
f^{1_B1_B}_{\mu\nu,\mathrm{ground}}
&=&
-f^{0_A0_A}_{\mu\nu,\mathrm{ground}}
=
\frac{1}{N}\sum_{\mathbf{k}}\frac{1}{2}
\left(1-\frac{U}{\omega_\mathbf{k}}\right)
e^{i(\mathbf{x}_\mu-\mathbf{x}_\nu)\cdot\mathbf{k}}
\,,
\\
f^{1_B0_A}_{\mu\nu,\mathrm{ground}}
&=&
{f}^{0_A1_B}_{\mu\nu,\mathrm{ground}}
=
\frac{1}{N}\sum_{\mathbf{k}}
\frac{JT_\mathbf{k}}{\omega_\mathbf{k}}\,
e^{i(\mathbf{x}_\mu-\mathbf{x}_\nu)\cdot\mathbf{k}}
\,.
\label{ground-10}
\end{eqnarray}
Somewhat similar to the Bose-Hubbard model, the symmetric combination 
(\ref{ground-11}) scales with $J^2$ for small $J$ while the other 
(\ref{ground-10}) starts linearly in $J$. 
Other correlators such as 
$\langle\hat{c}_{\mu\downarrow}^\dagger\hat{c}_{\nu\downarrow}\rangle$
can be obtained from these expressions.
For example, if $\mu$ and $\nu$ are in $\cal A$, we find, 
using $\hat{n}_{\mu\uparrow}+\hat{\bar n}_{\mu\uparrow}=1$ 
and $\hat{n}_{\nu\uparrow}+\hat{\bar n}_{\nu\uparrow}=1$
\bea
\langle\hat{c}_{\mu\downarrow}^\dagger\hat{c}_{\nu\downarrow}\rangle
=
f_{\mu\nu}^{1_A1_A}
+f_{\mu\nu}^{0_A1_A}
+f_{\mu\nu}^{1_A0_A}
+f_{\mu\nu}^{0_A0_A}
=
f_{\mu\nu}^{0_A0_A}
\,.
\ea

\subsection{Quantum depletion}

In the zeroth-order N\'eel state (\ref{Neel}), we have 
$\langle\hat{n}_{\mu\uparrow}\hat{n}_{\mu\downarrow}\rangle=0$.
Thus this quantity 
$\langle\hat{n}_{\mu\uparrow}\hat{n}_{\mu\downarrow}\rangle$
measures the deviation from this zeroth-order N\'eel state (\ref{Neel})
due to quantum charge fluctuations.
In order to calculate 
$\langle\hat{n}_{\mu\uparrow}\hat{n}_{\mu\downarrow}\rangle$, 
we also need some of the other sectors discussed after (\ref{sectors}).
Obviously, the correlators containing 
$\hat{c}_{\mu\uparrow}^\dagger\hat{c}_{\nu\uparrow}$ 
behave in the same way as those with 
$\hat{c}_{\mu\downarrow}^\dagger\hat{c}_{\nu\downarrow}$
after interchanging the sub-lattices $\cal A$ and $\cal B$.
Thus a completely analogous system of differential equations exists for 
the correlations of the form 
$\langle \hat{c}_{\mu \uparrow}^\dagger\hat{c}_{\mu \uparrow}
\hat{n}_{\mu\downarrow}\hat{n}_{\nu\downarrow}\rangle
=g_{\mu\nu}^{1_A1_B}$ etc. 
If we insert (\ref{number-operator}) in order to calculate 
$i\partial_t\langle\hat{n}_{\mu\uparrow}\hat{n}_{\mu\downarrow}\rangle$, 
we find that these two sectors are enough for deriving 
$\langle\hat{n}_{\mu\uparrow}\hat{n}_{\mu\downarrow}\rangle$. 
Assuming $\mu\in\cal B$ for simplicity, we find  
\begin{eqnarray}
i\partial_t\langle \hat{n}_{\mu s}\hat{n}_{\mu \bar{s}}\rangle
&=&
-\frac{J}{Z}\sum_{\kappa\neq\mu}T_{\kappa\mu}
\Big\{
g_{\mu\kappa}^{1_B1_A}+g_{\mu\kappa}^{1_B0_A}+
f_{\mu\kappa}^{1_B1_A}+f_{\mu\kappa}^{1_B0_A}
\nonumber\\
& &\quad
-g_{\kappa\mu}^{1_A1_B}-g_{\kappa\mu}^{0_A1_B}
-f_{\kappa\mu}^{1_A1_B}-f_{\kappa\mu}^{0_A1_B}
\Big\}
\,.
\end{eqnarray}
Setting the correlations with vanishing source terms to zero, we find
\begin{eqnarray}
i\partial_t\langle \hat{n}_{\mu s}\hat{n}_{\mu \bar{s}}\rangle
&=&
-\frac{J}{Z}\sum_{\kappa\neq\mu}T_{\kappa\mu}
\Big\{
f_{\mu\kappa}^{1_B0_A}
-f_{\kappa\mu}^{0_A1_B}
\Big\}\nonumber\\
& =&-\frac{1}{N}\sum_\mathbf{k}JT_\mathbf{k}\Big\{
f_{\mathbf{k}}^{1_B0_A}
-f_{\mathbf{k}}^{0_A1_B}
\Big\}=\frac{i}{N}\sum_\mathbf{k}\partial_t f_{\mathbf{k}}^{1_B1_B}
\,.
\end{eqnarray}
Thus, in the ground state, the quantum depletion reads 
\begin{eqnarray}
\langle\hat{n}_{\mu s}\hat{n}_{\mu \bar{s}}\rangle
=
\langle\hat{\bar{n}}_{\mu s}\hat{\bar{n}}_{\mu \bar{s}}\rangle
=
\frac{1}{N}\sum_\mathbf{k}\frac{1}{2}
\left(1-\frac{U}{\omega_\mathbf{k}}\right)
\,.
\end{eqnarray}
As one would expect, this quantity scales with $J^2$ for small $J$. 

\begin{center}
\begin{figure}[h]
\includegraphics{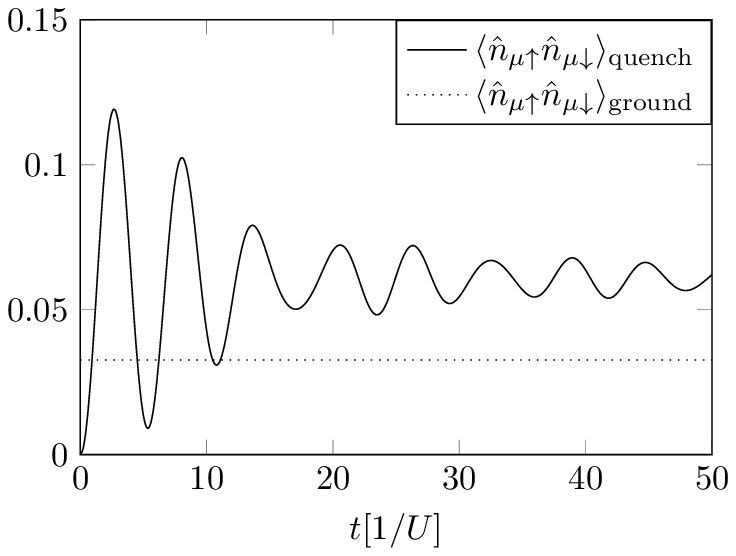}
\includegraphics{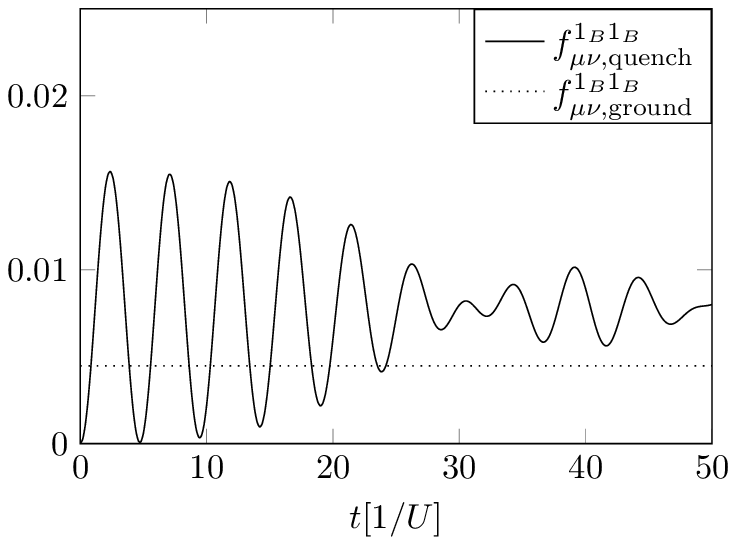}
\begin{center}
\includegraphics{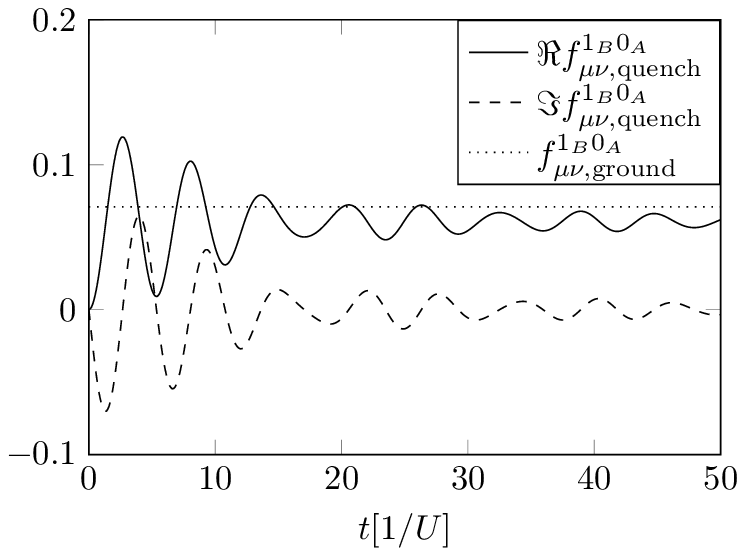}
\end{center}
\caption{Time-dependence of the quantum depletion, the nearest-neighbour 
correlation function $f^{1_B0_A}_{\mu\nu}$, and the next-to-nearest-neighbour 
correlation function $f^{1_B1_B}_{\mu\nu}$ in three dimensions 
after a quench within the Mott phase from $J/U=0$ to $J/U=0.5$  
in comparison to their ground state values. }\label{quenchfermi}
%
%We depict the correlation function $f^{1_B0_A}_{\mu\nu}$ for next neighbors,
%$f^{1_B1_B}_{\mu\nu}$ for second next neighbors 
%and the depletion 
%after a quench within the Mott phase from $J/U=0$ to $J/U=0.5$ 
%in three dimensions.
%The dotted lines are the ground stat values of the correlations 
%and the depletion, respectively.
\end{figure}
\end{center}

\subsection{Quench}

Now we consider a quantum quench, i.e., a sudden switch from $J=0$ 
to some finite value of $J$, where we find the following non-vanishing 
correlations
\begin{eqnarray}
f^{1_B1_B}_{\mu\nu,\mathrm{quench}}
=
-f^{0_A0_A}_{\mu\nu,\mathrm{quench}}
&=&
\frac{1}{N}\sum_{\mathbf{k}}
2 J^2 T_{\mathbf{k}}^2\,
\frac{1-\cos(\omega_\mathbf{k} t)}{\omega_\mathbf{k}^2}\,
e^{i(\mathbf{x}_\mu-\mathbf{x}_\nu)\cdot\mathbf{k}}
\,,
\\
f^{1_B0_A}_{\mu\nu,\mathrm{quench}}
=
\left({f}^{0_A1_B}_{\mu\nu,\mathrm{quench}}\right)^*
&=&
\frac{1}{N}\sum_{\mathbf{k}}
J T_\mathbf{k}U
\frac{1 - \cos(\omega_\mathbf{k} t)}{\omega_\mathbf{k}^{2}}
e^{i(\mathbf{x}_\mu-\mathbf{x}_\nu)\cdot\mathbf{k}}
\nn
& &
-\frac{i}{N}\sum_{\mathbf{k}}
J T_\mathbf{k}\,
\frac{\sin(\omega_\mathbf{k} t)}{\omega_\mathbf{k}}\,
e^{i(\mathbf{x}_\mu-\mathbf{x}_\nu)\cdot\mathbf{k}}
\,.
\end{eqnarray}
Again, these correlations equilibrate to a quasi-stationary value,
which is, however, not thermal.  
For some of these correlations, this quasi-stationary value lies 
even {\em below} the ground-state correlation, see Fig.~\ref{quenchfermi}. 
The probability to have two or zero particles at a site reads 
\begin{eqnarray}
\langle\hat{n}_{\mu s}\hat{n}_{\mu \bar{s}}\rangle_\mathrm{quench}
=
\langle\hat{\bar{n}}_{\mu s}\hat{\bar{n}}_{\mu \bar{s}}\rangle_\mathrm{quench}
=
\frac{1}{N}\sum_\mathbf{k}
2 J^2 T_{\mathbf{k}}^2\,
\frac{1-\cos(\omega_\mathbf{k} t)}{\omega_\mathbf{k}^2}
\,.
\end{eqnarray}
This quantity also equilibrates to a quasi-stationary value of order $1/Z$.
In analogy to the bosonic case, this quasi-stationary value could be explained 
by a small effective temperature -- but this small effective temperature then
does not work for the other observables, e.g., the correlations. 

\subsection{Spin modes}

So far, we have considered expectations values such as 
$\langle\hat{c}_{\mu a}^\dagger\hat{c}_{\nu b}
\hat{n}_{\mu\bar{a}}\hat{n}_{\nu\bar{b}}\rangle$, 
where -- apart from the number operators $\hat{n}_{\mu\bar{a}}$ and 
$\hat{n}_{\nu\bar{b}}$ -- one particle is annihilated at site $\nu$ 
and one is created at site $\mu$. 
These operator combinations correspond to a change of the occupation 
numbers and are thus called charge modes.
However, as already indicated in Section~\ref{Fermi-Hubbard Model}, 
there are also other modes which leave the total occupation number 
of all lattice sites unchanged. 
Examples are 
$\langle\hat{c}_{\mu s}^\dagger\hat{c}_{\mu\bar{s}}
\hat{c}_{\nu\bar{s}}^\dagger\hat{c}_{\nu s}\rangle$
or 
$\langle\hat{n}_{\mu a}\hat{n}_{\nu b}\rangle$
or combinations thereof. 
Many of these combinations can be expressed in terms of the 
effective spin operators in (\ref{spin-operators}) via 
$\langle\hat{S}_{\mu}^i\hat{S}_{\nu}^j\rangle$.
As one would expect from our study of the Bose-Hubbard model, 
the evolution of these spin modes vanishes to first order in $1/Z$ 
\bea
\partial_t\langle\hat{S}_{\mu}^i\hat{S}_{\nu}^j\rangle=\ord(1/Z^2)
\,,
\ea
consistent with the Heisenberg Hamiltonian (\ref{Heisenberg-Hamiltonian}).
In analogy to the $\langle\hat{n}_{\mu}\hat{n}_{\nu}\rangle$-correlator 
in the bosonic case, one has to go to second order $\ord(1/Z^2)$ in order
to calculate these quantities.
Fortunately, the charge modes discussed above do not couple to these spin 
modes to first order in $1/Z$ and hence we can omit them to this level
of accuracy. 

%%%%%%%%%%%%%%%%%%%%%%%%%%%%%%%%%%%%%%%%%%%%%%%%%%%%%%%%%%%%%%%%%%%%%%%%%%%%%%%
\section{Tilted Fermi-Hubbard Lattice}\label{fermitilt}
%%%%%%%%%%%%%%%%%%%%%%%%%%%%%%%%%%%%%%%%%%%%%%%%%%%%%%%%%%%%%%%%%%%%%%%%%%%%%%%

Motivated by the bosonic case, we now study particle-hole pair creation 
via tunnelling in a tilted lattice.
Again, we assume a spatially constant but possibly time-dependent force 
on the particles which acts on both spin species in the same way. 
Accordingly, we modify our Hamiltonian via 
\begin{eqnarray}
\label{Fermi-Hubbard-tilt}
\hat H
=
-\frac{J}{Z}\sum_{\mu\nu,s}T_{\mu\nu}\hat{c}_{\mu,s}^\dagger\hat{c}_{\nu,s}
+U\sum_{\mu}\hat{n}_\mu^\uparrow\hat{n}_\mu^\downarrow
+\sum_{\mu}V_\mu (\hat{n}_\mu^\uparrow+\hat{n}_\mu^\downarrow)
\,,
\end{eqnarray}
where $V_\mu(t)=\mathbf{E}(t)\cdot\mathbf{x}_\mu$ denotes the additional 
potential. 
Performing the same procedure as before, we find modified equations of 
motion for the charge modes
\begin{eqnarray}
\label{modify-charge-mode-1}
i\partial_t f^{0_A0_A}_{\mu\nu}
&=&
+\frac{J}{Z}\sum_{\kappa\neq\mu,\nu}
\left( 
T_{\mu\kappa}f^{1_B0_A}_{\kappa\nu}-T_{\kappa\nu}f^{0_A1_B}_{\mu\kappa}
\right)
\,,
\\
\label{modify-charge-mode-2}
i\partial_t  f^{0_A1_B}_{\mu\nu}
&=&
+\frac{J}{Z}\sum_{\kappa\neq\mu,\nu}
\left( 
T_{\mu\kappa}f^{1_B1_B}_{\kappa\nu}-T_{\kappa\nu}f^{0_A0_A}_{\mu\kappa}
\right)
\nonumber\\
& &
+(U+V_\nu-V_\mu)f^{0_A1_B}_{\mu\nu}-\frac{J}{Z}T_{\mu\nu}
\,,
\\
\label{modify-charge-mode-3}
i\partial_t f^{1_B0_A}_{\mu\nu}
&=&
-\frac{J}{Z}\sum_{\kappa\neq\mu,\nu} 
\left(
T_{\kappa\nu}f^{1_B1_B}_{\mu\kappa}-T_{\mu\kappa}f^{0_A0_A}_{\kappa\nu}
\right)
\nonumber\\
& &
-(U-V_\nu+V_\mu)f^{1_B0_A}_{\mu\nu}+\frac{J}{Z}T_{\mu\nu}
\,,
\\
\label{modify-charge-mode-4}
i\partial_t f^{1_B1_B}_{\mu\nu}
&=&
+\frac{J}{Z}\sum_{\kappa\neq\mu,\nu}
\left( 
T_{\mu\kappa} f^{0_A1_B}_{\kappa\nu}-T_{\kappa\nu}f^{1_B0_A}_{\mu\kappa}
\right)\,.
\end{eqnarray}
In complete analogy to the bosonic case 
it is possible to factorise the differential equations 
for the correlation functions.
We define {\em effective} particle and hole operators such that
we have for the correlations functions without source terms 
\begin{eqnarray}
\langle \hat{p}_{\mu,A}^\dagger \hat{h}_{\nu,B}\rangle
&=&
f^{1_A0_B}_{\mu\nu},\qquad
\langle \hat{h}_{\mu,B}^\dagger \hat{p}_{\nu,A}\rangle=f^{0_B1_A}_{\mu\nu},
\\
\langle \hat{h}_{\mu,B}^\dagger \hat{h}_{\nu,B}\rangle
&=&
f^{0_B0_B}_{\mu\nu},\qquad
\langle \hat{p}_{\mu,A}^\dagger \hat{p}_{\nu,A}\rangle=f^{1_A1_A}_{\mu\nu},
\\
\langle \hat{h}_{\mu,A}^\dagger \hat{h}_{\nu,B}\rangle
&=&
f^{0_A0_B}_{\mu\nu},\qquad
\langle \hat{p}_{\mu,B}^\dagger \hat{h}_{\nu,B}\rangle=f^{1_B0_B}_{\mu\nu},
\\
\langle \hat{h}_{\mu,B}^\dagger \hat{h}_{\nu,A}\rangle
&=&
f^{0_B0_A}_{\mu\nu},\qquad
\langle \hat{h}_{\mu,B}^\dagger \hat{p}_{\nu,B}\rangle=f^{0_B1_B}_{\mu\nu},
\\
\langle \hat{p}_{\mu,B}^\dagger \hat{p}_{\nu,A}\rangle
&=&
f^{1_B1_A}_{\mu\nu},\qquad
\langle \hat{h}_{\mu,A}^\dagger \hat{p}_{\nu,A}\rangle=f^{0_A1_A}_{\mu\nu}, 
\\
\langle \hat{p}_{\mu,A}^\dagger \hat{p}_{\nu,B}\rangle
&=&
f^{1_A1_B}_{\mu\nu},\qquad 
\langle \hat{p}_{\mu,A}^\dagger \hat{h}_{\nu,A}\rangle=f^{1_A0_A}_{\mu\nu},
\end{eqnarray}
and for the correlation functions with source terms
\begin{eqnarray}
\langle \hat{h}_{\mu,A}^\dagger \hat{p}_{\nu,B}\rangle
&=&
f^{0_A1_B}_{\mu\nu},\qquad
\langle \hat{h}_{\mu,A}^\dagger \hat{h}_{\nu,A}\rangle
=
f^{0_A0_A}_{\mu\nu}+\delta_{\mu\nu},
\\ 
\langle \hat{p}_{\mu,B}^\dagger \hat{h}_{\nu,A}\rangle
&=& 
f^{1_B0_A}_{\mu\nu},\qquad
\langle \hat{p}_{\mu,B}^\dagger \hat{p}_{\nu,B}\rangle=f^{1_B1_B}_{\mu\nu}
\,.
\end{eqnarray}
This allows us to effectively factorise the equations for the
correlation functions 
\begin{eqnarray}
i\partial_t \hat{p}_{\mu,B}
&=&
-\frac{J}{Z}\sum_{\kappa\neq\mu}T_{\mu\kappa}
\left(\hat{h}_{\kappa,A}+\hat{p}_{\kappa,A}\right)
+\left(\frac{U}{2}+V_\mu\right)\hat{p}_{\mu,B}
\\
i\partial_t \hat{h}_{\mu,A}
&=&
-\frac{J}{Z}\sum_{\kappa\neq\mu}T_{\mu\kappa}
\left(\hat{h}_{\kappa,B}+\hat{p}_{\kappa,B}\right)
+\left(-\frac{U}{2}+V_\mu\right)\hat{h}_{\mu,A}
\\
i\partial_t \hat{p}_{\mu,A}
&=&
\left(\frac{U}{2}+V_\mu\right)\hat{p}_{\mu,A}
\\
i\partial_t \hat{h}_{\mu,B}
&=&
\left(-\frac{U}{2}+V_\mu\right)\hat{h}_{\mu,B}\,.
\end{eqnarray}
Due to the initial conditions,
we can set the operators $\hat{h}_{\mu,B}$ and $\hat{p}_{\mu,A}$
to zero. 
After Fourier and Peierls transformation (\ref{Peierls}), 
we find the symmetric form
\begin{eqnarray}
i\partial_t \hat{p}_{\mathbf{k},B}
&=&
+\frac{U}{2}\hat{p}_{\mathbf{k},B}-JT_\mathbf{k}(t)\hat{h}_{\mathbf{k},A}\,,
\label{dirac1}
\\
i\partial_t \hat{h}_{\mathbf{k},A}
&=&
-\frac{U}{2}\hat{h}_{\mathbf{k},A}-JT_\mathbf{k}(t)\hat{p}_{\mathbf{k},B}
\label{dirac2}
\,,
\end{eqnarray}
% 
% Now we proceed in complete analogy to the bosonic case and define 
% {\em effective} particle and hole operators 
% %
% %{\green $\hat p_\mu = {\hat {\bar n}}_{\mu \uparrow} 
% %\hat c_{\mu \downarrow}$ and $\hat h_\nu = \hat  n_{\nu \downarrow} 
% %\hat c_{\nu \uparrow}$ ($\mu \in A$ and $\nu \in B$)  } 
% %
% such that 
% $\langle \hat{p}_{\mu}^\dagger \hat{p}_\nu\rangle=
% f^{0_A0_A}_{\mu\nu}+\delta_{\mu\nu}$, 
% $\langle \hat{h}_{\mu}^\dagger \hat{p}_\nu\rangle=f^{0_A1_B}_{\mu\nu}$,
% $\langle \hat{p}_{\mu}^\dagger \hat{h}_\nu\rangle= f^{1_B0_A}_{\mu\nu}$, 
% and  
% $\langle \hat{h}_{\mu}^\dagger \hat{h}_\nu\rangle=f^{1_B1_B}_{\mu\nu}$,
% which allows us to effectively factorise the above equations for the 
% charge modes via 
% %
% \begin{eqnarray}
% i\partial_t \hat{p}_\mu
% &=&
% \left(+\frac{U}{2}+V_\mu\right)\hat{p}_\mu
% +\frac{J}{Z}\sum_\kappa T_{\kappa\mu}\hat{h}_\kappa
% \,,
% \label{particle}
% \\
% i\partial_t \hat{h}_\mu
% &=&
% \left(-\frac{U}{2}+V_\mu\right)\hat{h}_\mu
% +\frac{J}{Z}\sum_\kappa T_{\kappa\mu}\hat{p}_\kappa
% \label{hole}
% \,.
% \end{eqnarray}
%
%
where the $T_\mathbf{k}(t)$ are time-dependent.
Now the line of reasoning is analogous to the Bose-Hubbard model. 
Initially, the operators evolve according to
\begin{eqnarray}
\hat{h}_{\mathbf{k},A}&=&e^{+iU t/2}\hat{A}_\mathbf{k}
\,,
\\
\hat{p}_{\mathbf{k},B}&=&e^{-iU t/2}\hat{B}_\mathbf{k}
\,,
\end{eqnarray}
with $\langle \hat{A}_\mathbf{k}^\dagger \hat{A}_\mathbf{p}\rangle=0$ and
$\langle\hat{B}_\mathbf{k}^\dagger\hat{B}_\mathbf{p}\rangle=
\delta_{\mathbf{k,p}}$.
At the end of the evolution, we find  
\begin{eqnarray}
\hat{h}_{\mathbf{k},A}
&=&
\left(\alpha_\mathbf{k} \hat{A}_\mathbf{k}
+\beta_\mathbf{k} \hat{B}_\mathbf{k}\right)e^{+iU t/2}
\\
\hat{p}_{\mathbf{k},B}
&=&
\left(\beta_\mathbf{k}^* \hat{A}_\mathbf{k}-
\alpha_\mathbf{k}^* \hat{B}_\mathbf{k}\right)e^{-iU t/2}
\,.
\end{eqnarray}
In contrast to the bosonic case 
(where $|\alpha_\mathbf{k}|^2-|\beta_\mathbf{k}|^2=1$), 
we have 
$|\alpha_\mathbf{k}|^2+|\beta_\mathbf{k}|^2=1$.  
%(instead of $|\alpha_\mathbf{k}|^2-|\beta_\mathbf{k}|^2=1$).
%
This difference reflects the fermionic nature of the quasi-particles 
and will have consequences for the case of resonant hopping
(see below).
The number of created particle-hole pairs then yields the depletion 
\begin{eqnarray}
\langle\hat{n}_{\mu s}\hat{n}_{\mu \bar{s}}\rangle
=
\langle\hat{\bar{n}}_{\mu s}\hat{\bar{n}}_{\mu \bar{s}}\rangle
=
\frac{1}{N}\sum_\mathbf{k} |\beta_\mathbf{k}|^2
\,.
\end{eqnarray}
Note that the equations (\ref{dirac1}) and (\ref{dirac2}) are analogous to 
the Dirac equation in 1+1 dimensions, if we consider a small effective 
electric field $\mathbf{E}$.
In this case, particle-hole pair creation will occur predominantly in the 
vicinity of those points in $\mathbf{k}$-space, where $T_\mathbf{k}$
vanishes, i.e., where the energy gap $\omega_\mathbf{k}$ in 
(\ref{omega-fermi}) assumes it minimum. 
Inserting $\mathbf{k}=\mathbf{k}_0+\delta\mathbf{k}$ with 
$T_{\mathbf{k}_0}=0$, we may approximate $T_\mathbf{k}(t)$ via 
\bea
T_\mathbf{k}(t)
\approx
[\delta\mathbf{k}+\mathbf{A}(t)]
\cdot
[\nabla_\mathbf{k}T_\mathbf{k}]_{\mathbf{k}_0} 
\,.
\ea
Inserting this approximation into the equations (\ref{dirac1}) and 
(\ref{dirac2}), we get 
\bea
i\partial_t 
\left(
\begin{array}{c} 
\hat{p}_{\mathbf{k},B}
\\
\hat{h}_{\mathbf{k},A}
\end{array} 
\right)
=
\left(
\frac{U}{2}\,\sigma^z-
J
[\nabla_\mathbf{k}T_\mathbf{k}]_{\mathbf{k}_0} 
\cdot
[\delta\mathbf{k}+\mathbf{A}(t)]
\sigma^x
\right)
\cdot 
\left(
\begin{array}{c} 
\hat{p}_{\mathbf{k},B}
\\
\hat{h}_{\mathbf{k},A}
\end{array} 
\right)
\,.
\ea
This is precisely the same form as a Dirac equation in 1+1 space-time 
dimensions if we identify the effective speed of light via 
\bea
\mathbf{c}_{\rm eff}=J[\nabla_\mathbf{k}T_\mathbf{k}]_{\mathbf{k}_0} 
\,,
\ea
and the effective mass according to 
\bea
m_{\rm eff}\mathbf{c}_{\rm eff}^2=\frac{U}{2}
\,.
\ea
Note, however, that $\mathbf{c}_{\rm eff}$ depends on $\mathbf{k}_0$
in general, i.e., the analogy only works if we single out a specific 
direction.
In contrast to the bosonic case, we do not find a full analogy valid 
for all $\mathbf{k}$-directions, since the dispersion relation is not 
isotropic near the minimum in the fermionic case. 
Nevertheless, we may use the analogy to the 1+1 dimensional Dirac 
equation in order to estimate the pair creation probability via 
\begin{eqnarray}
|\beta_{\mathbf{k}\approx\mathbf{k}_0}|^2
\sim
\exp\left\{-\pi
\frac{m_{\rm eff}^2\mathbf{c}_{\rm eff}^4}{\mathbf{c}_{\rm eff}\cdot\mathbf{E}}
\right\}
=
\exp\left\{-\pi
\frac{U^2}{4J[\nabla_\mathbf{k}T_\mathbf{k}]_{\mathbf{k}_0}\cdot\mathbf{E}}
\right\}
\,,
\end{eqnarray}
where we have assumed a slowly varying field $\mathbf{E}$.
This result should be relevant for the investigations of the dielectric 
break-down in the Fermi-Hubbard model, see, e.g., \cite{EOW10,OAA03,OA10}.

%%%%%%%%%%%%%%%%%%%%%%%%%%%%%%%%%%%%%%%%%%%%%%%%%%%%%%%%%%%%%%%%%%%%%%%%%%%%%%%
\section{Resonant Tunnelling}\label{restun}
%%%%%%%%%%%%%%%%%%%%%%%%%%%%%%%%%%%%%%%%%%%%%%%%%%%%%%%%%%%%%%%%%%%%%%%%%%%%%%%

In the previous Section, we have studied the case of small potential 
gradients, i.e., small effective electric fields
$V_\mu(t)=\mathbf{E}(t)\cdot\mathbf{x}_\mu$, for which 
we have obtained a quantitative analogy to the Sauter-Schwinger 
effect, which describes tunnelling from the negative continuum 
(i.e., the Dirac sea) to the positive continuum.  
Now let us investigate the case of strong potential gradients.
In this case, the lattice structure becomes important and resonance 
effects play a role. 
For simplicity, we assume a small hopping rate $J\ll U$ where we can solve 
the equations for the charge modes 
(\ref{modify-charge-mode-1}-\ref{modify-charge-mode-4}) via time-dependent 
perturbation theory. 
In this case, Eq.~(\ref{modify-charge-mode-2}) simplifies to 
\bea
\left(i\partial_t-U-V_\nu+V_\mu\right)
f^{0_A1_B}_{\mu\nu}
=
-\frac{J}{Z}T_{\mu\nu}
+\ord(J^2)
\,,
\ea
as the other terms 
$J(T_{\mu\kappa}f^{1_B1_B}_{\kappa\nu}-T_{\kappa\nu}f^{0_A0_A}_{\mu\kappa})$
are of higher order in $J$.
We see that this equation becomes resonant if $V_\mu-V_\nu=U$, i.e., 
if the energy gained by tunnelling from site $\mu$ to site $\nu$ 
compensates the gap $U+\ord(J^2)$.
In this resonance case, the correlation grows linearly with time 
$f^{0_A1_B}_{\mu\nu}=iJtT_{\mu\nu}/Z+\ord(J^2)$. 
Of course, Eq.~(\ref{modify-charge-mode-3}) has the same structure, but 
becomes resonant for the opposite case $V_\mu-V_\nu=-U$.
Either way, the other two correlators 
\bea
i\partial_t f^{0_A0_A}_{\mu\nu}
=
\frac{J}{Z}\sum_{\kappa\neq\mu,\nu}
\left( 
T_{\mu\kappa}f^{1_B0_A}_{\kappa\nu}-T_{\kappa\nu}f^{0_A1_B}_{\mu\kappa}
\right)
\,,
\ea
and similarly $f^{1_B1_B}_{\mu\nu}$, grow quadratically 
$f^{0_A0_A}_{\mu\nu}\propto J^2t^2$. 

The same perturbative analysis can be done for the bosonic case, 
if we start from equations (\ref{f12-Mott}-\ref{f11-Mott}).
Alternatively, we could employ the equations (\ref{diff0}-\ref{diff2}) 
in Fourier space 
\begin{eqnarray}
\left[i\partial_t-U+3 J T_\mathbf{k}(t)\right]
f_\mathbf{k}^{12}
&=&
-\sqrt{2}J T_\mathbf{k}(t)(f_\mathbf{k}^{11}+f_\mathbf{k}^{22}+1)
\,,
\nn
\left[i\partial_t+U-3 J T_\mathbf{k}(t)\right]
f_\mathbf{k}^{21}
&=&
+\sqrt{2}J T_\mathbf{k}(t)(f_\mathbf{k}^{11}+f_\mathbf{k}^{22}+1)
\,,
\nn
i\partial_t f_\mathbf{k}^{11}
=
i\partial_t f_\mathbf{k}^{22}
&=&
\sqrt{2}JT_\mathbf{k}(t)(f^{12}_\mathbf{k}-f^{21}_\mathbf{k})
\nonumber
\,,
\end{eqnarray}
where we have inserted the time-dependent hopping matrix $T_\mathbf{k}(t)$
after the Peierls transformation (\ref{Peierls}).
Since $T_\mathbf{k}$ is periodic in $\mathbf{k}$ (the $\mathbf{k}$-space 
is a periodic repetition of the Brillouin zone), the time-dependent 
hopping matrices $T_\mathbf{k}(t)$ are oscillating 
harmonically\footnote{For non-interacting particles, this behaviour is the
basis for the well-known Bloch oscillations.}  
for constant potential gradients.
Thus the above set of equations corresponds to parametric resonance and 
can be analysed with Floquet theory.
For simplicity, let us assume that the $T_\mathbf{k}(t)$ behave 
after the Peierls transformation as 
%
%For a $d$-dimensional lattice and the potential gradients $E_i=E_0$,  
%($i=1,...,d$) we find after the Peierls transformation
%
\begin{eqnarray}
T_\mathbf{k}(t)
=
\frac{1}{Z}
\left(e^{iE_0t}\chi_\mathbf{k}+e^{-iE_0t}\chi_\mathbf{k}^*\right)
%\quad\mathrm{with}
%\quad \alpha_\mathbf{k}=\sum_{i=1}^d e^{ik_i }
\,.
\end{eqnarray}
In order to solve equations (\ref{diff0}-\ref{diff2}) we make the Floquet
ansatz 
\begin{eqnarray}
f^{12}_\mathbf{k}
=
\sum_{n=-\infty}^{\infty}e^{i(\nu+n)E_0 t}f^{12}_n
\,,
\\
f^{11}_\mathbf{k}
=
f^{22}_\mathbf{k}
=
\sum_{n=-\infty}^{\infty}e^{i(\nu+n)E_0 t}f^{11}_n-\frac{1}{2}
\,,
\\
f^{21}_\mathbf{k}
=
\sum_{n=-\infty}^{\infty}e^{i(\nu+n)E_0 t}f^{21}_n
\,,
\end{eqnarray}
where $\nu$ denotes the Floquet exponent and the $f^{ab}_n$ 
are discrete Fourier coefficients
of the correlation functions $f_\mathbf{k}^{ab}$.
In order to satisfy equations (\ref{diff0}-\ref{diff2}), the  
functions $f^{ab}_n$ have to fulfill the linear system of equations  
\begin{eqnarray}
\label{linsys}
\hat{\mathbf{M}}\cdot\mathbf{f}=\mathbf{0}
\,,
\end{eqnarray}
where we defined the infinite-dimensional matrix 
\begin{eqnarray}
\hat{\mathbf{M}}
=\left( \begin{array}{ccccccc}
....&....&....& & & & \\
    &\chi_\mathbf{k}\mathbf{M}_{-1}&\mathbf{1}&\chi_\mathbf{k}^*\mathbf{M}_{-1}& & & \\
    & &\chi_\mathbf{k}\mathbf{M}_{0}&\mathbf{1}&\chi_\mathbf{k}^*\mathbf{M}_{0}& & \\
    & & &\chi_\mathbf{k}\mathbf{M}_{1}&\mathbf{1}&\chi_\mathbf{k}^*\mathbf{M}_{1} &\\
    &    &    &    &....&....&....
 \end{array}
\right)
\end{eqnarray}
with
\begin{eqnarray}
%\displaystyle
 \mathbf{M}_n=
\frac{J}{ZE_0}\left( \begin{array}{ccc}
-\frac{3}{\nu+n+{U}/{E_0}}&-\frac{2\sqrt{2}}{\nu+n+{U}/{E_0}}&0\\
\frac{2}{\nu+n}&0&-\frac{\sqrt{2}}{\nu+n}\\
0&\frac{2\sqrt{2}}{\nu+n-{U}/{E_0}}&
 \frac{3}{\nu+n-{U}/{E_0}}
\end{array}\right)
\,,
\end{eqnarray}
and the vector 
\begin{eqnarray}
\mathbf{f}=(...,f^{12}_{-1},f^{11}_{-1},f^{21}_{-1},f^{12}_0,f^{11}_0,f^{21}_0,f^{12}_1,f^{11}_1,f^{21}_1,...)^T
\,.
\end{eqnarray}
The set of equations (\ref{linsys}) has only nontrivial solutions 
if the determinant of the infinite-dimensional matrix vanishes, that is
\begin{eqnarray}
\Delta(\nu)=\mathrm{Det}(\hat{\mathbf{M}})=0
\,.
\end{eqnarray}
The above relation determines the Floquet exponent $\nu$ up to multiples of 
$2\pi$ and it can be shown that $\nu$ satisfies the equality \cite{Floquet}
\begin{eqnarray}
\sin^2(\pi\nu)=\sin^2\left(\frac{\pi U}{E_0}\right)\Delta(0)\,.
\end{eqnarray}
If the hopping rate is much smaller than the potential gradient, 
that is $J\ll E_0$, we may expand $\Delta(0)$ in powers of $J/E_0$.
Using the matrix-identity
\begin{eqnarray}
\mathrm{Det}(\hat{\mathbf{M}})=\exp(\tr\{\ln\hat{\mathbf{M}}\})
\,,
\end{eqnarray}
we find up to forth order in $J/E_0$
\begin{eqnarray}\label{resonance}
\sin^2\left(\pi\nu\right)
&=&
\sin^2\left(\frac{\pi U}{E_0}\right)
\Bigg[1+\frac{16 J^2|\chi_\mathbf{k}|^2\pi U}{Z^2 E_0(E_0^2 -U^2)}
\cot\left(\frac{\pi U}{E_0}\right)
\nonumber\\
& &+
\frac{8 J^4|\chi_\mathbf{k}|^4 \pi U}
{Z^4\sin^2\left(\frac{\pi U}{E_0}\right)}
\frac{1}{ E_0^2(E_0^2-U^2)^3(4E_0^2-U^2)}
\nonumber\\
& &\quad\times \Bigg\{8\pi U\left(4 E_0^4-5 E_0^2 U^2+U^4\right)
\cos\left(\frac{2\pi U}{E_0}\right)\nonumber\\
& &\quad\quad+E_0\left(-19E_0^4+76E_0^2U^2-33 U^4\right)
\sin\left(\frac{2\pi U}{E_0}\right)\Bigg\}\Bigg]
\,.
\end{eqnarray}
Two cases need to be distinguished.
In the first case, the right hand side of (\ref{resonance}) 
is between zero and unity,
the Floquet exponent is real and the correlation functions are bounded.
In the second case, the right hand side of (\ref{resonance}) 
is bigger than unity
or smaller than zero, the Floquet exponent acquires an imaginary part 
and the correlation functions grow exponentially, 
$f_\mathbf{k}^{ab}\sim \exp(\Im \nu t E_0) $, corresponding
to a Floquet resonance.

We find the first resonance being located at $U=E_0$ with
a width of 
$\Delta U=2(\Im \nu)_\mathrm{max}E_0=4\sqrt{2}J|\chi_\mathbf{k}|/Z$.
The second resonance is located at 
$U=2E_0+16J^2|\chi_\mathbf{k}|^2/(3E_0Z^2)$ and has
the width 
$\Delta U=2(\Im \nu)_\mathrm{max}E_0=
12\sqrt{2}J^2|\chi_\mathbf{k}|^2/(Z^2E_0)$.
In principle, there are resonances when $E_0/U$ adopts higher integer 
values but they become smaller for increasing $E_0/U$.

In contrast, the correlation functions in the Fermi-Hubbard model do not 
grow exponentially.
This distinction between the bosonic and the fermionic case can already be 
deduced from the difference of the conserved quantities (\ref{inv}) and 
(\ref{invfermi}).
While the relation (\ref{inv}) allows in principle arbitrary large 
values of the correlation functions, we find from (\ref{invfermi}) 
that $f_\mathbf{k}^{1_B1_B}$ cannot exceed unity and is therefore 
bounded.
%These findings are closely related to the quasi-particle picture 
%which is disscussed in appendix \ref{quasiparticles}.

%%%%%%%%%%%%%%%%%%%%%%%%%%%%%%%%%%%%%%%%%%%%%%%%%%%%%%%%%%%%%%%%%%%%%%%%%%%%%%%
\section{Conclusions \& Outlook}
%%%%%%%%%%%%%%%%%%%%%%%%%%%%%%%%%%%%%%%%%%%%%%%%%%%%%%%%%%%%%%%%%%%%%%%%%%%%%%%

For the Bose and the Fermi-Hubbard model, we studied the quantum correlations 
analytically by using the hierarchy of correlations obtained via a formal 
expansion into powers of $1/Z$.  
Starting deep in the Mott regime $J/U\to0$ with exactly one particle per 
lattice site, we derived the correlations in the ground state for a finite 
value of $J$ and after a quantum quench (i.e., suddenly switching on $J$). 
From these correlations, we can also infer the quantum depletion, i.e., 
the probability of having zero (``holon'') or two (``doublon'') particles 
at a given lattice site.  
It turns out that these observables approach a quasi-equilibrium state some 
time after the quench -- but this quasi-equilibrium state is {\em not} 
thermal.  
Furthermore, we derived the particle-hole (``doublon-holon'') pair creation 
probability via tunnelling in tilted lattices and found remarkable analogies  
to the Sauter-Schwinger effect (i.e., electron-positron pair creation out of 
the quantum vacuum by an external electric field) in the case of weak tilts. 
For strong tilts, one obtains resonant tunnelling reminiscent of the Bloch 
oscillations for non-interacting particles. 

For the Bose-Hubbard model, we also studied a quench from the Mott state to 
the super-fluid regime and calculated the growth of phase coherence. 
Going up to second order in $1/Z$, we derived the correlations of the 
number and parity operators, again in the ground state and after a quench. 
For the Fermi-Hubbard model, we found that the spin and charge modes
decouple to first order in $1/Z$.
The dynamics of the charge modes (particle-hole excitations) already 
contributes to first order in $1/Z$ whereas the time-evolution of the 
spin modes requires going up to the second order in $1/Z$, similar to the 
number and parity correlations in the bosonic case. 

Comparing our analytical results to numerical simulations for bosons on 
finite-size lattices in one and two spatial dimensions, 
we found qualitative agreement.
Thus, although our analytical approach is formally based on the limit $Z\gg1$,
we expect that our results apply -- at least qualitatively -- to lattices 
in three ($Z=6$), two ($Z=4$), or even one ($Z=2$) spatial dimension. 
There are only a few properties which strongly depend on the dimensionality 
of the system, one example being the maximum of the parity correlator well 
within the Mott regime, which occurs in one spatial dimension only. 
In view of the tremendous experimental progress regarding ultra-cold atoms 
in optical lattices, for example, most of our predictions should be testable 
experimentally. 

In this paper, we used the zero-temperature Mott phase as our initial state 
-- but the presented method can easily be applied to other initial states.
For example, finite initial temperatures can be incorporated as well because 
our approach is based on density matrices. 
Even at zero temperature, it should be interesting to study other initial 
states. 
In the bosonic case, we could start with $U=0$ (instead of $J=0$), 
i.e., in the super-fluid phase, where we may use 
$\rho^0_\mu=\ket{\alpha}_\mu\!\bra{\alpha}$ with the coherent state 
$\hat b_\mu\ket{\alpha}_\mu=\alpha\ket{\alpha}_\mu$, 
see, e.g., \cite{KN11}.
In this way, an order parameter $\langle\hat b_\kappa\rangle\not=0$ 
is introduced at the expense of making the total particle number 
ill-defined $[\hat N,\hat\rho_{\mu}]\not=0$.
As another possibility, one could assume non-integer filling 
$\langle\hat n_\mu\rangle\not\in\mathbb N$, where one has a non-vanishing 
super-fluid component even for arbitrarily small $J$.
For example, taking $\langle\hat n_\mu\rangle<1$, the initial state would be 
$\rho^0_\mu=\ket{\psi}_\mu\!\bra{\psi}$ with 
$\ket{\psi}=\alpha\ket{0}+\beta\ket{1}$.
In these cases, the time-dependence of $\rho^0_\mu$ will be non-trivial 
in general. 
In the fermionic case, an analogous initial state would be 
$\rho^0_\mu=\ket{\psi}_\mu\!\bra{\psi}$ with 
$\ket{\psi}=
\alpha\ket{0^\uparrow0^\downarrow}+\beta\ket{1^\uparrow1^\downarrow}$,  
which could describe a Bose-Einstein condensate of Cooper-like pairs, 
which may be stabilised by an attractive interaction $U<0$, for example.   

%%%%%%%%%%%%%%%%%%%%%%%%%%%%%%%%%%%%%%%%%%%%%%%%%%%%%%%%%%%%%%%%%%%%%%%%%%%%%%%
\section*{Acknowledgements} 
%%%%%%%%%%%%%%%%%%%%%%%%%%%%%%%%%%%%%%%%%%%%%%%%%%%%%%%%%%%%%%%%%%%%%%%%%%%%%%%

The authors acknowledge valuable discussions with M.~Vojta, A.~Rosch, 
W.~Hofstetter, and others (e.g., several members of the SFB-TR12).
This work was supported by the DFG (SFB-TR12). 

%%%%%%%%%%%%%%%%%%%%%%%%%%%%%%%%%%%%%%%%%%%%%%%%%%%%%%%%%%%%%%%%%%%%%%%%%%%%%%%

\section{Appendix: Derivation of the hierarchy}\label{hierarchyApp}

In this Appendix, we derive the hierarchical set of equations for the 
correlation functions.
%
%The Hilbert space associated to any site $\mu$ is quite general 
%and can be described in terms of a set of 
%orthogonal states $|i\rangle_\mu$. 
%
%Typically for a one component gas, the index $i$ labels the state 
%$\{ n_{-s}\dots n_{s}\}$ where $s$ is the spin of the particle and 
%where $n_{m_s}$ 
%the particle number with $z$-component $m_s=-s...s$ and taking positive 
%integer values for bosons and $0$ or $1$ for fermions, respectively.
%
The quantum evolution of the one-site density matrix can be derived
by tracing von Neumann's equation (\ref{Liouville})
over all lattice sites but $\mu$ and exploiting the invariance of the 
trace under cyclic permutations 
\begin{eqnarray}
\label{singlesite}
i\partial_t \hat{\rho}_\mu
&=&
\frac{1}{Z}\tr_{\not\mu}
\left\{\sum_{\alpha,\beta\neq\mu}\mathcal{L}_{\alpha\beta}\hat{\rho}
+\sum_{\alpha\neq \mu}\mathcal{L}^S_{\alpha\mu}\hat{\rho}\right\}
+
\tr_{\not\mu}\left\{\sum_{\alpha\neq\mu}
\mathcal{L}_\alpha\hat{\rho}+\mathcal{L}_\mu\hat{\rho}\right\}
\nonumber\\
&=&
\frac{1}{Z}\sum_{\alpha\neq \mu} 
\mathcal{L}^S_{\mu\alpha}\tr_{\alpha}\{\hat{\rho}_{\mu\kappa}\}
+\mathcal{L}_\mu\hat{\rho}_\mu\,.
\end{eqnarray}
Using the definition of the two-point correlations given in 
(\ref{correlated-parts}), we arrive at (\ref{one-site}).
Similarly, the differential equation for the two-particle density matrix can 
be deduced by tracing over all lattice sites but $\mu$ and $\nu$,
\begin{eqnarray}
i\partial_t \hat{\rho}_{\mu\nu}
&=&
i\left(\partial_t \hat{\rho}_{\mu\nu}^\mathrm{corr}+
\hat{\rho}_\mu \partial_t \hat{\rho}_\nu+
\hat{\rho}_\nu \partial_t \hat{\rho}_\mu\right)
\nonumber\\
&=&
\frac{1}{Z}\sum_{\alpha\neq \mu\nu}
\tr_\alpha\left\{\mathcal{L}_{\mu\kappa}^S\hat{\rho}_{\mu\nu\alpha}\right\}
+\frac{1}{Z}\sum_{\alpha\neq \mu\nu}
\tr_\alpha\left\{\mathcal{L}_{\kappa\nu}^S\hat{\rho}_{\mu\nu\alpha}\right\}
\nonumber\\
&&
+\frac{1}{Z}\mathcal{L}_{\mu\nu}^S\hat{\rho}_{\mu\nu}
+\mathcal{L}_\mu \hat{\rho}_{\mu\nu}+\mathcal{L}_\nu \hat{\rho}_{\mu\nu}\,.
\end{eqnarray}
With the definitions (\ref{correlated-parts}) and the 
time-evolution for the single-site density matrix (\ref{singlesite}),
we find for the two-point correlation functions (\ref{two-sites}).
The equations (\ref{one-site}) and (\ref{two-sites})
preserve the hierarchy in time if initially
$\hat{\rho}_\mu=\mathcal{O}(Z^0)$ and 
$\hat{\rho}_{\mu\nu}^\mathrm{corr}=\mathcal{O}(1/Z)$ holds.
In order to derive the full hierarchy,
we define the generating functional
\bea
{\cal F}(\hat\alpha)
=
{\cal F}(\{\hat\alpha_\mu\}) 
=
\ln\left[
\tr\left\{\hat\rho\bigotimes_\mu(\mathbf{1}_\mu+\hat\alpha_\mu)\right\}
\right]
\,,
\eea 
where $\hat\rho$ is the density matrix of the full lattice and 
\begin{eqnarray}
\hat\alpha_\mu=\sum_{m,n}\alpha_\mu^{m,n}|m\rangle_\mu\langle n|
\end{eqnarray}
are arbitrary operators acting on the Hilbert spaces associated to the 
lattice sites $\mu$ with the local basis $\{\ket{n}_\mu\}$.
%
%in an arbitrary basis $\{\ket{n}_\mu\}$.
%
The role of this functional is to generate all correlated density matrices 
via the derivatives with respect to these operators $\hat\alpha_\mu$ which 
are defined via 
\begin{eqnarray}
\frac{\partial{\cal F(\{\alpha\})}}{{\partial\hat\alpha_\mu}}
=
\sum_{m,n}|n\rangle_\mu\langle m|\,
\frac{\partial \cal F(\{\alpha\})}{\partial \alpha^{m,n}_\mu}
=
\sum_{m,n}|n\rangle_\mu\langle m|\,
\frac{\partial{\cal F}(\{\alpha\})}
{\partial\, {_\mu\!\bra{m}}\hat\alpha_\mu\ket{n}_\mu}
\,.
\end{eqnarray}
If we consider an ensemble ${\cal S}=\{\mu_1, \dots ,\mu_\ell\}$  
of $\ell$ different lattice sites $\mu_1\not= \dots \not=\mu_\ell$, 
we obtain the correlation operators via 
\bea
\hat\rho^{\rm corr}_{\cal S} 
=
\left.
\frac{\partial}{\partial\hat\alpha_{\mu_1}}
\frac{\partial}{\partial\hat\alpha_{\mu_2}}
\dots 
\frac{\partial}{\partial\hat\alpha_{\mu_\ell}}
{\cal F}(\hat \alpha)\right|_{\hat \alpha=0} 
\,.
\eea
These operators are related to the corresponding reduced density matrix 
operator $\rho_{\cal S}$ through the relation 
\bea
\hat\rho_{\cal S} 
= 
\hat\rho_{\mu_1 \dots \mu_\ell}
=
\sum_{\cup_i{\cal P}_i={\cal S}}\prod_i\hat\rho^{\rm corr}_{{\cal P}_i}
\ea
where the sum runs over all possible segmentations of the subset ${\cal S}$
into partitions ${\cal P}_i$ starting from the whole subset 
${\cal P}={\cal S}$ and ranging to single lattice sites 
${\cal P}_i=\{\mu\}$ where 
$\hat\rho^{\rm corr}_{{\cal P}_i=\{\mu\}}=\hat\rho_\mu$ is understood. 
For two and three lattice sites, the above equation reproduces 
Eq.~(\ref{correlated-parts}). 

Our derivation is based on the following scaling hierarchy of 
correlations:
\bea
\label{hierarchy1}
\hat\rho_{\cal S}^{\rm c}=\ord\left(Z^{1-|\cal S|}\right)
\ea
where $|\cal S|$ is the number $\ell$ of lattice sites in the set $\cal S$.
%
%
%
%\bea
%i \partial_t \hat\rho 
%=
%\left[\hat H,\hat\rho\right]
%= 
%\left(\sum_\mu \li_\mu + \frac{1}{Z} \sum_{\mu,\nu} \li_{\mu \nu}\right)
%\hat\rho
%\,,
%\eea
%
From the Liouville equation (\ref{Liouville}), 
the temporal evolution of ${\cal F}$ is given by 
\bea
\label{eqgf}
i \partial_t 
{\cal F}(\hat\alpha)
&=&
\sum_\mu\tr_\mu 
\left\{
\hat\alpha_\mu \li_\mu \frac{\partial{\cal F}}{\partial\hat\alpha_\mu}
\right\}
 \\
&+&
\frac{1}{Z} \sum_{\mu,\nu} \tr_{\mu\nu} 
\left\{
(\hat\alpha_\mu + \hat\alpha_\nu +\hat\alpha_\mu \hat\alpha_\nu) 
\li_{\mu \nu} 
\left(
\frac{\partial^2{\cal F}}{\partial\hat\alpha_\mu\partial\hat\alpha_\nu}
+
\frac{\partial{\cal F}}{\partial\hat\alpha_\mu}
\frac{\partial{\cal F}}{\partial\hat\alpha_\nu}
\right)
\right\}\nonumber
\,.
\eea
By taking successive derivatives and using the generalised Leibniz rule 
\bea
\frac{\partial}{\partial\hat\alpha_{\mu_1}}
\frac{\partial}{\partial\hat\alpha_{\mu_2}}
\dots 
\frac{\partial}{\partial\hat\alpha_{\mu_\ell}}
\left[{\cal F}(\hat \alpha)\right]^2
&=&
\sum_{{\cal P}\subseteq {\cal S}}^{{\cal P}\cup\bar{\cal P}={\cal S}}  
\left[
\left(
\prod_{\mu_i \in {\cal P}}
\frac{\partial}{\partial\hat\alpha_{\mu_i}}
\right)
{\cal F}(\hat \alpha)
\right]
\times
\\
&&
\times
\left[
\left(
\prod_{\mu_j \in \bar{\cal P}}
\frac{\partial}{\partial\hat\alpha_{\mu_j}}
\right)
{\cal F}(\hat \alpha)
\right]
\,,
%\\
%\partial_{\hat \alpha_{\mu_1}}\dots \partial_{\hat \alpha_{\mu_l}}
%{\cal F}(\hat \alpha){\cal F}(\hat \alpha)
%=
%\sum_{{\cal P}\subseteq {\cal S}}^{{\cal P}\cup\bar{\cal P}={\cal S}}  
%\left\{\left(\prod_{\mu_i \in {\cal P}} \partial_{\hat \alpha_{\mu_i}}\right) 
%{\cal F}(\hat \alpha)\right\}
%\nonumber\\
%\times\left\{\left(
%\prod_{\mu_j \in {\cal P}^c} \partial_{\hat\alpha_{\mu_j}} 
%\right)
%{\cal F}(\hat \alpha)\right\}
\eea
as well as the the property 
\bea
\frac{\partial^2{\cal F}(\hat \alpha)}{\partial\hat\alpha_{\mu}^2}
=
\frac{\partial}{\partial\hat \alpha_{\mu}}
\frac{\partial}{\partial\hat \alpha_{\mu}}
{\cal F}(\hat \alpha)
=
-
\frac{\partial{\cal F}(\hat \alpha)}{\partial\hat\alpha_{\mu}}
\frac{\partial{\cal F}(\hat \alpha)}{\partial\hat\alpha_{\mu}}
=
-
\left(\frac{\partial{\cal F}(\hat \alpha)}{\partial\hat\alpha_{\mu}}\right)^2
\,,
%
%\\
%\partial_{\hat \alpha_{\mu}} \partial_{\hat \alpha_{\mu}} 
%{\cal F}(\hat \alpha)
%&=&
%-\partial_{\hat \alpha_{\mu}} {\cal F}(\hat \alpha) 
%\partial_{\hat \alpha_{\mu}}
%{\cal F}(\hat \alpha)
% \\
% \frac{\partial}{\partial\hat \alpha_{\mu_1}}
% \frac{\partial}{\partial\hat \alpha_{\mu_2}}
% \dots 
% \frac{\partial}{\partial\hat \alpha_{\mu_\ell}}
% {\cal F}(\hat \alpha)
% &=&
% \sum_{{\cal P}\subseteq{\cal N}}^{{\cal P}\cup\bar{\cal P}={\cal N}} 
% \prod_{\mu_i\in{\cal P}}
% \frac{\partial{\cal F}(\hat \alpha)}{\partial\hat\alpha_{\mu_i}}
% \prod_{\mu_j\in\bar{\cal P}}
% \frac{\partial{\cal F}(\hat \alpha)}{\partial\hat\alpha_{\mu_j}}
\ea
we establish the following set of 
equations for the correlated density matrices: 
%
%\begin{widetext}
\begin{eqnarray}
\label{general}
i \partial_t\hat\rho^{\rm corr}_{\cal S}
&=&
\sum_{\mu \in {\cal S}} \li_\mu \hat\rho^{\rm corr}_{\cal S}
+
\frac{1}{Z}\sum_{\mu,\nu\in{\cal S}}
\li_{\mu \nu}\,\hat\rho^{\rm corr}_{\cal S} 
\nonumber\\
& &+ 
\frac{1}{Z}
\sum_{\kappa\notin{\cal S}} \sum_{\mu\in{\cal S}} 
\tr_{\kappa}
\Bigg[
\li^S_{\mu \kappa}
\hat\rho^{\rm corr}_{{\cal S}\cup {\kappa}}
+ 
\sum_{{\cal P}\subseteq{\cal S}\setminus\{\mu\}}
^{{\cal P}\cup\bar{\cal P}={\cal S}\setminus\{\mu\}}
\li^S_{\mu \kappa}
\hat\rho^{\rm corr}_{\{\mu\}\cup{\cal P}}\,
\hat\rho^{\rm corr}_{\{\kappa\}\cup\bar{\cal P}}
\Bigg]\nonumber\\
& &+
\frac{1}{Z}
\sum_{\mu,\nu\in{\cal S}} 
\sum_{{\cal P}\subseteq{\cal S}\setminus\{\mu,\nu\}}
^{{\cal P}\cup\bar{\cal P}={\cal S}\setminus\{\mu,\nu\}}
\Bigg\{
\li_{\mu \nu}\,
\hat\rho^{\rm corr}_{\{\mu\}\cup{\cal P}}\,
\hat\rho^{\rm corr}_{\{\nu\}\cup\bar{\cal P}}
\nonumber \\
& &-\tr_{\nu}\Bigg[
\li^S_{\mu \nu}(\hat\rho^{\rm corr}_{\{\mu,\nu\}\cup\bar{\cal P}} 
+
\sum_{{\cal Q}\subseteq\bar{\cal P}}
^{{\cal Q}\cup\bar{\cal Q}=\bar{\cal P}}
\hat\rho^{\rm corr}_{\{\mu\}\cup{\cal Q}}\,
\hat\rho^{\rm corr}_{\{\nu\}\cup\bar{\cal Q}}
)
\Bigg]
\hat\rho^{\rm corr}_{\{\nu\}\cup{\cal P}}
\Bigg\}
\,.
\end{eqnarray}
% \begin{eqnarray}
% \label{general}
% i \partial_t\hat\rho^{\rm corr}_{\cal S}
% =
% \sum_{\mu \in {\cal S}} \li_\mu \hat\rho^{\rm corr}_{\cal S}
% +
% \frac{1}{Z}\sum_{\mu,\nu\in{\cal S}}
% \li_{\mu \nu}\,\hat\rho^{\rm corr}_{\cal S} 
% + 
% \frac{1}{Z}
% \sum_{\kappa\notin{\cal S}} \sum_{\mu\in{\cal S}} 
% \tr_{\kappa}
% \Bigg[
% \li^S_{\mu \kappa}
% \hat\rho^{\rm corr}_{{\cal S}\cup {\kappa}}
% + 
% \nonumber \\
% \sum_{{\cal P}\subseteq{\cal S}\setminus\{\mu\}}
% ^{{\cal P}\cup\bar{\cal P}={\cal S}\setminus\{\mu\}}
% \li^S_{\mu \kappa}
% \hat\rho^{\rm corr}_{\{\mu\}\cup{\cal P}}\,
% \hat\rho^{\rm corr}_{\{\kappa\}\cup\bar{\cal P}}
% \Bigg]
% +
% \frac{1}{Z}
% \sum_{\mu,\nu\in{\cal S}} 
% \sum_{{\cal P}\subseteq{\cal S}\setminus\{\mu,\nu\}}
% ^{{\cal P}\cup\bar{\cal P}={\cal S}\setminus\{\mu,\nu\}}
% \Bigg\{
% \li_{\mu \nu}\,
% \hat\rho^{\rm corr}_{\{\mu\}\cup{\cal P}}\,
% \hat\rho^{\rm corr}_{\{\nu\}\cup\bar{\cal P}}
% -
% \nonumber \\
% \tr_{\nu}\Bigg[
% \li^S_{\mu \nu}(\hat\rho^{\rm corr}_{\{\mu,\nu\}\cup\bar{\cal P}} 
% +
% \sum_{{\cal Q}\subseteq\bar{\cal P}}
% ^{{\cal Q}\cup\bar{\cal Q}=\bar{\cal P}}
% \hat\rho^{\rm corr}_{\{\mu\}\cup{\cal Q}}\,
% \hat\rho^{\rm corr}_{\{\nu\}\cup\bar{\cal Q}}
% )
% \Bigg]
% \hat\rho^{\rm corr}_{\{\nu\}\cup{\cal P}}
% \Bigg\}
% \,.
% \end{eqnarray}
%\end{widetext}
%
For $\ell=1$ and $\ell=2$ we recover the equations 
(\ref{one-site}) and (\ref{two-sites}).
A careful inspection of this set of equations shows that the hierarchy 
in (\ref{hierarchy1}) is preserved in time: 
Imposing the scaling 
$\hat\rho^{\rm corr}_{\cal S}=\ord(Z^{1-|\cal S|})$ on the r.h.s.\ of the 
above equation, we find that the time derivative on the l.h.s.\ does also 
satisfy the hierarchy (\ref{hierarchy1}). 
Therefore, inserting (\ref{hierarchy1}) into (\ref{general}) and taking the 
limit $Z\rightarrow\infty$, we obtain the leading-order contributions
\bea
\label{general1}
i \partial_t\hat\rho^{\rm corr}_{\cal S}
&=&
\sum_{\mu \in {\cal S}} \li_\mu \hat\rho^{\rm corr}_{\cal S}
+ 
\frac{1}{Z}
\sum_{\kappa\notin{\cal S}} \sum_{\mu\in{\cal S}} 
\tr_{\kappa}\Bigg[
\sum_{{\cal P}\subseteq{\cal S}\setminus\{\mu\}}
^{{\cal P}\cup\bar{\cal P}={\cal S}\setminus\{\mu\}}
\li^S_{\mu \kappa}
\hat\rho^{\rm corr}_{\{\mu\}\cup{\cal P}}\,
\hat\rho^{\rm corr}_{\{\kappa\}\cup\bar{\cal P}}
\Bigg]
\nonumber \\
& &+\frac{1}{Z}
\sum_{\mu,\nu\in{\cal S}} 
\sum_{{\cal P}\subseteq{\cal S}\setminus\{\mu,\nu\}}
^{{\cal P}\cup\bar{\cal P}={\cal S}\setminus\{\mu,\nu\}}
\Bigg\{
\li_{\mu \nu}\,
\hat\rho^{\rm corr}_{\{\mu\}\cup{\cal P}}\,
\hat\rho^{\rm corr}_{\{\nu\}\cup\bar{\cal P}}
\nonumber\\ 
& &-
\tr_{\nu}\Bigg[
\li^S_{\mu \nu}
\sum_{{\cal Q}\subseteq\bar{\cal P}}
^{{\cal Q}\cup\bar{\cal Q}=\bar{\cal P}}
\hat\rho^{\rm corr}_{\{\mu\}\cup{\cal Q}}\,
\hat\rho^{\rm corr}_{\{\nu\}\cup\bar{\cal Q}}
\Bigg]
\hat\rho^{\rm corr}_{\{\nu\}\cup{\cal P}}
\Bigg\}
+\ord(Z^{-|\cal S|})
\,.
%\nonumber \\
\eea
For $\ell=1$ and $\ell=2$, we recover equations 
(\ref{one-site-approx}) and (\ref{two-sites-approx}).

In contrast to the exact expression (\ref{general}), the approximated 
leading-order equations (\ref{general1}) form a closed set.
The exact time evolution (\ref{general}) of the $|\cal S|$-point 
correlator $\partial_t\hat\rho^{\rm corr}_{\cal S}$ also depends on the 
higher-order correlation term $\hat\rho^{\rm corr}_{{\cal S}\cup{\kappa}}$ 
involving $|{\cal S}|+1$ points. 
The approximated  expression (\ref{general1}), on the other hand, only 
contains correlators of the same or lower rank. 
This facilitates the iterative solution of the problem sketched in 
Section~\ref{hierarchyofcorr}.
First one solves the zeroth-order equation (\ref{one-site-approx}) 
for $\hat\rho_\mu^0$.
Inserting this result $\hat\rho_\mu^0$ into the first-order (in $1/Z$) 
equation (\ref{two-sites-approx}) for $\hat\rho_{\mu\nu}^{\rm corr}$, 
we obtain a first-order result for $\hat\rho_{\mu\nu}^{\rm corr}$. 
This first-order result for $\hat\rho_{\mu\nu}^{\rm corr}$ can then be 
inserted into the equation for $\hat\rho_{\mu\nu\lambda}^{\rm corr}$
which is of second order $1/Z^2$.
Furthermore, we may use the first-order result for 
$\hat\rho_{\mu\nu}^{\rm corr}$ in order to obtain a better approximation 
for the on-site density matrix $\hat\rho_\mu^1$ which is valid to first 
order in $1/Z$ and contains the quantum depletion etc. 
Repeating this iteration, we may successively ``climb up'' to higher and 
higher orders in $1/Z$.

%%%%%%%%%%%%%%%%%%%%%%%%%%%%%%%%%%%%%%%%%%%%%%%%%%%%%%%%%%%%%%%%%%%%%%%%%%%%%%%
\section{Appendix: Staggered Magnetic Field}\label{staggered}
%%%%%%%%%%%%%%%%%%%%%%%%%%%%%%%%%%%%%%%%%%%%%%%%%%%%%%%%%%%%%%%%%%%%%%%%%%%%%%%

We assumed in Section \ref{chargemodes} that the initial state of the 
Fermi-Hubbard system is given by the N\'eel state.
However, for $J=0$ we have infinitely many states with same energy and 
vanishing total spin.
In order to single out the N\'eel state, we add a staggered magnetic field 
to the Hubbard Hamiltonian,
\begin{eqnarray}
\hat H
&=&
-\frac{J}{Z}\sum_{\mu,\nu}T_{\mu\nu}
\left(\hat{c}_{\mu\uparrow}^\dagger\hat{c}_{\mu\uparrow}+
\hat{c}_{\mu\downarrow}^\dagger\hat{c}_{\mu\downarrow}\right)
\nonumber\\
&&+
\sum_{\mu}\left(
U\hat{n}_\mu^\uparrow\hat{n}_\mu^\downarrow
-A_{\mu\downarrow}\hat{n}_\mu^\downarrow
-A_{\mu\uparrow} \hat{n}_\mu^\uparrow
\right)
\,.
\end{eqnarray}
If we choose the magnetic field as 
$A_{\mu\downarrow}(x_\mu\in \mathcal{A})=a$,
$A_{\mu\downarrow}(x_\mu\in \mathcal{B})=0$, 
$A_{\mu\uparrow}(x_\mu\in \mathcal{B})=a$, and 
$A_{\mu\uparrow}(x_\mu\in \mathcal{A})=0$,  
the N\'eel state is the unique ground state for $J=0$ at half filling. 
The Heisenberg equations (\ref{annihilation-operator})-(\ref{number-operator}) read now
\begin{eqnarray}
i\partial_t \hat{c}_{\mu s}
&=&
-\frac{J}{Z}\sum_{\kappa\neq \mu} T_{\mu\kappa}\hat{c}_{\kappa s}
+U\hat{c}_{\mu s} \hat{n}_{\mu \bar{s}}-A_{\mu s}\hat{c}_{\mu s}
\\
i\partial_t \hat{c}_{\mu s}^\dagger
&=&
\frac{J}{Z}\sum_{\kappa\neq \mu} T_{\mu\kappa}\hat{c}^\dagger_{\kappa s}
-U\hat{c}^\dagger_{\mu s} \hat{n}_{\mu \bar{s}}
+A_{\mu s}\hat{c}^\dagger_{\mu s}
\\
i\partial_t\hat{n}_{\mu s}
&=&
-i\partial_t\hat{\bar{n}}_{\mu s}
=
\frac{J}{Z}\sum_{\kappa\neq \mu}T_{\mu\kappa}\left(
\hat{c}_{\kappa s}^\dagger \hat{c}_{\mu s}
-\hat{c}_{\mu s}^\dagger \hat{c}_{\kappa s}
\right)\,,
\end{eqnarray}
To first order in $1/Z$, we find the closed set of differential equations, 
cf.~Eqs.~(\ref{corr1})-(\ref{sectors}), 
\begin{eqnarray}
i\partial_t \langle \hat{c}_{\mu a}^\dagger\hat{c}_{\nu b}\hat{n}_{\mu\bar{a}}\hat{n}_{\nu\bar{b}}\rangle&=&
\frac{J}{Z}\sum_{\kappa\neq \mu,\nu}T_{\mu\kappa}
\langle\hat{c}_{\kappa a}^\dagger\hat{c}_{\nu b}(\hat{n}_{\kappa\bar{a}}+\hat{\bar{n}}_{\kappa\bar{a}})
\hat{n}_{\nu\bar{b}}\rangle\langle\hat{n}_{\mu \bar{a}}\rangle_0\nonumber\\
& &-\frac{J}{Z}\sum_{\kappa\neq \mu,\nu}T_{\nu\kappa}
\langle\hat{c}_{\mu a}^\dagger\hat{c}_{\kappa b}\hat{n}_{\mu\bar{a}}(\hat{n}_{\kappa\bar{b}}+\hat{\bar{n}}_{\kappa\bar{b}})
\rangle\langle\hat{n}_{\nu \bar{b}}\rangle_0\nonumber\\
& &+(A_{\mu a}-A_{\nu b})\langle \hat{c}_{\mu a}^\dagger\hat{c}_{\nu b}\hat{n}_{\mu\bar{a}}\hat{n}_{\nu\bar{b}}\rangle\nonumber\\
& &+\frac{J}{Z}T_{\mu\nu}\langle \hat{c}_{\nu a}^\dagger\hat{c}_{\nu b}\hat{n}_{\mu\bar{a}}\hat{n}_{\nu\bar{b}}\rangle_0
-\frac{J}{Z}T_{\mu\nu}\langle \hat{c}_{\mu a}^\dagger\hat{c}_{\mu b}\hat{n}_{\mu\bar{a}}\hat{n}_{\nu\bar{b}}\rangle_0
\end{eqnarray}
\begin{eqnarray}
i\partial_t \langle \hat{c}_{\mu a}^\dagger\hat{c}_{\nu b}\hat{n}_{\mu\bar{a}}\hat{\bar{n}}_{\nu\bar{b}}\rangle&=&
\frac{J}{Z}\sum_{\kappa\neq \mu,\nu}T_{\mu\kappa}
\langle\hat{c}_{\kappa a}^\dagger\hat{c}_{\nu b}(\hat{n}_{\kappa\bar{a}}+\hat{\bar{n}}_{\kappa\bar{a}})
\hat{\bar{n}}_{\nu\bar{b}}\rangle\langle\hat{n}_{\mu \bar{a}}\rangle_0\nonumber\\
& &-\frac{J}{Z}\sum_{\kappa\neq \mu,\nu}T_{\nu\kappa}
\langle\hat{c}_{\mu a}^\dagger\hat{c}_{\kappa b}\hat{n}_{\mu\bar{a}}(\hat{n}_{\kappa\bar{b}}+\hat{\bar{n}}_{\kappa\bar{b}})
\rangle\langle\hat{\bar{n}}_{\nu \bar{b}}\rangle_0\nonumber\\
& &-(U-A_{\mu a}+A_{\nu b})\langle \hat{c}_{\mu a}^\dagger\hat{c}_{\nu b}\hat{n}_{\mu \bar{a}}\hat{\bar{n}}_{\nu \bar{b}}\rangle\nonumber\\
& &+\frac{J}{Z}T_{\mu\nu}\langle \hat{c}_{\nu a}^\dagger\hat{c}_{\nu b}\hat{n}_{\mu\bar{a}}\hat{\bar{n}}_{\nu\bar{b}}\rangle_0
-\frac{J}{Z}T_{\mu\nu}\langle \hat{c}_{\mu a}^\dagger\hat{c}_{\mu b}\hat{n}_{\mu\bar{a}}\hat{\bar{n}}_{\nu\bar{b}}\rangle_0
\end{eqnarray}
\begin{eqnarray}
i\partial_t \langle \hat{c}_{\mu a}^\dagger\hat{c}_{\nu b}\hat{\bar{n}}_{\mu\bar{a}}\hat{n}_{\nu\bar{b}}\rangle&=&
\frac{J}{Z}\sum_{\kappa\neq \mu,\nu}T_{\mu\kappa}
\langle\hat{c}_{\kappa a}^\dagger\hat{c}_{\nu b}(\hat{n}_{\kappa\bar{a}}+\hat{\bar{n}}_{\kappa\bar{a}})
\hat{n}_{\nu\bar{b}}\rangle\langle\hat{\bar{n}}_{\mu \bar{a}}\rangle_0\nonumber\\
& &-\frac{J}{Z}\sum_{\kappa\neq \mu,\nu}T_{\nu\kappa}
\langle\hat{c}_{\mu a}^\dagger\hat{c}_{\kappa b}\hat{\bar{n}}_{\mu\bar{a}}(\hat{n}_{\kappa\bar{b}}+\hat{\bar{n}}_{\kappa\bar{b}})
\rangle\langle\hat{n}_{\nu \bar{b}}\rangle_0\nonumber\\
& &+(U+A_{\mu a}-A_{\nu b})\langle \hat{c}_{\mu a}^\dagger\hat{c}_{\nu b}\hat{\bar{n}}_{\mu \bar{a}}\hat{n}_{\nu \bar{b}}\rangle\nonumber\\
& &+\frac{J}{Z}T_{\mu\nu}\langle \hat{c}_{\nu a}^\dagger\hat{c}_{\nu b}\hat{\bar{n}}_{\mu\bar{a}}\hat{n}_{\nu\bar{b}}\rangle_0
-\frac{J}{Z}T_{\mu\nu}\langle \hat{c}_{\mu a}^\dagger\hat{c}_{\mu b}\hat{\bar{n}}_{\mu\bar{a}}\hat{n}_{\nu\bar{b}}\rangle_0
\end{eqnarray}
\begin{eqnarray}
i\partial_t \langle \hat{c}_{\mu a}^\dagger\hat{c}_{\nu b}\hat{\bar{n}}_{\mu\bar{a}}\hat{\bar{n}}_{\nu\bar{b}}\rangle&=&
\frac{J}{Z}\sum_{\kappa\neq \mu,\nu}T_{\mu\kappa}
\langle\hat{c}_{\kappa a}^\dagger\hat{c}_{\nu b}(\hat{n}_{\kappa\bar{a}}+\hat{\bar{n}}_{\kappa\bar{a}})
\hat{\bar{n}}_{\nu\bar{b}}\rangle\langle\hat{\bar{n}}_{\mu \bar{a}}\rangle_0\nonumber\\
& &-\frac{J}{Z}\sum_{\kappa\neq \mu,\nu}T_{\nu\kappa}
\langle\hat{c}_{\mu a}^\dagger\hat{c}_{\kappa b}\hat{\bar{n}}_{\mu\bar{a}}(\hat{n}_{\kappa\bar{b}}+\hat{\bar{n}}_{\kappa\bar{b}})
\rangle\langle\hat{\bar{n}}_{\nu \bar{b}}\rangle_0\nonumber\\
& &+(A_{\mu a}-A_{\nu b})\langle \hat{c}_{\mu a}^\dagger\hat{c}_{\nu b}\hat{\bar{n}}_{\mu\bar{a}}\hat{\bar{n}}_{\nu\bar{b}}\rangle\nonumber\\
& &+\frac{J}{Z}T_{\mu\nu}\langle \hat{c}_{\nu a}^\dagger\hat{c}_{\nu b}\hat{\bar{n}}_{\mu\bar{a}}\hat{\bar{n}}_{\nu\bar{b}}\rangle_0
-\frac{J}{Z}T_{\mu\nu}\langle \hat{c}_{\mu a}^\dagger\hat{c}_{\mu b}\hat{\bar{n}}_{\mu\bar{a}}\hat{\bar{n}}_{\nu\bar{b}}\rangle_0\,.
\end{eqnarray}
If $x_{\mu}\in A$ and $x_{\nu}\in B$ we denote the correlations 
$\langle \hat{c}_{\mu \downarrow}^\dagger\hat{c}_{\mu \downarrow}\hat{n}_{\mu\uparrow}\hat{n}_{\nu\uparrow}\rangle=f_{\mu\nu}^{1_A1_B}$,
$\langle \hat{c}_{\mu \downarrow}^\dagger\hat{c}_{\mu \downarrow}\hat{\bar{n}}_{\mu\uparrow}\hat{n}_{\nu\uparrow}\rangle=f_{\mu\nu}^{0_A1_B}$, etc.
Inserting the zeroth-order N\'eel state, 
we find four equations which fully decouple, cf.~Eq.~(\ref{hom1}),
\begin{eqnarray}
i\partial_t f^{1_A0_B}_{\mu\nu}&=&-(U-a) f^{1_A0_B}_{\mu\nu}\\
i\partial_t f^{0_B1_A}_{\mu\nu}&=&(U-
a) f^{0_B1_A}_{\mu\nu}\\
i\partial_t f^{0_B0_B}_{\mu\nu}&=&0\\
i\partial_t f^{1_A1_A}_{\mu\nu}&=&0\,.
\end{eqnarray}
In general, these four correlations are sources in the following four 
pairs of coupled equations, cf.~Eqs.~(\ref{hom2})-(\ref{hom12}), 
\begin{eqnarray}
i\partial_t f^{0_A0_B}_{\mu\nu}&=&\frac{J}{Z}\sum_{\kappa\neq{\mu,\nu}}T_{\mu\kappa}\left(f^{1_B0_B}_{\kappa\nu}+f^{0_B0_B}_{\kappa\nu}\right)+a f^{0_A0_B}_{\mu\nu}\\
i\partial_t f^{1_B0_B}_{\mu\nu}&=&\frac{J}{Z}\sum_{\kappa\neq{\mu,\nu}}T_{\mu\kappa}\left(f^{0_A0_B}_{\kappa\nu}+f^{1_A0_B}_{\kappa\nu}\right)
-Uf^{1_B0_B}_{\mu\nu}\,,\\
i\partial_t f^{0_B0_A}_{\mu\nu}&=&-\frac{J}{Z}\sum_{\kappa\neq{\mu,\nu}}T_{\kappa\nu}\left(f^{0_B1_B}_{\mu\kappa}+f^{0_B0_B}_{\mu\kappa}\right)-af^{0_B0_A}_{\mu\nu}\\
i\partial_t f^{0_B1_B}_{\mu\nu}&=&-\frac{J}{Z}\sum_{\kappa\neq{\mu,\nu}}T_{\kappa\nu}\left(f^{0_B0_A}_{\mu\kappa}+f^{0_B1_A}_{\mu\kappa}\right)
+Uf^{0_B1_B}_{\mu\nu}\,,\\
i\partial_t f^{1_B1_A}_{\mu\nu}&=&\frac{J}{Z}\sum_{\kappa\neq{\mu,\nu}}T_{\mu\kappa}\left(f^{0_A1_A}_{\kappa\nu}+f^{1_A1_A}_{\kappa\nu}\right)-af^{1_B1_A}_{\mu\nu}\\
i\partial_t f^{0_A1_A}_{\mu\nu}&=&\frac{J}{Z}\sum_{\kappa\neq{\mu,\nu}}T_{\mu\kappa}\left(f^{0_B1_A}_{\kappa\nu}+f^{1_B1_A}_{\kappa\nu}\right)
+Uf^{0_A1_A}_{\mu\nu}\,,\\
i\partial_t f^{1_A1_B}_{\mu\nu}&=&-\frac{J}{Z}\sum_{\kappa\neq{\mu,\nu}}T_{\kappa\nu}\left(f^{1_A0_A}_{\mu\kappa}+f^{1_A1_A}_{\mu\kappa}\right)+af^{1_A1_B}_{\mu\nu}\\
i\partial_t f^{1_A0_A}_{\mu\nu}&=&-\frac{J}{Z}\sum_{\kappa\neq{\mu,\nu}}T_{\kappa\nu}\left(f^{1_A0_B}_{\mu\kappa}+f^{1_A1_B}_{\mu\kappa}\right)
-Uf^{1_A0_A}_{\mu\nu}\,.
\end{eqnarray}
The eigenfrequencies of these equations read now
\begin{eqnarray}
\label{eigenmodesstaggerd}
\omega_\mathbf{k}^\pm=\frac{U+a\pm\sqrt{4J^2T_\mathbf{k}^2+(U-a)^2}}{2}\,.
\end{eqnarray}
In contrast to the eigenmodes (\ref{eigenmodes}) we see that the modes 
(\ref{eigenmodesstaggerd}) have a gap in the limit $J\rightarrow 0$.
This enables us to switch on $J$ and then switch off $a$ adiabatically
in order to have a well-defined initial state without correlations. 
Furthermore we have the four coupled equations, cf.~Eqs.~(\ref{charge1})-(\ref{charge4}),
\begin{eqnarray}
i\partial_t f^{0_A0_A}_{\mu\nu}&=&\frac{J}{Z}\sum_{\kappa\neq{\mu,\nu}}\left\{T_{\mu\kappa}\left(f^{0_B0_A}_{\kappa\nu}+f^{1_B0_A}_{\kappa\nu}\right)
-T_{\kappa\nu}\left(f^{0_A0_B}_{\mu\kappa}+f^{0_A1_B}_{\mu\kappa}\right)\right\}\\
i\partial_t f^{0_A1_B}_{\mu\nu}&=&\frac{J}{Z}\sum_{\kappa\neq{\mu,\nu}}\left\{T_{\mu\kappa}\left(f^{1_B1_B}_{\kappa\nu}+f^{0_B1_B}_{\kappa\nu}\right)
-T_{\kappa\nu}\left(f^{0_A0_A}_{\mu\kappa}+f^{0_A1_A}_{\mu\kappa}\right)\right\}\nonumber\\
& &+(U+a) f^{0_A1_B}_{\mu\nu}-\frac{J}{Z}T_{\mu\nu}\\
i\partial_t f^{1_B0_A}_{\mu\nu}&=&\frac{J}{Z}\sum_{\kappa\neq{\mu,\nu}}\left\{T_{\mu\kappa}\left(f^{0_A0_A}_{\kappa\nu}+f^{1_A0_A}_{\kappa\nu}\right)
-T_{\kappa\nu}\left(f^{1_B1_B}_{\mu\kappa}+f^{1_B0_B}_{\mu\kappa}\right)\right\}\nonumber\\
& &-(U+a) f^{1_B0_A}_{\mu\nu}+\frac{J}{Z}T_{\mu\nu}\\
i\partial_t f^{1_B1_B}_{\mu\nu}&=&\frac{J}{Z}\sum_{\kappa\neq{\mu,\nu}}\left\{T_{\mu\kappa}\left(f^{0_A1_B}_{\kappa\nu}+f^{1_A1_B}_{\kappa\nu}\right)
-T_{\kappa\nu}\left(f^{1_B0_A}_{\mu\kappa}+f^{1_B1_A}_{\mu\kappa}\right)\right\}\,.
\end{eqnarray}
The eigenfrequencies of these equations read now
\begin{eqnarray}
\omega_\mathbf{k}^\pm=\pm\sqrt{4J^2T_\mathbf{k}^2+(U+a)^2}\,.
\end{eqnarray}

%%%%%%%%%%%%%%%%%%%%%%%%%%%%%%%%%%%%%%%%%%%%%%%%%%%%%%%%%%%%%%%%%%%%%%%%%%%%%%%
\section{Appendix: Second-order Equations}\label{secondorder}
%%%%%%%%%%%%%%%%%%%%%%%%%%%%%%%%%%%%%%%%%%%%%%%%%%%%%%%%%%%%%%%%%%%%%%%%%%%%%%%

The differential equation for the three-point correlator reads
\begin{eqnarray}
i\partial_t\hat{\rho}_{\alpha\beta\gamma}^\mathrm{corr}
&=&
\hat{\mathcal{L}}_\alpha\hat{\rho}_{\alpha\beta\gamma}^\mathrm{corr}
+\frac{1}{Z}\hat{\mathcal{L}}^S_{\alpha\beta}
\left(
\hat{\rho}_\alpha\hat{\rho}^\mathrm{corr}_{\beta\gamma}
+\hat{\rho}_\beta\hat{\rho}^\mathrm{corr}_{\alpha\gamma}
\right)
\nonumber\\
& &+
\frac{1}{Z}\sum_{\kappa\neq\{\alpha,\beta,\gamma\}}
\mathrm{tr}_\kappa
\left\{
\hat{\mathcal{L}}^S_{\alpha\kappa}
\left(
\hat{\rho}^\mathrm{corr}_{\alpha\beta\gamma}\hat{\rho}_{\kappa}+
\hat{\rho}^\mathrm{corr}_{\alpha\beta}\hat{\rho}^\mathrm{corr}_{\kappa\gamma}
+\hat{\rho}^\mathrm{corr}_{\alpha\gamma}
\hat{\rho}^\mathrm{corr}_{\kappa\beta}
+\hat{\rho}_{\alpha}\hat{\rho}^\mathrm{corr}_{\kappa\beta\gamma}
\right)
\right\}
\nonumber\\
& &-
\frac{1}{Z}\hat{\rho}_{\alpha\gamma}^\mathrm{corr} 
\mathrm{tr}_\alpha 
\left\{
\hat{\mathcal{L}}^S_{\alpha\beta}
\hat{\rho}_\beta\hat{\rho}_\alpha
\right\}
-\frac{1}{Z}\hat{\rho}_{\alpha\beta}^\mathrm{corr} 
\mathrm{tr}_\alpha
\left\{
\hat{\mathcal{L}}^S_{\alpha\gamma}\hat{\rho}_\gamma\hat{\rho}_\alpha
\right\}
\nonumber\\
& &-
\frac{1}{Z}\hat{\rho}_\alpha
\mathrm{tr}_\alpha
\left\{
\hat{\mathcal{L}}_{\alpha\beta}^S 
\hat{\rho}_{\beta\gamma}^\mathrm{corr}\hat{\rho}_\alpha+
\hat{\mathcal{L}}_{\alpha\beta}^S 
\hat{\rho}_\beta\hat{\rho}_{\alpha\gamma}^\mathrm{corr}
\right\}
\nonumber\\
& &-
\frac{1}{Z}\hat{\rho}_\alpha
\mathrm{tr}_\alpha
\left\{
\hat{\mathcal{L}}_{\alpha\gamma}^S 
\hat{\rho}_{\gamma\beta}^\mathrm{corr}\hat{\rho}_\alpha+
\hat{\mathcal{L}}_{\alpha\gamma}^S 
\hat{\rho}_\gamma\hat{\rho}_{\alpha\beta}^\mathrm{corr}
\right\}
\nonumber\\
& &
+(\alpha\rightarrow\beta,\beta\rightarrow\gamma,\gamma\rightarrow\alpha)
%\nonumber\\
%& &
+(\alpha\rightarrow\gamma,\gamma\rightarrow\beta,\beta\rightarrow\alpha)
\nonumber\\
& &
+\mathcal{O}(1/Z^3)
\,.
\end{eqnarray}
In the following we use the matrix elements of
$\hat{\rho}_{\mu\nu}^\mathrm{corr}$ and $\hat{\rho}_\mu$ in order $1/Z$
and define
\begin{eqnarray}
\hat{\rho}_{\alpha\beta\gamma}^\mathrm{corr}
=
\sum_{a,a',b,b',c,c'}\rho_{\alpha\beta\gamma}^{aa'bb'cc'}
|a\rangle_\alpha\langle a'|\otimes
|b\rangle_\beta \langle b'|\otimes
|c\rangle_\gamma\langle c'|
\,.
\end{eqnarray}
All three-point correlations can be deduced by permutation and 
complex conjugation from the following set of differential equations
\begin{eqnarray}\label{rho001001}
i\partial_t \rho_{\alpha\beta\gamma}^{001001}
&=&
\frac{J}{Z}\sum_{\kappa\neq\alpha,\beta,\gamma}T_{\beta\kappa}
\left(\rho_{\alpha\kappa\gamma}^{001001}+\sqrt{2}\rho_{\alpha\kappa\gamma}^{002101}\right)
\nonumber\\
& &-\frac{J}{Z}\sum_{\kappa\neq\alpha,\beta,\gamma}T_{\gamma\kappa}
\left(
\rho_{\alpha\beta\kappa}^{001001}+\sqrt{2}\rho_{\alpha\beta\kappa}^{001012}
\right)
\nonumber\\& &
+\frac{J}{Z}\sum_{\kappa\neq\alpha,\beta,\gamma}T_{\alpha\kappa}
f^{11}_{\alpha\beta}
\left(f^{11}_{\kappa\gamma}+\sqrt{2}f^{12}_{\kappa\gamma}\right)
\nonumber\\
& &-\frac{J}{Z}\sum_{\kappa\neq\alpha,\beta,\gamma}T_{\alpha\kappa}
f^{11}_{\alpha\gamma}
\left(f^{11}_{\kappa\beta}+\sqrt{2}f^{21}_{\kappa\beta}\right)
\end{eqnarray}

\begin{eqnarray}
i\partial_t \rho_{\alpha\beta\gamma}^{001012}&=&-U\rho_{\alpha\beta\gamma}^{001012}
+\frac{J}{Z}\sum_{\kappa\neq\alpha,\beta,\gamma}T_{\beta\kappa}
\left(\rho_{\alpha\kappa\gamma}^{001012}+\sqrt{2}\rho_{\alpha\kappa\gamma}^{002112}\right)\nonumber\\
& &+\frac{J}{Z}\sum_{\kappa\neq\alpha,\beta,\gamma}T_{\gamma\kappa}
\left(\sqrt{2}\rho_{\alpha\beta\kappa}^{001001}+2\rho_{\alpha\beta\kappa}^{001012}\right)\nonumber\\& &+
\frac{J}{Z}\sum_{\kappa\neq\alpha,\beta,\gamma}T_{\alpha\kappa}f^{11}_{\alpha\beta}
\left(f^{21}_{\kappa\gamma}+\sqrt{2}f^{22}_{\kappa\gamma}\right)\nonumber\\
& &-\frac{J}{Z}\sum_{\kappa\neq\alpha,\beta,\gamma}T_{\alpha\kappa}f^{21}_{\alpha\gamma}
\left(f^{11}_{\kappa\beta}+\sqrt{2}f^{21}_{\kappa\beta}\right)+\frac{J}{Z}T_{\alpha\gamma}\sqrt{2}f^{11}_{\alpha\beta}
\end{eqnarray}

\begin{eqnarray}
i\partial_t \rho_{\alpha\beta\gamma}^{002101}&=&U\rho_{\alpha\beta\gamma}^{002101}-
 \frac{J}{Z}\sum_{\kappa\neq\alpha,\beta,\gamma}T_{\beta\kappa}
\left(\sqrt{2}\rho_{\alpha\kappa\gamma}^{001001}+2\rho_{\alpha\kappa\gamma}^{002101}\right)\nonumber\\
& &-\frac{J}{Z}\sum_{\kappa\neq\alpha,\beta,\gamma}T_{\gamma\kappa}
\left(\rho_{\alpha\beta\kappa}^{002101}+\sqrt{2}\rho_{\alpha\beta\kappa}^{002112}\right)\nonumber\\& &+
\frac{J}{Z}\sum_{\kappa\neq\alpha,\beta,\gamma}T_{\alpha\kappa}f^{12}_{\alpha\beta}
\left(f^{11}_{\kappa\gamma}+\sqrt{2}f^{12}_{\kappa\gamma}\right)\nonumber\\
& &-\frac{J}{Z}\sum_{\kappa\neq\alpha,\beta,\gamma}T_{\alpha\kappa}f^{11}_{\alpha\gamma}
\left(f^{12}_{\kappa\beta}+\sqrt{2}f^{22}_{\kappa\beta}\right)-\frac{J}{Z}T_{\alpha\beta}\sqrt{2}f^{11}_{\alpha\gamma}
\end{eqnarray}

\begin{eqnarray}\label{rho002112}
i\partial_t \rho_{\alpha\beta\gamma}^{002112}&=&-
\frac{J}{Z}\sum_{\kappa\neq\alpha,\beta,\gamma}T_{\beta\kappa}
\left(\sqrt{2}\rho_{\alpha\kappa\gamma}^{001012}+2\rho_{\alpha\kappa\gamma}^{002112}\right)\nonumber\\
& &+\frac{J}{Z}\sum_{\kappa\neq\alpha,\beta,\gamma}T_{\gamma\kappa}
\left(\sqrt{2}\rho_{\alpha\beta\kappa}^{002101}+2\rho_{\alpha\beta\kappa}^{002112}\right)\nonumber\\& &+
\frac{J}{Z}\sum_{\kappa\neq\alpha,\beta,\gamma}T_{\alpha\kappa}f^{12}_{\alpha\beta}
\left(f^{21}_{\kappa\gamma}+\sqrt{2}f^{22}_{\kappa\gamma}\right)\nonumber\\
& &-\frac{J}{Z}\sum_{\kappa\neq\alpha,\beta,\gamma}T_{\alpha\kappa}f^{21}_{\alpha\gamma}
\left(f^{12}_{\kappa\beta}+\sqrt{2}f^{22}_{\kappa\beta}\right)\nonumber\\
& &-\frac{J}{Z}T_{\alpha\beta}\sqrt{2}f^{21}_{\alpha\gamma}+\frac{J}{Z}T_{\alpha\gamma}\sqrt{2}f^{12}_{\alpha\beta}
\end{eqnarray}

\begin{eqnarray}\label{rho221001}
i\partial_t \rho_{\alpha\beta\gamma}^{221001}&=&
\frac{J}{Z}\sum_{\kappa\neq\alpha,\beta,\gamma}T_{\beta\kappa}
\left(\rho_{\alpha\kappa\gamma}^{221001}+\sqrt{2}\rho_{\alpha\kappa\gamma}^{222101}\right)\nonumber\\
& &-\frac{J}{Z}\sum_{\kappa\neq\alpha,\beta,\gamma}T_{\gamma\kappa}
\left(\rho_{\alpha\beta\kappa}^{221001}+\sqrt{2}\rho_{\alpha\beta\kappa}^{221012}\right)\nonumber\\& &-
\frac{J}{Z}\sum_{\kappa\neq\alpha,\beta,\gamma}T_{\alpha\kappa}f^{21}_{\alpha\beta}
\left(f^{11}_{\kappa\gamma}+\sqrt{2}f^{12}_{\kappa\gamma}\right)\nonumber\\
& &+\frac{J}{Z}\sum_{\kappa\neq\alpha,\beta,\gamma}T_{\alpha\kappa}\sqrt{2}f^{12}_{\alpha\gamma}
\left(f^{11}_{\kappa\beta}+\sqrt{2}f^{21}_{\kappa\beta}\right)\nonumber\\
& &-\frac{J}{Z}T_{\alpha\gamma}\sqrt{2}f^{21}_{\alpha\beta}+\frac{J}{Z}T_{\alpha\beta}\sqrt{2}f^{12}_{\alpha\gamma}
\end{eqnarray}

\begin{eqnarray}
i\partial_t \rho_{\alpha\beta\gamma}^{221012}&=&-U\rho_{\alpha\beta\gamma}^{221012}
+\frac{J}{Z}\sum_{\kappa\neq\alpha,\beta,\gamma}T_{\beta\kappa}
\left(\rho_{\alpha\kappa\gamma}^{221012}+\sqrt{2}\rho_{\alpha\kappa\gamma}^{222112}\right)\nonumber\\
& &+\frac{J}{Z}\sum_{\kappa\neq\alpha,\beta,\gamma}T_{\gamma\kappa}
\left(\sqrt{2}\rho_{\alpha\beta\kappa}^{221001}+2\rho_{\alpha\beta\kappa}^{221012}\right)\nonumber\\& &-
\frac{J}{Z}\sum_{\kappa\neq\alpha,\beta,\gamma}T_{\alpha\kappa}\sqrt{2}f^{21}_{\alpha\beta}
\left(f^{21}_{\kappa\gamma}+\sqrt{2}f^{22}_{\kappa\gamma}\right)\nonumber\\
& &+\frac{J}{Z}\sum_{\kappa\neq\alpha,\beta,\gamma}T_{\alpha\kappa}\sqrt{2}f^{22}_{\alpha\gamma}
\left(f^{11}_{\kappa\beta}+\sqrt{2}f^{21}_{\kappa\beta}\right)+\frac{J}{Z}T_{\alpha\beta}\sqrt{2}f^{22}_{\alpha\gamma}
\end{eqnarray}

\begin{eqnarray}
i\partial_t \rho_{\alpha\beta\gamma}^{222101}&=&U\rho_{\alpha\beta\gamma}^{222101}-
\frac{J}{Z}\sum_{\kappa\neq\alpha,\beta,\gamma}T_{\beta\kappa}
\left(\sqrt{2}\rho_{\alpha\kappa\gamma}^{221001}+2\rho_{\alpha\kappa\gamma}^{222101}\right)\nonumber\\
& &-\frac{J}{Z}\sum_{\kappa\neq\alpha,\beta,\gamma}T_{\gamma\kappa}
\left(\rho_{\alpha\beta\kappa}^{222101}+\sqrt{2}\rho_{\alpha\beta\kappa}^{222112}\right)\nonumber\\& &-
\frac{J}{Z}\sum_{\kappa\neq\alpha,\beta,\gamma}T_{\alpha\kappa}\sqrt{2}f^{22}_{\alpha\beta}
\left(f^{11}_{\kappa\gamma}+\sqrt{2}f^{12}_{\kappa\gamma}\right)\nonumber\\
& &+\frac{J}{Z}\sum_{\kappa\neq\alpha,\beta,\gamma}T_{\alpha\kappa}\sqrt{2}f^{12}_{\alpha\gamma}
\left(f^{12}_{\kappa\beta}+\sqrt{2}f^{22}_{\kappa\beta}\right)-\frac{J}{Z}T_{\alpha\gamma}\sqrt{2}f^{22}_{\alpha\beta}
\end{eqnarray}

\begin{eqnarray}\label{rho222112}
i\partial_t \rho_{\alpha\beta\gamma}^{222112}&=&-
\frac{J}{Z}\sum_{\kappa\neq\alpha,\beta,\gamma}T_{\beta\kappa}
\left(\sqrt{2}\rho_{\alpha\kappa\gamma}^{221012}+2\rho_{\alpha\kappa\gamma}^{222112}\right)\nonumber\\
& &+\frac{J}{Z}\sum_{\kappa\neq\alpha,\beta,\gamma}T_{\gamma\kappa}
\left(\sqrt{2}\rho_{\alpha\beta\kappa}^{222101}+2\rho_{\alpha\beta\kappa}^{222112}\right)\nonumber\\& &-
\frac{J}{Z}\sum_{\kappa\neq\alpha,\beta,\gamma}T_{\alpha\kappa}\sqrt{2}f^{22}_{\alpha\beta}
\left(f^{21}_{\kappa\gamma}+\sqrt{2}f^{22}_{\kappa\gamma}\right)\nonumber\\
& &+\frac{J}{Z}\sum_{\kappa\neq\alpha,\beta,\gamma}T_{\alpha\kappa}\sqrt{2}f^{22}_{\alpha\gamma}
\left(f^{12}_{\kappa\beta}+\sqrt{2}f^{22}_{\kappa\beta}\right)\nonumber\\
\end{eqnarray}

\begin{eqnarray}\label{rho111001}
i\partial_t \rho_{\alpha\beta\gamma}^{111001}&=&
\frac{J}{Z}\sum_{\kappa\neq\alpha,\beta,\gamma}T_{\beta\kappa}
\left(\rho_{\alpha\kappa\gamma}^{111001}+\sqrt{2}\rho_{\alpha\kappa\gamma}^{112101}\right)\nonumber\\
& &-\frac{J}{Z}\sum_{\kappa\neq\alpha,\beta,\gamma}T_{\gamma\kappa}
\left(\rho_{\alpha\beta\kappa}^{111001}+\sqrt{2}\rho_{\alpha\beta\kappa}^{111012}\right)\nonumber\\
& &-\frac{J}{Z}\sum_{\kappa\neq\alpha,\beta,\gamma}T_{\alpha\kappa}\left(f^{11}_{\alpha\beta}-\sqrt{2}f^{21}_{\alpha\beta}\right)
\left(f^{11}_{\kappa\gamma}+\sqrt{2}f^{12}_{\kappa\gamma}\right)\nonumber\\
& &+\frac{J}{Z}\sum_{\kappa\neq\alpha,\beta,\gamma}T_{\alpha\kappa}\left(f^{11}_{\alpha\gamma}-\sqrt{2}f^{12}_{\alpha\gamma}\right)
\left(f^{11}_{\kappa\beta}+\sqrt{2}f^{21}_{\kappa\beta}\right)\nonumber\\
& &+\frac{\sqrt{2}J}{Z}T_{\alpha\gamma}f^{21}_{\alpha\beta}-\frac{\sqrt{2}J}{Z}T_{\alpha\beta}f^{12}_{\alpha\gamma}
\end{eqnarray}

\begin{eqnarray}
i\partial_t \rho_{\alpha\beta\gamma}^{111012}&=&-U\rho_{\alpha\beta\gamma}^{111012}
+\frac{J}{Z}\sum_{\kappa\neq\alpha,\beta,\gamma}T_{\beta\kappa}
\left(\rho_{\alpha\kappa\gamma}^{111012}+\sqrt{2}\rho_{\alpha\kappa\gamma}^{112112}\right)\nonumber\\
& &+\frac{J}{Z}\sum_{\kappa\neq\alpha,\beta,\gamma}T_{\gamma\kappa}
\left(\sqrt{2}\rho_{\alpha\beta\kappa}^{111001}+2\rho_{\alpha\beta\kappa}^{111012}\right)\nonumber\\
& &-
\frac{J}{Z}\sum_{\kappa\neq\alpha,\beta,\gamma}T_{\alpha\kappa}
\left(f^{11}_{\alpha\beta}-\sqrt{2}f^{21}_{\alpha\beta}\right)
\left(f^{21}_{\kappa\gamma}+\sqrt{2}f^{22}_{\kappa\gamma}\right)\nonumber\\
& &+\frac{J}{Z}\sum_{\kappa\neq\alpha,\beta,\gamma}T_{\alpha\kappa}
\left(f^{21}_{\alpha\gamma}-\sqrt{2}f^{22}_{\alpha\gamma}\right)
\left(f^{11}_{\kappa\beta}+\sqrt{2}f^{21}_{\kappa\beta}\right)\nonumber\\
& &-\frac{\sqrt{2}J}{Z}T_{\alpha\gamma}f^{11}_{\alpha\beta}-\frac{\sqrt{2}J}{Z}T_{\alpha\beta}f^{22}_{\alpha\gamma}
\end{eqnarray}

\begin{eqnarray}
i\partial_t \rho_{\alpha\beta\gamma}^{112101}&=&U\rho_{\alpha\beta\gamma}^{112101}-
\frac{J}{Z}\sum_{\kappa\neq\alpha,\beta,\gamma}T_{\beta\kappa}
\left(\sqrt{2}\rho_{\alpha\kappa\gamma}^{111001}+2\rho_{\alpha\kappa\gamma}^{112101}\right)\nonumber\\
& &-\frac{J}{Z}\sum_{\kappa\neq\alpha,\beta,\gamma}T_{\gamma\kappa}
\left(\rho_{\alpha\beta\kappa}^{112101}+\sqrt{2}\rho_{\alpha\beta\kappa}^{112112}\right)\nonumber\\
& &-\frac{J}{Z}\sum_{\kappa\neq\alpha,\beta,\gamma}T_{\alpha\kappa}
\left(f^{11}_{\alpha\beta}-\sqrt{2}f^{22}_{\alpha\beta}\right)
\left(f^{11}_{\kappa\gamma}+\sqrt{2}f^{12}_{\kappa\gamma}\right)\nonumber\\
& &+\frac{J}{Z}\sum_{\kappa\neq\alpha,\beta,\gamma}T_{\alpha\kappa}
\left(f^{11}_{\alpha\gamma}-\sqrt{2}f^{12}_{\alpha\gamma}\right)
\left(f^{12}_{\kappa\beta}+\sqrt{2}f^{22}_{\kappa\beta}\right)\nonumber\\
& &+\frac{\sqrt{2}J}{Z}T_{\alpha\gamma}f^{22}_{\alpha\beta}+\frac{\sqrt{2}J}{Z}T_{\alpha\beta}f^{11}_{\alpha\gamma}
\end{eqnarray}

\begin{eqnarray}\label{rho112112}
i\partial_t \rho_{\alpha\beta\gamma}^{112112}&=&-
\frac{J}{Z}\sum_{\kappa\neq\alpha,\beta,\gamma}T_{\beta\kappa}
\left(\sqrt{2}\rho_{\alpha\kappa\gamma}^{111012}+2\rho_{\alpha\kappa\gamma}^{112112}\right)\nonumber\\
& &+\frac{J}{Z}\sum_{\kappa\neq\alpha,\beta,\gamma}T_{\gamma\kappa}
\left(\sqrt{2}\rho_{\alpha\beta\kappa}^{112101}+2\rho_{\alpha\beta\kappa}^{112112}\right)\nonumber\\
& &-\frac{J}{Z}\sum_{\kappa\neq\alpha,\beta,\gamma}T_{\alpha\kappa}
\left(f^{12}_{\alpha\beta}-\sqrt{2}f^{22}_{\alpha\beta}\right)
\left(f^{21}_{\kappa\gamma}+\sqrt{2}f^{22}_{\kappa\gamma}\right)\nonumber\\
& &+\frac{J}{Z}\sum_{\kappa\neq\alpha,\beta,\gamma}T_{\alpha\kappa}
\left(f^{21}_{\alpha\gamma}-\sqrt{2}f^{22}_{\alpha\gamma}\right)
\left(f^{12}_{\kappa\beta}+\sqrt{2}f^{22}_{\kappa\beta}\right)\nonumber\\
& &-\frac{\sqrt{2}J}{Z}T_{\alpha\gamma}f^{12}_{\alpha\beta}+\frac{\sqrt{2}J}{Z}T_{\alpha\beta}f^{21}_{\alpha\gamma}
\end{eqnarray}

\begin{eqnarray}
i\partial_t \rho_{\alpha\beta\gamma}^{200101}&=&U \rho_{\alpha\beta\gamma}^{200101}
-\frac{J}{Z}\sum_{\kappa\neq\alpha,\beta,\gamma}T_{\beta\kappa}
\left(\rho_{\alpha\kappa\gamma}^{200101}+\sqrt{2}\rho_{\alpha\kappa\gamma}^{201201}\right)\nonumber\\
& &-\frac{J}{Z}\sum_{\kappa\neq\alpha,\beta,\gamma}T_{\gamma\kappa}
\left(\rho_{\alpha\beta\kappa}^{200101}+\sqrt{2}\rho_{\alpha\beta\kappa}^{200112}\right)\nonumber\\
& &+\frac{J}{Z}\sum_{\kappa\neq\alpha,\beta,\gamma}T_{\alpha\kappa}
\left(f^{12}_{\alpha\beta}-\sqrt{2}f^{11}_{\alpha\beta}\right)
\left(f^{11}_{\kappa\gamma}+\sqrt{2}f^{12}_{\kappa\gamma}\right)\nonumber\\
& &+\frac{J}{Z}\sum_{\kappa\neq\alpha,\beta,\gamma}T_{\alpha\kappa}
\left(f^{12}_{\alpha\gamma}-\sqrt{2}f^{11}_{\alpha\gamma}\right)
\left(f^{11}_{\kappa\beta}+\sqrt{2}f^{12}_{\kappa\beta}\right)\nonumber\\
& &-\frac{\sqrt{2}J}{Z}T_{\alpha\beta}f^{11}_{\alpha\gamma}-\frac{\sqrt{2}J}{Z}T_{\alpha\gamma}f^{11}_{\alpha\beta}
\end{eqnarray}

\begin{eqnarray}
i\partial_t \rho_{\alpha\beta\gamma}^{200112}&=&
-\frac{J}{Z}\sum_{\kappa\neq\alpha,\beta,\gamma}T_{\beta\kappa}
\left(\rho_{\alpha\kappa\gamma}^{200112}+\sqrt{2}\rho_{\alpha\kappa\gamma}^{201212}\right)\nonumber\\
& &+\frac{J}{Z}\sum_{\kappa\neq\alpha,\beta,\gamma}T_{\gamma\kappa}
\left(\sqrt{2}\rho_{\alpha\beta\kappa}^{200101}+2\rho_{\alpha\beta\kappa}^{200112}\right)\nonumber\\
& &+\frac{J}{Z}\sum_{\kappa\neq\alpha,\beta,\gamma}T_{\alpha\kappa}
\left(f^{12}_{\alpha\beta}-\sqrt{2}f^{11}_{\alpha\beta}\right)
\left(f^{21}_{\kappa\gamma}+\sqrt{2}f^{22}_{\kappa\gamma}\right)\nonumber\\
& &+\frac{J}{Z}\sum_{\kappa\neq\alpha,\beta,\gamma}T_{\alpha\kappa}
\left(f^{22}_{\alpha\gamma}-\sqrt{2}f^{21}_{\alpha\gamma}\right)
\left(f^{11}_{\kappa\beta}+\sqrt{2}f^{12}_{\kappa\beta}\right)\nonumber\\
& &+\frac{\sqrt{2}J}{Z}T_{\alpha\gamma}f^{12}_{\alpha\beta}-\frac{\sqrt{2}J}{Z}T_{\alpha\beta}f^{21}_{\alpha\gamma}
\end{eqnarray}

\begin{eqnarray}
i\partial_t \rho_{\alpha\beta\gamma}^{201201}&=&
\frac{J}{Z}\sum_{\kappa\neq\alpha,\beta,\gamma}T_{\beta\kappa}
\left(\sqrt{2}\rho_{\alpha\kappa\gamma}^{200101}+2\rho_{\alpha\kappa\gamma}^{201201}\right)\nonumber\\
& &-\frac{J}{Z}\sum_{\kappa\neq\alpha,\beta,\gamma}T_{\gamma\kappa}
\left(\rho_{\alpha\beta\kappa}^{201201}+\sqrt{2}\rho_{\alpha\beta\kappa}^{201212}\right)\nonumber\\
& &+\frac{J}{Z}\sum_{\kappa\neq\alpha,\beta,\gamma}T_{\alpha\kappa}
\left(f^{22}_{\alpha\beta}-\sqrt{2}f^{21}_{\alpha\beta}\right)
\left(f^{11}_{\kappa\gamma}+\sqrt{2}f^{12}_{\kappa\gamma}\right)\nonumber\\
& &+\frac{J}{Z}\sum_{\kappa\neq\alpha,\beta,\gamma}T_{\alpha\kappa}
\left(f^{12}_{\alpha\gamma}-\sqrt{2}f^{11}_{\alpha\gamma}\right)
\left(f^{21}_{\kappa\beta}+\sqrt{2}f^{22}_{\kappa\beta}\right)\nonumber\\
& &+\frac{\sqrt{2}J}{Z}T_{\alpha\beta}f^{12}_{\alpha\gamma}-\frac{\sqrt{2}J}{Z}T_{\alpha\gamma}f^{21}_{\alpha\beta}
\end{eqnarray}

\begin{eqnarray}
i\partial_t \rho_{\alpha\beta\gamma}^{201212}&=&-U \rho_{\alpha\beta\gamma}^{201212}
+\frac{J}{Z}\sum_{\kappa\neq\alpha,\beta,\gamma}T_{\beta\kappa}
\left(\sqrt{2}\rho_{\alpha\kappa\gamma}^{200112}+2\rho_{\alpha\kappa\gamma}^{201212}\right)\nonumber\\
& &+\frac{J}{Z}\sum_{\kappa\neq\alpha,\beta,\gamma}T_{\gamma\kappa}
\left(\sqrt{2}\rho_{\alpha\beta\kappa}^{201201}+2\rho_{\alpha\beta\kappa}^{201212}\right)\nonumber\\
& &+\frac{J}{Z}\sum_{\kappa\neq\alpha,\beta,\gamma}T_{\alpha\kappa}
\left(f^{22}_{\alpha\beta}-\sqrt{2}f^{21}_{\alpha\beta}\right)
\left(f^{21}_{\kappa\gamma}+\sqrt{2}f^{22}_{\kappa\gamma}\right)\nonumber\\
& &+\frac{J}{Z}\sum_{\kappa\neq\alpha,\beta,\gamma}T_{\alpha\kappa}
\left(f^{21}_{\alpha\gamma}-\sqrt{2}f^{22}_{\alpha\gamma}\right)
\left(f^{21}_{\kappa\beta}+\sqrt{2}f^{22}_{\kappa\beta}\right)\nonumber\\
& &+\frac{\sqrt{2}J}{Z}T_{\alpha\beta}f^{22}_{\alpha\gamma}+\frac{\sqrt{2}J}{Z}T_{\alpha\gamma}f^{22}_{\alpha\beta}
\end{eqnarray}

\begin{eqnarray}
i\partial_t \rho_{\alpha\beta\gamma}^{310101}&=&3U \rho_{\alpha\beta\gamma}^{310101}
-\frac{J}{Z}\sum_{\kappa\neq\alpha,\beta,\gamma}T_{\beta\kappa}
\left(\rho_{\alpha\kappa\gamma}^{310101}+\sqrt{2}\rho_{\alpha\kappa\gamma}^{311201}\right)\nonumber\\
& &-\frac{J}{Z}\sum_{\kappa\neq\alpha,\beta,\gamma}T_{\gamma\kappa}
\left(\rho_{\alpha\beta\kappa}^{310101}+\sqrt{2}\rho_{\alpha\beta\kappa}^{310112}\right)\nonumber\\
& &-\frac{\sqrt{3}J}{Z}\sum_{\kappa\neq{\alpha,\beta,\gamma}}f^{12}_{\alpha\beta}\left(f^{11}_{\kappa\gamma}+\sqrt{2}f^{12}_{\kappa\gamma}\right)\nonumber\\
& &-\frac{\sqrt{3}J}{Z}\sum_{\kappa\neq{\alpha,\beta,\gamma}}f^{12}_{\alpha\gamma}\left(f^{11}_{\kappa\beta}+\sqrt{2}f^{12}_{\kappa\beta}\right)\nonumber\\
& &-\frac{\sqrt{3}J}{Z}T_{\alpha\beta}f^{12}_{\alpha\gamma}-\frac{\sqrt{3}J}{Z}T_{\alpha\gamma}f^{12}_{\alpha\beta}
\end{eqnarray}

\begin{eqnarray}
i\partial_t \rho_{\alpha\beta\gamma}^{310112}&=&2U \rho_{\alpha\beta\gamma}^{310112}
-\frac{J}{Z}\sum_{\kappa\neq\alpha,\beta,\gamma}T_{\beta\kappa}
\left(\rho_{\alpha\kappa\gamma}^{310112}+\sqrt{2}\rho_{\alpha\kappa\gamma}^{311212}\right)\nonumber\\
& &+\frac{J}{Z}\sum_{\kappa\neq\alpha,\beta,\gamma}T_{\gamma\kappa}
\left(\sqrt{2}\rho_{\alpha\beta\kappa}^{310101}+2\rho_{\alpha\beta\kappa}^{310112}\right)\nonumber\\
& &-\frac{\sqrt{3}J}{Z}\sum_{\kappa\neq{\alpha,\beta,\gamma}}f^{12}_{\alpha\beta}\left(f^{21}_{\kappa\gamma}+\sqrt{2}f^{22}_{\kappa\gamma}\right)\nonumber\\
& &-\frac{\sqrt{3}J}{Z}\sum_{\kappa\neq{\alpha,\beta,\gamma}}f^{22}_{\alpha\gamma}\left(f^{11}_{\kappa\beta}+\sqrt{2}f^{12}_{\kappa\beta}\right)-\frac{\sqrt{3}J}{Z}T_{\alpha\beta}f^{22}_{\alpha\gamma}
\end{eqnarray}

\begin{eqnarray}
i\partial_t \rho_{\alpha\beta\gamma}^{311201}&=&2U \rho_{\alpha\beta\gamma}^{311201}
+\frac{J}{Z}\sum_{\kappa\neq\alpha,\beta,\gamma}T_{\beta\kappa}
\left(\sqrt{2}\rho_{\alpha\kappa\gamma}^{310101}+2\rho_{\alpha\kappa\gamma}^{311201}\right)\nonumber\\
& &-\frac{J}{Z}\sum_{\kappa\neq\alpha,\beta,\gamma}T_{\gamma\kappa}
\left(\rho_{\alpha\beta\kappa}^{311201}+\sqrt{2}\rho_{\alpha\beta\kappa}^{311212}\right)\nonumber\\
& &-\frac{\sqrt{3}J}{Z}\sum_{\kappa\neq{\alpha,\beta,\gamma}}f^{22}_{\alpha\beta}\left(f^{11}_{\kappa\gamma}+\sqrt{2}f^{12}_{\kappa\gamma}\right)\nonumber\\
& &-\frac{\sqrt{3}J}{Z}\sum_{\kappa\neq{\alpha,\beta,\gamma}}f^{12}_{\alpha\gamma}\left(f^{21}_{\kappa\beta}+\sqrt{2}f^{22}_{\kappa\beta}\right)-\frac{\sqrt{3}J}{Z}T_{\alpha\gamma}f^{22}_{\alpha\beta}
\end{eqnarray}

\begin{eqnarray}
i\partial_t \rho_{\alpha\beta\gamma}^{311212}&=&U \rho_{\alpha\beta\gamma}^{311212}
+\frac{J}{Z}\sum_{\kappa\neq\alpha,\beta,\gamma}T_{\beta\kappa}
\left(\sqrt{2}\rho_{\alpha\kappa\gamma}^{310112}+2\rho_{\alpha\kappa\gamma}^{311212}\right)\nonumber\\
& &+\frac{J}{Z}\sum_{\kappa\neq\alpha,\beta,\gamma}T_{\gamma\kappa}
\left(\sqrt{2}\rho_{\alpha\beta\kappa}^{311201}+2\rho_{\alpha\beta\kappa}^{311212}\right)\nonumber\\
& &-\frac{\sqrt{3}J}{Z}\sum_{\kappa\neq{\alpha,\beta,\gamma}}f^{22}_{\alpha\beta}\left(f^{21}_{\kappa\gamma}+\sqrt{2}f^{22}_{\kappa\gamma}\right)\nonumber\\
& &-\frac{\sqrt{3}J}{Z}\sum_{\kappa\neq{\alpha,\beta,\gamma}}f^{22}_{\alpha\gamma}\left(f^{21}_{\kappa\beta}+\sqrt{2}f^{22}_{\kappa\beta}\right)
\end{eqnarray}
By separating the two-point correlations in terms of order $1/Z$ 
and of order $1/Z^2$ according to 
$\hat{\rho}_{\mu\nu}^\mathrm{corr}=
\hat{\rho}_{\mu\nu}^{\mathrm{corr}(1)}+
\hat{\rho}_{\mu\nu}^{\mathrm{corr}(2)}$
we find with
\begin{eqnarray}
\hat{\rho}_{\mu\nu}^{\mathrm{corr}(2)}
=
\sum_{m,m',n,n'}\rho_{\mu\nu}^{mm'nn'}
|m\rangle_\mu\langle m'|\otimes
|n\rangle_\nu \langle n'|
\end{eqnarray}
the following set of differential equations
\begin{eqnarray}\label{rho1001}
i\partial_t \rho_{\mu\nu}^{1001}&=&\frac{J}{Z}\sum_{\kappa\neq\mu,\nu}T_{\mu\kappa}\left(\rho_{\kappa\nu}^{1001}+\sqrt{2}\rho_{\kappa\nu}^{2101}+\sqrt{3}\rho_{\kappa\nu}^{3201}\right)\nonumber\\
& &-\frac{J}{Z}\sum_{\kappa\neq\mu,\nu}T_{\nu\kappa}\left(\rho_{\mu\kappa}^{1001}+\sqrt{2}\rho_{\mu\kappa}^{1012}+\sqrt{3}\rho_{\mu\kappa}^{1023}\right)\nonumber\\
& &-\frac{J}{Z}\sum_{\kappa\neq\mu,\nu}T_{\mu\kappa}\Big(\rho^{\mu\kappa\nu}_{001001}+\sqrt{2}\rho^{\mu\kappa\nu}_{002101}-
\rho^{\mu\kappa\nu}_{111001}\nonumber\\
& &\qquad\qquad-\sqrt{2}\rho^{\mu\kappa\nu}_{112101}+\sqrt{2}\rho^{\mu\kappa\nu}_{200101}+2\rho^{\mu\kappa\nu}_{201201}\Big)\nonumber \\
& &-\frac{J}{Z}\sum_{\kappa\neq\mu,\nu}T_{\nu\kappa}\Big(\rho^{\nu\mu\kappa}_{111001}+\sqrt{2}\rho^{\nu\mu\kappa}_{111012}-
\rho^{\nu\mu\kappa}_{001001}\nonumber\\
& &\qquad\qquad-\sqrt{2}\rho^{\nu\mu\kappa}_{001012}-\sqrt{2}\rho^{\nu\mu\kappa}_{021010}-2\rho^{\nu\mu\kappa}_{021021}\Big)\nonumber\\
& &-3f_0\frac{\sqrt{2}J}{Z}\sum_{\kappa\neq\mu,\nu}\left(T_{\nu \kappa}f^{12}_{\mu\kappa}-T_{\mu\kappa}f^{21}_{\kappa\nu}\right) 
\end{eqnarray}

\begin{eqnarray}
i\partial_t \rho_{\mu\nu}^{2112}&=&-\frac{J}{Z}\sum_{\kappa\neq\mu,\nu}T_{\mu\kappa}\left(\sqrt{2}\rho_{\kappa\nu}^{1012}+2\rho_{\kappa\nu}^{2112}+\sqrt{6}\rho_{\kappa\nu}^{3212}\right)\nonumber\\
& &+\frac{J}{Z}\sum_{\kappa\neq\mu,\nu}T_{\nu\kappa}\left(\sqrt{2}\rho_{\mu\kappa}^{2101}+2\rho_{\mu\kappa}^{2112}+\sqrt{6}\rho_{\mu\kappa}^{2123}\right)\nonumber\\
& &-\frac{J}{Z}\sum_{\kappa\neq\mu,\nu}T_{\mu\kappa}\bigg(\sqrt{2}\rho^{\mu\kappa\nu}_{111012}+2\rho^{\mu\kappa\nu}_{112112}-
\sqrt{2}\rho^{\mu\kappa\nu}_{221012}-2\rho^{\mu\kappa\nu}_{222112}\nonumber\\
& &\qquad\qquad+\sqrt{3}\rho^{\mu\kappa\nu}_{310112}+\sqrt{6}\rho^{\mu\kappa\nu}_{311212}-\rho^{\mu\kappa\nu}_{200112}-\sqrt{2}\rho^{\mu\kappa\nu}_{201212}\bigg) \nonumber\\
& &-\frac{J}{Z}\sum_{\kappa\neq\mu,\nu}T_{\nu\kappa}\bigg(\sqrt{2}\rho^{\nu\mu\kappa}_{222101}+2\rho^{\nu\mu\kappa}_{222112}-
\sqrt{2}\rho^{\nu\mu\kappa}_{112101}-2\rho^{\nu\mu\kappa}_{112112}\nonumber\\
& &\qquad\qquad+\rho^{\nu\mu\kappa}_{022110}+\sqrt{2}\rho^{\nu\mu\kappa}_{022121}-\sqrt{3}\rho^{\nu\mu\kappa}_{132110}-\sqrt{6}\rho^{\nu\mu\kappa}_{132121}\bigg)\nonumber\\ 
& &-3f_0\frac{\sqrt{2}J}{Z}\sum_{\kappa\neq\mu,\nu}\left(T_{\nu \kappa}f^{12}_{\mu\kappa}-T_{\mu\kappa}f^{21}_{\kappa\nu}\right) 
\end{eqnarray}

\begin{eqnarray}
i\partial_t \rho_{\mu\nu}^{1012}&=&-U\rho_{\mu\nu}^{1012}+\frac{J}{Z}\sum_{\kappa\neq\mu,\nu}T_{\mu\kappa}\left(\rho_{\kappa\nu}^{1012}
+\sqrt{2}\rho_{\kappa\nu}^{2112}+\sqrt{3}\rho_{\kappa\nu}^{3212}\right)\nonumber\\
& &+\frac{J}{Z}\sum_{\kappa\neq\mu,\nu}T_{\nu\kappa}\left(\sqrt{2}\rho_{\mu\kappa}^{1001}+2\rho_{\mu\kappa}^{1012}+\sqrt{6}\rho_{\mu\kappa}^{1032}\right)\nonumber\\
& &-\frac{J}{Z}\sum_{\kappa\neq\mu,\nu}T_{\mu\kappa}\Big(\rho^{\mu\kappa\nu}_{001012}+\sqrt{2}\rho^{\mu\kappa\nu}_{002112}-
\rho^{\mu\kappa\nu}_{111012}\nonumber\\
& &\qquad\qquad-\sqrt{2}\rho^{\mu\kappa\nu}_{112112}+\sqrt{2}\rho^{\mu\kappa\nu}_{200112}+2\rho^{\mu\kappa\nu}_{201212}\Big) \nonumber\\
& &-\frac{J}{Z}\sum_{\kappa\neq\mu,\nu}T_{\nu\kappa}\bigg(\sqrt{2}\rho^{\nu\mu\kappa}_{221001}+2\rho^{\nu\mu\kappa}_{221012}-
\sqrt{2}\rho^{\nu\mu\kappa}_{111001}-2\rho^{\nu\mu\kappa}_{111012}\nonumber\\
& &\qquad\qquad+\rho^{\nu\mu\kappa}_{021010}+\sqrt{2}\rho^{\nu\mu\kappa}_{021021}-\sqrt{3}\rho^{\nu\mu\kappa}_{131010}-\sqrt{6}\rho^{\nu\mu\kappa}_{131021}\bigg)\nonumber \\
& &-3f_0\frac{J}{Z}\sum_{\kappa\neq\mu,\nu}\left(T_{\nu\kappa}\left(f^{21}_{\mu\kappa}+\sqrt{2}f^{22}_{\mu\kappa}\right)
+\sqrt{2}T_{\mu\kappa}\left(f^{11}_{\kappa\nu}+\sqrt{2}f^{21}_{\kappa\nu}\right)\right)\nonumber\\
& &+4f_0\frac{\sqrt{2}J}{Z}T_{\mu\nu}
\end{eqnarray}

\begin{eqnarray}\label{rho2101}
i\partial_t \rho_{\mu\nu}^{2101}&=&U\rho_{\mu\nu}^{2101}-\frac{J}{Z}\sum_{\kappa\neq\mu,\nu}T_{\nu\kappa}\left(\rho_{\mu\kappa}^{2101}
+\sqrt{2}\rho_{\mu\kappa}^{2112}+\sqrt{3}\rho_{\mu\kappa}^{2123}\right)\nonumber\\
& &-\frac{J}{Z}\sum_{\kappa\neq\mu,\nu}T_{\mu\kappa}\left(\sqrt{2}\rho_{\kappa\nu}^{1001}+2\rho_{\kappa\nu}^{2101}+
\sqrt{6}\rho_{\kappa\nu}^{3201}\right)\nonumber\\
& &+\frac{J}{Z}\sum_{\kappa\neq\mu,\nu}T_{\nu\kappa}\Big(\rho^{\nu\mu\kappa}_{002101}+\sqrt{2}\rho^{\nu\mu\kappa}_{002112}-
\rho^{\nu\mu\kappa}_{112101}\nonumber\\
& &\qquad\qquad-\sqrt{2}\rho^{\nu\mu\kappa}_{112112}+\sqrt{2}\rho^{\nu\mu\kappa}_{022110}+2\rho^{\nu\mu\kappa}_{022121}\Big) \nonumber\\
& &+\frac{J}{Z}\sum_{\kappa\neq\mu,\nu}T_{\mu\kappa}\bigg(\sqrt{2}\rho^{\mu\kappa\nu}_{221001}+2\rho^{\mu\kappa\nu}_{222101}-
\sqrt{2}\rho^{\mu\kappa\nu}_{111001}-2\rho^{\mu\kappa\nu}_{112101}\nonumber\\
& &\qquad\qquad+\rho^{\mu\kappa\nu}_{200101}+\sqrt{2}\rho^{\mu\kappa\nu}_{201201}-\sqrt{3}\rho^{\mu\kappa\nu}_{310101}-
\sqrt{6}\rho^{\mu\kappa\nu}_{311201}\bigg)\nonumber \\
& &+3f_0\frac{J}{Z}\sum_{\kappa\neq\mu,\nu}\left(T_{\mu\kappa}\left(f^{12}_{\kappa\nu}+\sqrt{2}f^{22}_{\kappa\nu}\right)
+\sqrt{2}T_{\nu\kappa}\left(f^{11}_{\mu\kappa}+\sqrt{2}f^{12}_{\mu\kappa}\right)\right)\nonumber\\
& &-4f_0\frac{\sqrt{2}J}{Z}T_{\mu\nu}
\end{eqnarray}

\begin{eqnarray}
i\partial_t \rho^{1023}_{\mu\nu}&=&-2U\rho^{1023}_{\mu\nu}+\frac{J}{Z}\sum_{\kappa\neq\mu,\nu}T_{\mu\kappa}
\left(\rho^{1023}_{\kappa\nu}+\sqrt{2}\rho^{2123}_{\kappa\nu}\right)\nonumber\\
& &-\frac{J}{Z}\sum_{\kappa\neq\mu,\nu}T_{\nu\kappa}\left(-\sqrt{3}\rho^{\nu\mu\kappa}_{221001}-\sqrt{6}\rho^{\nu\mu\kappa}_{221012}+\sqrt{2}\rho^{\nu\mu\kappa}_{131010}
+2\rho^{\nu\mu\kappa}_{131021}\right)\nonumber\\
& &+f_0\frac{\sqrt{3}J}{Z}T_{\mu\nu}+f_0\frac{\sqrt{3}J}{Z}\sum_{\kappa\neq\mu,\nu}T_{\nu\kappa}(f^{11}_{\mu\kappa}+\sqrt{2}f^{21}_{\mu\kappa})
\end{eqnarray}

\begin{eqnarray}
i\partial_t \rho^{3201}_{\mu\nu}&=&2U\rho^{3201}_{\mu\nu}-\frac{J}{Z}\sum_{\kappa\neq\mu,\nu}T_{\nu\kappa}
\left(\rho^{3201}_{\mu\kappa}+\sqrt{2}\rho^{3212}_{\mu\kappa}\right)\nonumber\\
& &-\frac{J}{Z}\sum_{\kappa\neq\mu,\nu}T_{\mu\kappa}\left(\sqrt{3}\rho^{\mu\kappa\nu}_{221001}+\sqrt{6}\rho^{\mu\kappa\nu}_{222101}-\sqrt{2}\rho^{\mu\kappa\nu}_{310101}
-2\rho^{\mu\kappa\nu}_{311201}\right)\nonumber\\
& &-f_0\frac{\sqrt{3}J}{Z}T_{\mu\nu}-f_0\frac{\sqrt{3}J}{Z}\sum_{\kappa\neq\mu,\nu}T_{\mu\kappa}(f^{11}_{\nu\kappa}+\sqrt{2}f^{12}_{\nu\kappa})
\end{eqnarray}

\begin{eqnarray}
i\partial_t \rho^{3212}_{\mu\nu}&=&U\rho^{3212}_{\mu\nu}+\frac{J}{Z}\sum_{\kappa\neq\mu,\nu}T_{\nu\kappa}
\left(\sqrt{2}\rho^{3201}_{\mu\kappa}+2\rho^{3212}_{\mu\kappa}\right)\nonumber\\
& &-\frac{J}{Z}\sum_{\kappa\neq\mu,\nu}T_{\mu\kappa}\left(\sqrt{3}\rho^{\mu\kappa\nu}_{221012}+\sqrt{6}\rho^{\mu\kappa\nu}_{222112}-\sqrt{2}\rho^{\mu\kappa\nu}_{310112}
-2\rho^{\mu\kappa\nu}_{311212}\right)\nonumber\\
& &-f_0\frac{\sqrt{3}J}{Z}\sum_{\kappa\neq\mu,\nu}T_{\mu\kappa}(f^{21}_{\nu\kappa}+\sqrt{2}f^{22}_{\nu\kappa})
\end{eqnarray}

\begin{eqnarray}
i\partial_t \rho^{2123}_{\mu\nu}&=&-U\rho^{2123}_{\mu\nu}-\frac{J}{Z}\sum_{\kappa\neq\mu,\nu}T_{\nu\kappa}
\left(\sqrt{2}\rho^{1023}_{\kappa\mu}+2\rho^{2123}_{\kappa\mu}\right)\nonumber\\
& &-\frac{J}{Z}\sum_{\kappa\neq\mu,\nu}T_{\nu\kappa}\left(-\sqrt{3}\rho^{\nu\mu\kappa}_{222101}-\sqrt{6}\rho^{\nu\mu\kappa}_{222112}+\sqrt{2}\rho^{\nu\mu\kappa}_{132110}
+2\rho^{\nu\mu\kappa}_{132121}\right)\nonumber\\
& &+f_0\frac{\sqrt{3}J}{Z}\sum_{\kappa\neq\mu,\nu}T_{\nu\kappa}(f^{12}_{\mu\kappa}+\sqrt{2}f^{22}_{\mu\kappa})
\end{eqnarray}

\begin{eqnarray}\label{1111}\label{rho1111}
i\partial_t\rho^{1111}_{\mu\nu}&=&-\frac{J}{Z}\sum_\kappa T_{\mu\kappa}\bigg(\rho^{101101}_{\kappa\nu\mu}+
\sqrt{2}\rho^{211101}_{\kappa\nu\mu}-\sqrt{2}\rho^{101112}_{\kappa\nu\mu}-2\rho^{211112}_{\kappa\nu\mu}\nonumber\\
& &\qquad\qquad-\rho^{101101}_{\mu\nu\kappa}-\sqrt{2}\rho^{101112}_{\mu\nu\kappa}+\sqrt{2}\rho^{211101}_{\mu\nu\kappa}+2\rho^{101112}_{\mu\nu\kappa}\bigg)\nonumber\\
& &-\frac{J}{Z}\sum_\kappa T_{\nu\kappa}\bigg(\rho^{111001}_{\mu\kappa\nu}+\sqrt{2}\rho^{112101}_{\mu\kappa\nu}-\sqrt{2}\rho^{111012}_{\mu\kappa\nu}-2\rho^{112112}_{\mu\kappa\nu}\nonumber\\
& &\qquad\qquad-\rho^{111001}_{\mu\nu\kappa}-\sqrt{2}\rho^{111012}_{\mu\nu\kappa}+\sqrt{2}\rho^{112101}_{\mu\nu\kappa}+2\rho^{112112}_{\mu\nu\kappa}\bigg)\nonumber\\
& &+\frac{J}{Z}2\sqrt{2}T_{\mu\nu}(f^{12}_{\mu\nu}-f^{21}_{\mu\nu})
\end{eqnarray}
\begin{eqnarray}
i\partial_t\rho^{2222}_{\mu\nu}&=&-\frac{J}{Z}\sum_{\kappa}T_{\mu\kappa}\bigg(\sqrt{2} \rho^{102212}_{\kappa\nu\mu}+
2 \rho^{212212}_{\kappa\nu\mu}-\sqrt{2}\rho^{212201}_{\mu\nu\kappa}-2\rho^{212212}_{\mu\nu\kappa}\bigg)\nonumber\\
& &-\frac{J}{Z}\sum_{\kappa}T_{\nu\kappa}\bigg(\sqrt{2} \rho^{221012}_{\mu\kappa\nu}+
2 \rho^{222112}_{\mu\kappa\nu}-\sqrt{2}\rho^{222101}_{\mu\nu\kappa}-2\rho^{222112}_{\mu\nu\kappa}\bigg)
\end{eqnarray}
\begin{eqnarray}
i\partial_t\rho^{0000}_{\mu\nu}&=&-\frac{J}{Z}\sum_{\kappa}T_{\mu\kappa}\bigg(- \rho^{100001}_{\kappa\nu\mu}-\sqrt{2} 
\rho^{210001}_{\kappa\nu\mu}+\rho^{100001}_{\mu\nu\kappa}+\sqrt{2}\rho^{100012}_{\mu\nu\kappa}\bigg)\nonumber\\
& &-\frac{J}{Z}\sum_{\kappa}T_{\nu\kappa}\bigg(- \rho^{001001}_{\mu\kappa\nu}-\sqrt{2} \rho^{002101}_{\mu\kappa\nu}+
\rho^{001001}_{\mu\nu\kappa}+\sqrt{2}\rho^{001012}_{\mu\nu\kappa}\bigg)
\end{eqnarray}
\begin{eqnarray}
i\partial_t\rho^{0011}_{\mu\nu}&=&-\frac{J}{Z}\sum_{\kappa}T_{\mu\kappa}\bigg(- \rho^{101101}_{\kappa\nu\mu}-\sqrt{2} 
\rho^{211101}_{\kappa\nu\mu}+\rho^{101101}_{\mu\nu\kappa}+\sqrt{2}\rho^{101112}_{\mu\nu\kappa}\bigg)\nonumber\\
& &-\frac{J}{Z}\sum_{\kappa}T_{\nu\kappa}\bigg(\rho^{001001}_{\mu\kappa\nu}+\sqrt{2} \rho^{002101}_{\mu\kappa\nu}-
\sqrt{2}\rho^{001012}_{\mu\kappa\nu}-2 \rho^{002112}_{\mu\kappa\nu}\nonumber\\
& &\qquad\qquad-
\rho^{001001}_{\mu\nu\kappa}-\sqrt{2}\rho^{001012}_{\mu\nu\kappa}+\rho^{002101}_{\mu\nu\kappa}+2\rho^{002112}_{\mu\nu\kappa}\bigg)\nonumber\\
& &-\sqrt{2}\frac{J}{Z}T_{\mu\nu}(f^{12}_{\mu\nu}-f^{21}_{\mu\nu})
\end{eqnarray}
\begin{eqnarray}
i\partial_t\rho^{1122}_{\mu\nu}&=&-\frac{J}{Z}\sum_{\kappa}T_{\mu\kappa}\bigg(\rho^{102201}_{\kappa\nu\mu}+\sqrt{2} 
\rho^{212201}_{\kappa\nu\mu}-\sqrt{2}\rho^{102212}_{\kappa\nu\mu}-2\rho^{212212}_{\kappa\nu\mu}\nonumber\\
& &\qquad\qquad-\rho^{102201}_{\mu\nu\kappa}-\sqrt{2}\rho^{102212}_{\mu\nu\kappa}+\sqrt{2}\rho^{212201}_{\mu\nu\kappa}+2\rho^{212212}_{\mu\nu\kappa}\bigg)\nonumber\\
& &-\frac{J}{Z}\sum_{\kappa}T_{\nu\kappa}\bigg(\sqrt{2}\rho^{111012}_{\mu\kappa\nu}+2\rho^{112112}_{\mu\kappa\nu}
-\sqrt{2}\rho^{112101}_{\mu\nu\kappa}-2\rho^{112112}_{\mu\nu\kappa}\bigg)\nonumber\\
& &-\sqrt{2}\frac{J}{Z}T_{\mu\nu}(f^{12}_{\mu\nu}-f^{21}_{\mu\nu})
\end{eqnarray}
\begin{eqnarray}\label{rho0022}
i\partial_t\rho^{0022}_{\mu\nu}&=&-\frac{J}{Z}\sum_{\kappa}T_{\mu\kappa}\bigg(- \rho^{102201}_{\kappa\nu\mu}-\sqrt{2} 
\rho^{102201}_{\kappa\nu\mu}+\rho^{102201}_{\mu\nu\kappa}+\sqrt{2}\rho^{102212}_{\mu\nu\kappa}\bigg)\nonumber\\
& &-\frac{J}{Z}\sum_{\kappa}T_{\nu\kappa}\bigg(\sqrt{2}\rho^{001012}_{\mu\kappa\nu}+2 \rho^{002112}_{\mu\kappa\nu}-
\sqrt{2}\rho^{002101}_{\mu\nu\kappa}-2 \rho^{002112}_{\mu\nu\kappa}\bigg)\nonumber\\
& &+\sqrt{2}\frac{J}{Z}T_{\mu\nu}(f^{12}_{\mu\nu}-f^{21}_{\mu\nu})
\end{eqnarray}

\subsection{Renormalised frequencies}

The two-point correlations to first order in $1/Z$ are determined by the 
differential equations
\begin{eqnarray}
i\partial_tf_\mathbf{k}^{12}
&=&+(U-3 J T_\mathbf{k})f_\mathbf{k}^{12}
-\sqrt{2}J T_\mathbf{k}(f_\mathbf{k}^{11}+f_\mathbf{k}^{22})
+\mathrm{source\,term}
\,,\label{f_12}
\\
i\partial_tf_\mathbf{k}^{21}
&=&-(U-3 J T_\mathbf{k})f_\mathbf{k}^{21}
+\sqrt{2}J T_\mathbf{k}(f_\mathbf{k}^{11}+f_\mathbf{k}^{22})
+\mathrm{source\,term}
\,,\label{f_21}
\\
i\partial_t f_\mathbf{k}^{11}&=&i\partial_t f_\mathbf{k}^{22}
=
\sqrt{2}JT_\mathbf{k}(f^{12}_\mathbf{k}-f^{21}_\mathbf{k})
\label{f_22}
\,.
\end{eqnarray}
The $1/Z^2$-contribution of the correlations 
$f^{12}_\mathbf{k},f^{21}_\mathbf{k},f^{11}_\mathbf{k}$
and $f^{22}_\mathbf{k}$ can be deduced from (\ref{rho1001}-\ref{rho2101}).
Defining the Fourier transform
\begin{eqnarray}
\rho_{\mu\nu}^{mm'nn'}
=
\frac{1}{N}\sum_{\mathbf{k}}\rho_{\mathbf{k}}^{mm'nn'}
e^{i\mathbf{k}\cdot(\mathbf{x}_\mu-\mathbf{x}_\nu)}\,.
\end{eqnarray}
we find from equations (\ref{rho1001}-\ref{rho2101})
\begin{eqnarray}
i\partial_t \rho^{2101}_\mathbf{k}&=&+U\rho^{2101}_\mathbf{k}-3JT_\mathbf{k}
(\rho^{2101}_\mathbf{k}-3f_0f^{12}_\mathbf{k})\nonumber\\
& &-\sqrt{2}JT_\mathbf{k}\left(
\rho_\mathbf{k}^{1001}+\rho_\mathbf{k}^{2112}
-3f_0 f^{11}_\mathbf{k}-3f_0f^{22}_\mathbf{k}\right)\nonumber\\
& &+\mathrm{source\,terms}\label{f2101}\,,\\
i\partial_t \rho^{1012}_\mathbf{k}&=&-U\rho^{1012}_\mathbf{k}+3JT_\mathbf{k}
(\rho^{1012}_\mathbf{k}-3f_0f^{21}_\mathbf{k})\nonumber\\
& &+\sqrt{2}JT_\mathbf{k}\left(
\rho_\mathbf{k}^{1001}+\rho_\mathbf{k}^{2112}
-3f_0 f^{11}_\mathbf{k}-3f_0f^{22}_\mathbf{k}\right)\nonumber\\
& &+\mathrm{source\,terms}\label{f1012}\,,\\
i\partial_t \rho^{1001}_\mathbf{k}&=&i\partial_t \rho^{2112}_\mathbf{k}=\sqrt{2}J T_p\left(\rho_\mathbf{k}^{2101}
-\rho_\mathbf{k}^{1012}-3f_0 f^{12}_\mathbf{k}+3f_0f^{21}_\mathbf{k}\right)\nonumber\\
& &+\mathrm{source\,terms}\label{f1001}\,.
\end{eqnarray}
As a next step we add equations (\ref{f_12}) and (\ref{f2101}), (\ref{f_21}) and (\ref{f1012}),
(\ref{f_22}) and (\ref{f1001}) and define
\begin{eqnarray}
\tilde{\rho}_\mathbf{k}^{2101}&=&f^{12}_\mathbf{k}+\rho_\mathbf{k}^{2101}\\
\tilde{\rho}_\mathbf{k}^{1012}&=&f^{21}_\mathbf{k}+\rho_\mathbf{k}^{1012}\\
\tilde{\rho}_\mathbf{k}^{2112}&=&f^{22}_\mathbf{k}+\rho_\mathbf{k}^{2112}\\
\tilde{\rho}_\mathbf{k}^{1001}&=&f^{11}_\mathbf{k}+\rho_\mathbf{k}^{1001}\,.
\end{eqnarray}
From this follows a system of differential equations which is valid up to $\mathcal{O}(1/Z^2)$,
\begin{eqnarray}
i\partial_t\tilde{\rho}_\mathbf{k}^{2101}
&=&+\left[U-3 J T_\mathbf{k}\left(1-3f_0\right)\right]\tilde{\rho}_\mathbf{k}^{2101}
-\sqrt{2}J T_\mathbf{k}\left(1-3f_0\right)\left(\tilde{\rho}_\mathbf{k}^{1001}+\tilde{\rho}_\mathbf{k}^{2112}\right)\nonumber\\
& &+\mathrm{source\,terms}\label{2101}
\,,
\\
i\partial_t\tilde{\rho}_\mathbf{k}^{1012}
&=&-\left[U-3 J T_\mathbf{k}(1-3f_0)\right]\tilde{\rho}_\mathbf{k}^{1012}
+\sqrt{2}J T_\mathbf{k}(1-3f_0)\left(\tilde{\rho}_\mathbf{k}^{1001}+\tilde{\rho}_\mathbf{k}^{2112}\right)\nonumber\\
& &+\mathrm{source\,terms}
\,,
\\
i\partial_t \tilde{\rho}_\mathbf{k}^{1001}&=&i\partial_t \tilde{\rho}_\mathbf{k}^{2112}
=
\sqrt{2}JT_\mathbf{k}(1-3f_0)\left(\tilde{\rho}_\mathbf{k}^{2101}-\tilde{\rho}_\mathbf{k}^{1012}\right)\nonumber\\
& &+\mathrm{source\,terms}
\,.\label{1001}
\end{eqnarray}
The homogeneous part of equations (\ref{f_12})-(\ref{f_22}) is related to the homogeneous part 
of (\ref{2101})-(\ref{1001}) via the substitution $T_\mathbf{k}\rightarrow T_\mathbf{k}(1-3f_0)$ from
which follows immediately the renormalised frequency (\ref{renomega}).

\subsection{Parity-parity and particle-number correlations}

The parity-parity and the particle-number correlations are determined in 
$\mathcal{O}(1/Z^2)$ 
by the differential equations (\ref{rho1111}-\ref{rho0022}).
Since the right-hand sides of (\ref{rho1111}-\ref{rho0022}) 
involve three-point correlations, we have solve
the equations (\ref{rho001001})-(\ref{rho112112}).
The calculations can be simplified by observing that it is possible 
to express the
right-hand sides of (\ref{rho1111}-\ref{rho0022}) by total time-derivatives 
using (\ref{rho001001}), (\ref{rho002112}), (\ref{rho221001}), 
(\ref{rho222112}), 
(\ref{rho111001}), (\ref{rho112112}) and (\ref{f_12})-(\ref{f_22}).
We find the exact expressions
\begin{eqnarray}\label{rho1111ex}
\rho^{1111}_{\mu\nu}&=&-\frac{1}{N^2}\sum_{\mathbf{k,p,q}}\left(\rho_{\mathbf{kpq}}^{111001}+
\rho_{\mathbf{kpq}}^{112112}\right)\left(e^{i\mathbf{q\cdot x}_\mu+i\mathbf{p\cdot x}_\mu+i\mathbf{k\cdot x}_\nu}
+e^{i\mathbf{q\cdot x}_\nu+i\mathbf{k\cdot x}_\mu+i\mathbf{p\cdot x}_\nu}\right)\nonumber\\
& &-\frac{2}{N^2}\sum_{\mathbf{p,q}}
\left(f^{11}_\mathbf{q} f^{11}_\mathbf{p}+ f^{12}_\mathbf{q} f^{21}_\mathbf{p}\right)e^{i(\mathbf{p}+\mathbf{q})\cdot
(\mathbf{x}_\mu-\mathbf{x}_\nu)}
\end{eqnarray}
\begin{eqnarray}\label{rho2222ex}
\rho^{2222}_{\mu\nu}&=&\frac{1}{N^2}\sum_{\mathbf{k,p,q}}\rho_{\mathbf{kpq}}^{222112}
\left(e^{i\mathbf{q\cdot x}_\mu+i\mathbf{p\cdot x}_\mu+i\mathbf{k\cdot x}_\nu}+
e^{i\mathbf{q\cdot x}_\nu+i\mathbf{k\cdot x}_\mu+i\mathbf{p\cdot x}_\nu}\right)\nonumber\\
& &-\frac{1}{N^2}\sum_{\mathbf{p,q}}f^{11}_\mathbf{p}f^{11}_\mathbf{q}e^{i(\mathbf{p}+\mathbf{q})\cdot
(\mathbf{x}_\mu-\mathbf{x}_\nu)}
\end{eqnarray}
\begin{eqnarray}\label{rho0000ex}
\rho^{0000}_{\mu\nu}&=&\frac{1}{N^2}\sum_{\mathbf{k,p,q}}\rho_{\mathbf{kpq}}^{001001}
\left(e^{i\mathbf{q\cdot x}_\mu+i\mathbf{p\cdot x}_\mu+i\mathbf{k\cdot x}_\nu}
+e^{i\mathbf{q\cdot x}_\nu+i\mathbf{k\cdot x}_\mu+i\mathbf{p\cdot x}_\nu}\right)\nonumber\\
& &-\frac{1}{N^2}\sum_{\mathbf{p,q}}f^{11}_\mathbf{p}f^{11}_\mathbf{q}e^{i(\mathbf{p}+\mathbf{q})\cdot
(\mathbf{x}_\mu-\mathbf{x}_\nu)}
\end{eqnarray}
\begin{eqnarray}\label{rho0011ex}
\rho^{0011}_{\mu\nu}&=&\frac{1}{N^2}\sum_{\mathbf{k,p,q}}\rho_{\mathbf{kpq}}^{111001}
e^{i\mathbf{p\cdot x}_\mu+i\mathbf{k\cdot x}_\nu+i\mathbf{q\cdot x}_\mu}
\nonumber\\& &-\frac{1}{N^2}\sum_{\mathbf{k,p,q}}\left(\rho_{\mathbf{kpq}}^{001001}+\rho_{\mathbf{kpq}}^{002112}\right)
e^{i\mathbf{q\cdot x}_\nu+i\mathbf{k\cdot x}_\mu+i\mathbf{p\cdot x}_\nu}\nonumber\\
& &+\frac{1}{N^2}\sum_{\mathbf{p,q}}\left(f^{11}_\mathbf{p}f^{11}_\mathbf{q}+f^{12}_\mathbf{p} f^{21}_\mathbf{q}\right)e^{i(\mathbf{p}+\mathbf{q})
\cdot(\mathbf{x}_\mu-\mathbf{x}_\nu)}
\end{eqnarray}
\begin{eqnarray}\label{rho1122ex}
\rho^{1122}_{\mu\nu}&=&-\frac{1}{N^2}\sum_{\mathbf{k,p,q}}\left(\rho_{\mathbf{kpq}}^{221001}+\rho_{\mathbf{kpq}}^{222112}\right)
e^{i\mathbf{p\cdot x}_\mu+i\mathbf{k\cdot x}_\nu+i\mathbf{q\cdot x}_\mu}\nonumber\\
& &+\frac{1}{N^2}\sum_{\mathbf{k,p,q}}\rho_{\mathbf{kpq}}^{112112}e^{i\mathbf{q\cdot x}_\nu+i\mathbf{k\cdot x}_\mu+i\mathbf{p\cdot x}_\nu}
\end{eqnarray}
\begin{eqnarray}\label{rho0022ex}
\rho^{0022}_{\mu\nu}&=&\frac{1}{N^2}\sum_{\mathbf{k,p,q}}\rho_{\mathbf{kpq}}^{221001}e^{i\mathbf{p\cdot x}_\mu+i\mathbf{k\cdot x}_\nu+i\mathbf{q\cdot x}_\mu}
+\frac{1}{N^2}\sum_{\mathbf{k,p,q}}\rho_{\mathbf{kpq}}^{002112}e^{i\mathbf{q\cdot x}_\nu+i\mathbf{k\cdot x}_\mu+i\mathbf{p\cdot x}_\nu}\nonumber\\
& &-\frac{1}{N^2}\sum_{\mathbf{p,q}}f^{12}_\mathbf{p} f^{21}_\mathbf{q} e^{i(\mathbf{p}+\mathbf{q})\cdot(\mathbf{x}_\mu-\mathbf{x}_\nu)}\,,
\end{eqnarray}
where we defined the Fourier transforms
\begin{eqnarray}
\rho^{aa'bb'cc'}_{\alpha\beta\gamma}=\frac{1}{N^2}\sum_{\mathbf{k,p,q}}
\rho^{aa'bb'cc'}_{\mathbf{kpq}}e^{i\mathbf{k\cdot x}_\alpha+i\mathbf{p\cdot x}_\beta+i\mathbf{q\cdot x}_\gamma}\,.
\end{eqnarray}
After solving the differential equations for the three-point-correlations
and inserting the solutions in (\ref{rho1111ex})-(\ref{rho0022ex})
we find the parity-parity and particle-number correlations which are given
in Section \ref{Z2}.

%%%%%%%%%%%%%%%%%%%%%%%%%%%%%%%%%%%%%%%%%%%%%%%%%%%%%%%%%%%%%%%%%%%%%%%%%%%%%%%
\section*{References}
%%%%%%%%%%%%%%%%%%%%%%%%%%%%%%%%%%%%%%%%%%%%%%%%%%%%%%%%%%%%%%%%%%%%%%%%%%%%%%%


\begin{thebibliography}{99}
\bibitem{H63}
J.~Hubbard, Proc.~Roy.~Soc.~A {\bf 276}, 238 (1963).
\bibitem{H64a}
J.~Hubbard, Proc.~R.~Soc.~Lond.~A, {\bf 277}, 237 (1964).
\bibitem{H64b}
J.~Hubbard, Proc.~R.~Soc.~Lond.~A, {\bf 281}, 401 (1964). 
\bibitem{T98}
H.~Tasaki, J.~Phys.: Condens.~Matter {\bf 10}, 4353 (1998)
%Boson localization and the superfluid-insulator transition
\bibitem{KMS05}
M.~K\"ohl, H.~Moritz, T.~St\"oferle, K.~G\"unter, and T.~Esslinger, Phys.~Rev.~Lett. {\bf 94}, 080403 (2005).
\bibitem{JSG08}
R.~J\"ordens, N.~Strohmaier, K.~G\"unter, H.~Moritz, T.~Esslinger, Nature {\bf 455}, 204 (2008).
\bibitem{SHW08}
U.~Schneider, L.~Hackerm\"uller, S.~Will, T.~Best, I.~Bloch, T.~A.~Costi, R.~W.~Helmes, D.~Rasch,~A.~Rosch,
Science, {\bf 322}, 1520 (2008).
\bibitem{FWGF89}
M.~P.~A.~Fisher, P.~B.~Weichman, G.~Grinstein, and D.~S.~Fisher,
Phys.~Rev.~B {\bf 40}, 546 (1989).
\bibitem{G02} M.~Greiner, O.~Mandel, T.~Esslinger, T.~W.~H\"ansch, and I.~Bloch, Nature {\bf 415}, 39 (2002).
%Quantum phase transition from a superfluid to a Mott insulator 
%in a gas of ultracold atoms
\bibitem{B10}W.~S.~Bakr, A.~Peng, M.~E.~Tai, R.~Ma1, J.~Simon, J.~I.~Gillen, S.~F\"olling, L.~Pollet, and M.~Greiner, Science {\bf 329}, 547 (2010).
%Probing the superfluid-to-Mott insulator transition at the single-atom level
\bibitem{LW68}
E.~H.~Lieb and F.~Y.~Wu, Phys.~Rev.~Lett.~{\bf 20}, 1445 (1968).
\bibitem{EFGKK05}
F.~H.~L.~Essler, H.~Frahm, F.~G\"ohmann, A.~Kl\"umper, V.~E.~Korepin, \textit{The One-Dimensional Hubbard Model},
Cambridge University Press (2005).
\bibitem{AP98}
L.~Amico and V.~Penna, Phys.~Rev.~Lett. {\bf 80}, 2189 (1998).  
\bibitem{GK96}
A.~Georges, G.~Kotliar, W.~Krauth and M.~J.~Rozenberg, Rev.~Mod.~Phys.~{\bf 68}, 13 (1996). 
\bibitem{LHS88}
H.~Q.~Lin, J.~E.~Hirsch, and D.~J.~Scalapino, Phys.~Rev.~B {\bf37}, 7359 (1988). 
\bibitem{KLA07}
C.~Kollath, A.~M.~L\"auchli, and E.~Altman, Phys.~Rev.~Lett.~{\bf 98}, 180601 (2007).  
\bibitem{SBG91}
R.~T.~Scalettar, G.~G.~Batrouni, and G.~T.~Zimanyi, Phys.~Rev.~Lett.~{\bf 66}, 3144 (1991).
\bibitem{G63}
M.~C.~Gutzwiller, 
Phys.~Rev.~Lett.~{\bf 10}, 159 (1963).
%Effect of Correlation on the Ferromagnetism of Transition Metals
\bibitem{CBPE12}
M.~Cheneau, P.~Barmettler,
D.~Poletti, M.~Endres, P.~Schau\ss,
T.~Fukuhara, C.~Gross, I.~Bloch,
C.~Kollath, and S.~Kuhr, Nature {\bf 481}, 484 (2012).
%Light-cone-like spreading of correlations in a quantum many-body system
\bibitem{E11}
M.~Endres, M.~Cheneau, T.~Fukuhara, C.~Weitenberg, 
P.~Schau\ss, C.~Gross, L.~Mazza, M.~C.~Ba\~nuls, L.~Pollet, I.~Bloch, S.~Kuhr, Science {\bf 334} 200 (2011).
%Observation of Correlated Particle-Hole Pairs and String Order in Low-Dimensional Mott Insulators
%fermi hubbard model
\bibitem{J98}
D.~Jaksch, C.~Bruder, J.~I.~Cirac, C.~W.~Gardiner, and P.~Zoller, Phys. Rev. Lett. {\bf 81}, 3108 (1998).
%Cold Bosonic Atoms in Optical Lattices
\bibitem{Z03}W.~Zwerger,J. Opt. B: Quantum Semiclass. Opt {\bf 5}, S9 (2003).
%Mott–Hubbard transition of cold atoms in optical lattices 
\bibitem{sachdev} 
S.~Sachdev, \textit{Quantum phase transitions},
(Cambridge University Press, Cambridge, England, 2001).
\bibitem{B05}I.~Bloch, Nature Physics, {\bf 1}, 23 (2005).
%Ultracold quantum gases in optical lattices
\bibitem{S07}C.~Sias, A.~Zenesini, H.~Lignier, S.~Wimberger, D.~Ciampini, O.~Morsch, and E.~Arimondo, Phys.~Rev.~Lett. {\bf 98}, 120403 (2007).
%Resonantly Enhanced Tunneling of Bose-Einstein Condensates 
%in Periodic Potentials
\bibitem{RSN97}M.~Raizen, C.~Salomon, and Q.~Niu, Physics Today, {\bf 50}, 30 (1997).
%New Light on Quantum Transport
\bibitem{SUXF06}
R.~Sch\"utzhold, M.~Uhlmann, Y.~Xu and U.~R.~Fischer,
Phys.~Rev.~Lett.~{\bf 97}, 200601 (2006).
\bibitem{FSU08}
U.~R.~Fischer, R.~Sch\"utzhold, M.~Uhlmann, Phys.~Rev.~A {\bf 77}, 043615
(2008).
\bibitem{FM94}
J.~K.~Freericks and H.~Monien, Europhys.~Lett.~{\bf 26} 545, (1994).
%Phase diagram of the Bose-Hubbard Model
\bibitem{FM96}
J.~K.~Freericks and H.~Monien,
Phys.~Rev.~B {\bf 53}, 2691 (1996).
%Strong-coupling expansions for the pure and disordered Bose-Hubbard model
\bibitem{DZ06}
B.~Damski and J.~Zakrzewski
Phys.~Rev.~A {\bf 74}, 043609 (2006).
\bibitem{HSH09}
A.~Hubener, M.~Snoek, and W.~Hofstetter, Phys.~Rev.~B {\bf 80}, 245109 (2009).
%Magnetic phases of two-component ultracold bosons in an optical lattice
\bibitem{LBHH11}
Y.~Li, M.~R.~Bakhtiari, L.~He, W.~Hofstetter, Phys. Rev. B {\bf 84}, 144411 (2011)
%Tunable anisotropic magnetism in trapped two-component Bose gases
\bibitem{LBHH12}
Y.~Li, M.~R.~Bakhtiari, L.~He, W.~Hofstetter, Phys. Rev. A {\bf 85}, 023624 (2012) 
%Pomeranchuk effect and spin-gradient cooling of Bose-Bose mixtures in an optical lattice
\bibitem{RK91}
D.~S.~Rokhsar, and B.~G.~Kotliar,  
Phys.~Rev.~B {\bf 44}, 10328 (1991).
%Gutzwiller projection for bosons
\bibitem{CKMR07}
M.~Christandl, R.~Koenig, G.~Mitchison,
R.~Renner, Comm.~Math.~Phys., {\bf 273}, 473 (2007).
\bibitem{K62}
R.~Kubo, J.~Phys.~Soc.~Japan {\bf 17}, 1100 (1962).
\bibitem{B75}
R.~Balescu, 
\textit{Equilibrium and Nonequilibrium Statistical Mechanics} 
(Wiley, New York, 1975).
\bibitem{BPCK12}
P.~Barmettler, D.~Poletti, M.~Cheneau, and C.~Kollath, arXiv:1202.5558v1 (2012).
%Propagation front of correlations in an interacting Bose gas
\bibitem{KJSW90} H.~R.~Krishnamurthy, C.~Jayaprakash, S.~Sarker, and W.~Wenzel  
Phys. Rev. Lett. {\bf 64}, 950 (1990).
%Mott-Hubbard metal-insulator transition in nonbipartite lattices
%{\bf dieelectrical breakdown}
\bibitem{KN11}
K.~V.~Krutitsky, and P.~Navez,
Phys.~Rev.~A {\bf 84}, 033602 (2011).
%Excitation dynamics in a lattice Bose gas within the time-dependent Gutzwiller mean-field approach
\bibitem{NS10}
P.~Navez, R.~Sch\"utzhold, Phys. Rev. A
{\bf 82},  063603 (2010).
\bibitem{S11}
M.~Snoek, EPL {\bf 95}, 30006 (2011). 

\bibitem{G01}
M.~Greiner, I.~Bloch, O.~Mandel, T.~W.~H\"ansch, and T.~Esslinger, Phys.~Rev.~Lett.~{\bf 87}, 160405 (2001).
%Exploring Phase Coherence in a 2D Lattice of Bose-Einstein Condensates
%{\bf fermi hubbard}
\bibitem{KWE11}
M.~Kollar, F.~Alexander Wolf, and M.~Eckstein, Phys. Rev. B {\bf 84}, 054304 (2011).
%Generalized Gibbs ensemble prediction of prethermalization plateaus and their relation to nonthermal steady states in integrable systems
\bibitem{GME11}
C.~Gogolin, M.~P.~M\"uller, and J.~Eisert, Phys.~Rev.~Lett.~{\bf 106}, 040401 (2011).
%Absence of Thermalization in Nonintegrable Systems
\bibitem{EHKKMWW}
M.~Eckstein, A.~Hackl, S.~Kehrein, M.~Kollar, M.~Moeckel, P.~Werner, F.~A.~Wolf, Eur. Phys. J. Spec.~Top.~{\bf 180}, 217 (2009).
%New theoretical approaches for correlated systems in nonequilibrium 
\bibitem{KISD11}
T.~Kitagawa, A.~Imambekov, J.~Schmiedmayer, and E.~Demler, New J. Phys. {\bf 13}, 073018 (2011).
%The dynamics and prethermalization of one-dimensional quantum systems probed through the full distributions of quantum noise
\bibitem{BCH11}
M.~C.~Ba$\tilde{\mathrm{n}}$uls, J.~I.~Cirac, and M.~B.~Hastings, Phys.~Rev.~Lett.~{\bf 106}, 050405 (2011).
%Strong and Weak Thermalization of Infinite Nonintegrable Quantum Systems
%%%%%%fermi gas
\bibitem{MK09}
M.~Moeckel, and S.~Kehrein, Ann.~Phys, {\bf 324}, 2146 (2009).
\bibitem{CFMSE08}
M.~Cramer et~al., Phys.~Rev.~Lett.~{\bf 101}, 063001 (2008).
%Exploring Local Quantum Many-Body Relaxation by Atoms in Optical Superlattices
\bibitem{FCMSE10}
A.~Flesch et~al., Phys. Rev. A {\bf 78}, 033608 (2008).
%Probing local relaxation of cold atoms in optical superlattices
\bibitem{CDEO08}
M.~Cramer, C.~M.~Dawson, J.~Eisert, and T.~J.~Osborne, Phys. Rev. Lett. {\bf 100}, 030602 (2008).
%Exact Relaxation in a Class of Nonequilibrium Quantum Lattice Systems
%\bibitem{SM02}
%P.~Schmidt, and H.~Monien, arXiv:cond-mat/0202046v1
%Nonequilibrium dynamical mean-field theory of a strongly correlated system
\bibitem{U09}
G.~S.~Uhrig, Phys. Rev. A {\bf 80}, 061602 (2009).
%Interaction quenches of Fermi gases
%%%%bose hubbard
\bibitem{MK08}
M.~Moeckel, and S.~Kehrein, Phys.~Rev.~Lett.~{\bf 100}, 175702 (2008).
%Interaction Quench in the Hubbard Model
\bibitem{EKW10}
M.~Eckstein, M.~Kollar, and P.~Werner, Phys.~Rev.~B, {\bf 81}, 115131 (2010).
%Interaction quench in the Hubbard model: Relaxation of the spectral function and the optical conductivity\\
\bibitem{MK10}
M.~Moeckel, and S.~Kehrein, New J. Phys. {\bf 12}, 055016 (2010).
%Crossover from adiabatic to sudden interaction quenches in the Hubbard model: prethermalization and non-equilibrium dynamics 
\bibitem{SGJ10}
N.~Strohmaier, D.~Greif, R.~J\"ordens, L.~Tarruell, H.~Moritz, and T.~Esslinger, Phys.~Rev.~Lett.~{\bf 104}, 080401 (2010).
%Observation of Elastic Doublon Decay in the Fermi-Hubbard Model
\bibitem{EKW09}
M.~Eckstein, M.~Kollar, and P.~Werner, Phys.~Rev.~Lett.~{\bf 103}, 056403 (2009).
%Thermalization after an Interaction Quench in the Hubbard Model
%%%%%general discussion
\bibitem{M01}
O.~Morsch, J.~H.~M\"uller, M.~Cristiani, D.~Ciampini, and E.~Arimondo, 
Phys.~Rev.~Lett.~{\bf 87}, 140402 (2001). 
\bibitem{M02}
M.~Cristiani, O.~Morsch, J.`H.~M\"uller, D.~Ciampini, and E.~Arimondo,
Phys.~Rev.~A {\bf 65}, 063612 (2002).
\bibitem{D96}
M.~B.~Dahan, E.~Peik, J.~Reichel, Y.~Castin, and C.~Salomon, Phys.~Rev.~Lett.~{\bf 76}, 4508 (1996).
%{\bf Bloch oscillations, theory}
\bibitem{QNS11} F.~Queisser, P.~Navez, and R.~Sch\"utzhold, Phys.~Rev.~A {\bf 85}, 033625 (2012).
\bibitem{WWMK05} D.~Witthaut, M.~Werder, S.~Mossmann, and H.~J.~Korsch, 
Phys.~Rev.~E {\bf 71}, 036625 (2005).
%Bloch oscillations of Bose-Einstein condensates: Breakdown and revival
\bibitem{SSG02}S.~Sachdev, K.~Sengupta, and S.~M.~Girvin, 
Phys.~Rev.~B {\bf 66}, 075128 (2002).
%Mott insulators in strong electric fields
\bibitem{W05}S.~Wimberger, R.~Mannella, O.~Morsch, E.~Arimondo, A.~R.~Kolovsky, and A.~Buchleitner, Phys.~Rev.~A {\bf 72}, 063610 (2005).
%Nonlinearity-induced destruction of resonant tunneling 
%in the Wannier-Stark problem
\bibitem{KK03}
A.~R.~Kolovsky and H.~J.~Korsch, Phys.~Rev.~A {\bf 67}, 063601 (2003).
%Bloch oscillations of cold atoms in two-dimensional optical lattices
\bibitem{KK04}
A.~R.~Kolovsky and H.~J.~Korsch, Int.~J.~Mod.~Phys.~B {\bf 18}, 1235 (2004).
%Bloch oscillations of cold atoms in optical lattices
\bibitem{K04} A.~R.~Kolovsky, Phys.~Rev.~A {\bf 70}, 015604 (2004).
%Bloch oscillations in the Mott-insulator regime
\bibitem{K03} A.~R.~Kolovsky, Phys.~Rev.~Lett. {\bf 90}, 213002 (2003).
%New Bloch Period for Interacting Cold Atoms in 1D Optical Lattices
\bibitem{KB03} A.~R.~Kolovsky and A.~Buchleitner, Phys.~Rev.~E {\bf 68}, 056213 (2003).
%Floquet-Bloch operator for the Bose-Hubbard model with static field
\bibitem{KKG09}A.~R.~Kolovsky, H.~J.~Korsch and E.-M.~Graefe, 
Phys.~Rev.~A {\bf 80} 023617 (2009).
%Bloch oscillations of Bose-Einstein condensates: 
%Quantum counterpart of dynamical instability
\bibitem{CN99}D.-I.~Choi and Q.~Niu, Phys.~Rev.~Lett. {\bf 82}, 2022 (1999).
%Bose-Einstein Condensates in an Optical Lattice
%{\bf experiment}
\bibitem{M12}R.~Ma, M.~E.~Tai, P.~M.~Preiss, W.~S.~Bakr, J.~Simon, and M.~Greiner, Phys.~Rev.~Lett.~{\bf 107}, 095301 (2011).
%Photon-Assisted Tunneling in a Biased, Strongly Correlated Bose Gas
\bibitem{Si11}J.~Simon, W.~S.~Bakr, R.~Ma, M.~E.~Tai, P.~M.~Preiss, and M.~Greiner, Nature {\bf 472}, 307 (2011). 
%Quantum simulation of antiferromagnetic spin chains in an optical lattice
%sauter schwinger
\bibitem{S31}
F.~Sauter, Zeits f.~Physik {\bf 69}, 742 (1931).
\bibitem{S51}
J.~Schwinger, Phys.~Rev.~82, {\bf 664} (1951).
\bibitem{K65}
L.~V.~Keldysh, Sov.~Phys.~JETP {\bf 02}, 1307 (1965).
\bibitem{BI70}
E.~Brezin and C.~Itzykson, Phys.~Rev.~D {\bf2}, 1191 (1970).
\bibitem{D09}
G.~V.~Dunne, Eur.~Phys.~J.~D, {\bf 55}, 327 (2009).
\bibitem{SGD08}
R.~Sch\"utzhold, H.~Gies, and G.~Dunne, Phys.~Rev.~Lett.~{\bf 101}, 130404 (2008).
\bibitem{KLY08}
S.~P.~Kim, H.~K.~Lee, and Y.~Yoon, Phys.~Rev.~D {\bf 78}, 105013 (2008).
%Effective action of QED in electric field backgrounds
\bibitem{GG96}
S.~P.~Gavrilov and D.~M.~Gitman, Phys.~Rev.~D {\bf 53}, 7162 (1996).
\bibitem{W09}P.~W\"urtz, T.~Langen, T.~Gericke, A.~Koglbauer, and H.~Ott, Phys.~Rev.~Lett. {\bf 103}, 080404 (2009).
%Experimental Demonstration of Single-Site Addressability in a 
%Two-Dimensional Optical Lattice
\bibitem{GZHC09}N.~Gemelke, X.~Zhang, C.-L.~Hung, and C.~Chin, Nature {\bf 460}, 995 (2009).
%In situ observation of incompressible Mott-insulating domains in
%ultracold atomic gases
\bibitem{S10}J.~F.~Sherson, C.~Weitenberg, M.~Endres, M.~Cheneau, I.~Bloch, S.~Kuhr, Nature {\bf 467}, 68 (2010).
%Single-atom-resolved fluorescence imaging of an atomic Mott insulator
\bibitem{EKO}
V.~F.~Elesin, V.~A.~Kashurnikov, and L.~A.~Openov,
JETP~Lett. {\bf 60}, 177 (1994).
\bibitem{KPS}
V.~A.~Kashurnikov, A.~I.~Podlivaev, and B.~V.~Svistunov,
Pis'ma~Zh.~Eksp.~Teor.~Fiz. {\bf 61}, 375 (1995)
[JETP~Lett. {\bf 61}, 381 (1995)].
\bibitem{KPPS}
V.~A.~Kashurnikov, A.~I.~Podlivaev, N.~V.~Prokof'ev, and B.~V.~Svistunov,
Phys.~Rev.~B {\bf 53}, 13091 (1996).
\bibitem{KS}
V.~A.~Kashurnikov and B.~V.~Svistunov,
Phys.~Rev.~B {\bf 53}, 11776 (1996).
\bibitem{RB}
R.~Roth and K.~Burnett,
Phys.~Rev.~A {\bf 67}, 031602(R) (2003);
{\it ibid}. {\bf 68}, 023604 (2003).
\bibitem{RB04}
R.~Roth and K.~Burnett,
J.~Phys.~B {\bf 37}, 3893 (2004).
\bibitem{KT08}
K.~V.~Krutitsky, M.~Thorwart, R.~Egger, and R.~Graham, Phys.~Rev.~A {\bf 77}, 053609 (2008).
\bibitem{HSTR}
M.~Hild, F.~Schmitt, I.~T\"urschmann, and R.~Roth,
Phys.~Rev.~A {\bf 76}, 053614 (2007).
\bibitem{ZD}
J.~M.~Zhang and R.~X.~Dong,
Eur.~J.~Phys. {\bf 31}, 591 (2010).
\bibitem{F91}
E.~Fradkin, \textit{Field Theories Of Condensed Matter Systems}, Addison-Wesley (1991).
%%%%Z scaling
\bibitem{MV89}
W.~Metzner and D.~Vollhardt, Phys.~Rev.~Lett.~{\bf 62}, 324 (1989).
\bibitem{F99}
P.~Fazekas, \textit{Lecture Notes On electron Correlation and Magnetism}, World Scientific (1999).
\bibitem{A94}
A.~Auerbach, \textit{Interacting Electrons And Quantum Magnetism}, Springer (1994).
%{\bf Bose Hubbard model}
\bibitem{EOW10}M.~Eckstein, T.~Oka, and P.~Werner, Phys.~Rev.~Lett. {\bf 105}, 146404 (2010).
%Dielectric Breakdown of Mott Insulators in Dynamical Mean-Field Theory
\bibitem{OAA03}T.~Oka, R.~Arita, and H.~Aoki, Phys.~Rev.~Lett. {\bf 91}, 066406 (2003).
%Breakdown of a Mott Insulator: A Nonadiabatic Tunneling Mechanism
\bibitem{OA10}T.~Oka and H.~Aoki, Phys.~Rev.~B {\bf 81}, 033103 (2010).
%Dielectric breakdown in a Mott insulator
%{\bf thermalization}
%%%%factor of two
%{\bf  Floquet theory }
\bibitem{Floquet}
Z.~X.~Wang and D.~R.~Guo, {\it Special Functions} (World Scientific, 1989).

\end{thebibliography}
\end{document}